\documentclass[fleqn,usenatbib]{mnras}

\usepackage[T1]{fontenc}
\usepackage{ae,aecompl}
\usepackage{threeparttable}

\usepackage{graphicx}
\graphicspath{{figures/}}
\usepackage{amsmath}
\usepackage{amssymb}
\usepackage{newtxtext,newtxmath}
\usepackage{mathtools}
\usepackage{pdflscape}	
\usepackage{subcaption}
\usepackage{booktabs}
\usepackage[dvipsnames]{xcolor}
\usepackage{longtable}

\title[Impact of glitches on young pulsar rotation]{The impact of glitches on young pulsar rotational evolution}

\author[M.~E.~Lower et al.]{\parbox{\textwidth}{M.~E.~Lower,$^{1,2}$\thanks{E-mail: mlower@swin.edu.au}
S.~Johnston,$^{2}$\thanks{E-mail: simon.johnston@csiro.au}
L.~Dunn,$^{3,4}$
R.~M.~Shannon,$^{1,5}$
M.~Bailes,$^{1,5}$
S.~Dai,$^{6,2}$
M.~Kerr,$^{7}$
R.~N.~Manchester,$^{2}$
A.~Melatos,$^{3,4}$
L.~S.~Oswald,$^{8,9}$
A.~Parthasarathy,$^{10}$
C.~Sobey,$^{11}$
and P.~Weltevrede$^{12}$
}
\\ \\
$^{1}$Centre for Astrophysics and Supercomputing, Swinburne University of Technology, PO Box 218, Hawthorn, VIC 3122, Australia\\
$^{2}$Australia Telescope National Facility, CSIRO, Space and Astronomy, PO Box 76, Epping, NSW 1710, Australia\\
$^{3}$School of Physics, University of Melbourne, Parkville, VIC 3010, Australia\\ 
$^{4}$OzGrav: The ARC Centre of Excellence for Gravitational-wave Discovery, Parkville, VIC 3010, Australia\\
$^{5}$OzGrav: The ARC Centre of Excellence for Gravitational-wave Discovery, Hawthorn VIC 3122, Australia\\
$^{6}$Western Sydney University, Locked Bag 1797, Penrith South DC, NSW 1797, Australia\\
$^{7}$Space Science Division, Naval Research Laboratory, Washington, DC 20375, USA\\
$^{8}$Department of Astrophysics, University of Oxford, Denys Wilkinson Building, Keble Road, Oxford OX1 3RH, UK\\
$^{9}$Magdalen College, University of Oxford, Oxford OX1 4AU, UK\\
$^{10}$Max-Planck-Institut f{\"u}r Radioastronomie, Auf dem H{\"u}gel 69, D-53121 Bonn, Germany\\
$^{11}$CSIRO, Space and Astronomy, PO Box 1130 Bentley, WA 6102, Australia\\
$^{12}$Jodrell Bank Centre for Astrophysics, The University of Manchester, Alan Turing Building, Manchester, M13 9PL, United Kingdom
}

\date{Accepted XXXX. Received YYYY; in original form ZZZZ}

\begin{document}
\label{firstpage}
\pagerange{\pageref{firstpage}--\pageref{lastpage}}
\maketitle

\begin{abstract}
We report on a timing programme of 74 young pulsars that have been observed by the Parkes 64-m radio telescope over the past decade. Using modern Bayesian timing techniques, we have measured the properties of 124 glitches in 52 of these pulsars, of which 74 are new. We demonstrate that the glitch sample is complete to fractional increases in spin-frequency greater than $\Delta\nu^{90\%}_{g}/\nu \approx 8.1 \times 10^{-9}$. We measure values of the braking index, $n$, in 33 pulsars. In most of these pulsars, their rotational evolution is dominated by episodes of spin-down with $n > 10$, punctuated by step changes in the spin-down rate at the time of a large glitch. The step changes are such that, averaged over the glitches, the long-term $n$ is small. We find a near one-to-one relationship between the inter-glitch value of $n$ and the change in spin-down of the previous glitch divided by the inter-glitch time interval. We discuss the results in the context of a range of physical models.
\end{abstract}



\section{Introduction} \label{sec:intro}

Pulsars are highly magnetised, rotating neutron stars that are observed as sources of electromagnetic pulses with a periodicity that matches their rotation rates.
The spin frequency, $\nu$, of a pulsar decreases over time as it loses rotational kinetic energy through various processes. 
This phenomenon can be described in terms of $\nu$ and the spin-down rate ($\dot{\nu}$) as a simple power-law of the form
\begin{equation}\label{eqn:spin_law}
    \dot{\nu} = -\kappa \nu^{n}.
\end{equation}
Here the coefficient $\kappa$ depends on the magnetic dipole moment amplitude, angle between magnetic and spin axes and moment of inertia of the neutron star, while $n$ is the braking index of the dominant torque acting on the pulsar over time.
Under the assumption that $\kappa$ remains constant over time, the braking index can be found by re-arranging the time derivative of Equation~\ref{eqn:spin_law} such that
\begin{equation}\label{eqn:brake}
    n = \frac{\nu\ddot{\nu}}{\dot{\nu}^{2}},
\end{equation}
where $\ddot{\nu}$ is the second derivative of pulsar spin frequency.
Braking indices are often measured by either performing local fits to $\nu$ and $\dot{\nu}$ across many years of observations and fitting the slope of the resulting $\dot{\nu}$ measurements~\citep[e.g.][]{Lyne1996}, or through long-term coherent pulsar timing with a single model \citep[e.g.][]{Parthasarathy2019}.
In the latter method, the rotation phase of a pulsar over time is described as a truncated Taylor series
\begin{equation}\label{eqn:spin_phase}
    \phi(t) = \phi_{0} + \nu(t - t_{0}) + \frac{1}{2}\dot{\nu}(t - t_{0})^{2} + \frac{1}{6}\ddot{\nu}(t - t_{0})^{3} + ...,
\end{equation}
where $\phi_{0}$ is the rotation phase at some reference time $t_{0}$, and the braking index is inferred from measurements of $\nu$ and its derivatives via Equation \ref{eqn:brake}.
Approximating a pulsar as a rotating dipole in a vacuum results in the star spinning down purely through dipole radiation with a braking index of $n = 3$~\citep[e.g.][]{Shapiro1983}.
However, pulsars are neither simple bar magnets, nor do they exist in a vacuum.
Their magnetospheres are filled with energetic plasma, some of which is accelerated away from the neutron star on open field lines as powerful particle winds.
Spin-down dominated by such outflows would result in a braking index satisfying $n \approx 1$ for the hypothetical case of a split-monopole outflow \citep{Michel1969}, and $n \lesssim 3$ for a co-rotating magnetosphere modelled as an extended dipole \citep{Melatos1997} or a superposed vacuum and force-free structure \citep{Contopoulos2006}.
It has also been suggested that pulsars spinning down due to gravitational-wave radiation may have $n = 5$ \citep{Bonazzola1996, Yue2007} or $n = 7$ \citep{Owen1998, Alford2014}. 
None of these models take into account the interior structure of pulsars or the coupling between the crust of the star and its magnetosphere, which almost certainly affect the observed rotational-evolution of neutron stars.

A small sample of young (characteristic age, $\tau_{c} < 30$\,kyr) pulsars have been found to possess values of $n$ between these theoretical values and are claimed to represent the long-term rotational evolution of these pulsars on timescales that are much longer than their inter-glitch intervals (see Tables 1 and 4 of \citealt{Espinoza2017}). 
However, many pulsars have measured $\ddot{\nu}$, and hence $n$, that span a large range of values \citep[e.g.][]{Namkham2019}.
These are often referred to as being `anomalous' in the literature as they deviate significantly from the small values expected from radiative mechanisms.
Additionally, many of these measurements can be attributed to various stochastic processes arising from either the magnetosphere or internal dynamics. 
Hence, the large inferred values of $n$ do not necessarily reflect the true long-term rotational evolution of these pulsars over many decades.
For example, the braking indices inferred from simple fits of Equation \ref{eqn:spin_phase} to the arrival times of a large samples of pulsars without accounting for these stochastic processes tend to be almost equally split between positive and negative values, induced by the presence of low-frequency stochastic variations in the pulse arrival times, often referred to as timing noise \citep{Hobbs2010}.
Glitches, sudden spin-up events that can be caused by some form of stress build-up and release process (see \citealt{Haskell2015} for a review of glitch mechanisms), can result in discontinuities in the otherwise smooth spin-down of pulsars.
Many glitching pulsars have been found to exhibit steep, positive gradients in $\dot{\nu}$ (i.e. a large $\ddot{\nu}$) in-between subsequent glitches \citep[e.g.][]{Yu2013}.
Unlike the braking indices inferred from arbitrary cubic fits, these large `inter-glitch' braking indices are consistently found with values of $n$ between $\sim$10-200 \citep{Johnston1999}.
They are often associated with a particular form of post-glitch behaviour, namely large step-changes in $\dot{\nu}$ followed by an a `linear' $\dot{\nu}$ recovery, which was first noticed in the timing of PSR~J0835$-$4510 (B0833$-$45: the Vela pulsar, \citealt{Cordes1988}). 
This phenomenon has been interpreted as possible evidence for the creeping of thermally unpinned superfluid vortices between pinning sites inside neutron stars \citep{Alpar1984a, Alpar1984b, Alpar1993, Haskell2020}, though such theoretical models are often difficult to falsify (see Section 7 of \citealt{Haskell2015}).

However, recent work by~\citet{Parthasarathy2019} challenges the assertions that large $n$ must result from either stochastic or glitch-based processes. Using a modern Bayesian inference framework, they discovered a sample of 19 young pulsars with high spin-down energies ($\dot{E}$), that possess large, predominately positive $n$ despite accounting for various types of timing noise. The robustness of these measurements were reinforced in a follow-on study~\citep{Parthasarathy2020}, where timing models that included long-term exponential glitch recoveries were found to be inconsistent with the data, while the inclusion of up to a decade of additional timing data had little effect on the recovered values of $n$. 
They further demonstrated these measurements could not be explained by the presence of unaccounted glitch recovery effects.
The lack of observed glitches in these pulsars indicates that the underlying mechanism responsible for the large $n$ must be stable over decade-long timescales.

One idea to explain these large values of $n$ is to relax the assumption that $\kappa$ in Equation~\ref{eqn:spin_law} remains constant with time. Physically, this could correspond to one of (or potentially a combination of) e.g. changes in the neutron star moment of inertia over time~\citep{Ho2012}, evolution of the magnetic and spin axes towards (or away from) alignment~\citep{Goldreich1970, Tauris1998, Melatos2000}, or changes in the surface magnetic field strength~\citep{Vigano2013, Ho2015}. 
Observationally, variations in $\kappa$ with time manifest themselves in the braking index itself evolving on $\sim$kyr timescales, a process that has been exploited in some population synthesis studies to explain the broad distribution pulsar of spin periods and period-derivatives \citep[e.g.][]{Johnston2017}.
Hence the robust measurement of $n$ of a large sample of pulsars could allow us to place constraints on the mechanisms responsible for the long-term rotational evolution of pulsars.

For this work, we analysed a group of 74 young pulsars that have been observed over the last decade as part of the young pulsar timing programme on the Parkes radio telescope. We derive the parameters for 124 glitches in 52 of these pulsars, and combine our measurements with those of \citet{Parthasarathy2019, Parthasarathy2020} to explore the rotational evolution of the pulsars.
The structure of the paper is as follows.
In Section \ref{sec:obs}, we briefly outline the observing and data processing steps, while the glitch search and inference frameworks that we employed are detailed in Section \ref{sec:inf}.
We present a timing noise limited catalogue of 124 pulsar glitches from our pulsar sample as well as our inferred upper-limits on the minimum glitch size across the sample and statistical analyses of the overall glitch properties in Section~\ref{sec:glitch}.
We also briefly outline the updated pulsar properties obtained as part of our model selection studies.
In Section~\ref{sec:brake} we highlight our braking index measurements for 33 pulsars and explore differences in the implied versus observed long-term rotational evolution of these pulsars.
Conclusions and potential future directions are summarized in Section~\ref{sec:conc}.

\section{Observations}\label{sec:obs}

The young pulsar timing project (P574) has been running on the CSIRO Parkes 64-m radio telescope (also known as \textit{Murriyang}) with an approximately monthly cadence since the beginning of 2007. 
Originally intended to aid in the detection of pulsed gamma-ray emission with the \textit{Fermi} satellite's Large Area Telescope ~\citep{Smith2008, Weltevrede2010}, the sample has changed somewhat over the years and now consists of some 260 pulsars \citep{Johnston2021a}. 
For the purposes of this paper we do not re-analyse the timing of those pulsars already reported in \citet{Parthasarathy2019, Parthasarathy2020}, nor those pulsars that were added to the project in 2014~\citep{Namkham2019}. 
This leaves a total of 74 objects, mainly pulsars with a high spin-down energy, $\dot{E} \gtrsim 10^{34}$\,ergs\,s$^{-1}$.

The pulsars in our sample were observed in the 20-cm band using the multi-beam, H-OH and UWL receivers~\citep{Staveley-Smith1996, Granet2011, Hobbs2020}.
All observations were folded in real-time using the polyphase digital filterbank signal processors to form {\sc psrfits} format archive files~\citep{Hotan2004}, each with 1024\,phase bins and 1024\,frequency channels covering 256\,MHz of bandwidth. Each folded archive was then excised of radio frequency interference, before being flux and polarization calibrated using the tools in {\sc psrchive}~\citep{Hotan2004,vanStraten2012}. After averaging the individual observations in time, frequency and polarization to obtain a one-dimensional profile of flux versus pulse phase, pulse times of arrival (ToAs) are then generated by cross-correlating the averaged pulse profiles in the Fourier-domain with a smoothed, high S/N template~\citep{Taylor1992}. A more comprehensive description of the observations and data processing can be found in \citet{Johnston2018} and \citet{Johnston2021a}.

We also make use of extended data sets available for 27 pulsars from the Parkes Observatory Pulsar Data Archive~\citep{Hobbs2011} that we pre-pended to the beginning of the P574 data. These `legacy' data comprise observations undertaken prior to 2007 using the multi-beam and H-OH receivers at 20-cm wavelengths with the analogue and digital signal processors as presented in \citet{Wang2007} and \citet{Yu2013}.

\section{Pulsar inference framework}\label{sec:inf}

In order to measure the properties of pulsars we must first obtain (approximately) phase connected timing solutions. For young pulsars this is not always a simple task as the presence of timing noise and glitches within the timing data often result in a loss of phase coherence in the pulse arrival times. Hence a complete catalogue of glitches in our pulsars is needed in order to characterise their timing properties.

\subsection{Glitch detection and phase connection}

For a large fraction of the pulsars in our sample, we were able to identify when a glitch had occurred via visual inspection of their timing residuals.
Preliminary fitting to the glitches using {\sc tempo2}~\citep{Hobbs2006, Edwards2006} allowed us to assign pulse numbers to each ToA -- i.e. determine the integer number of rotations that have occurred since the first observation.
However, in cases where a particularly large glitch had occurred, obtaining a preliminary fit became impossible.
To solve this problem we employed two separate methods: estimates of change in spin and spin-down from local measurements of $\nu$ and $\dot{\nu}$ obtained from stride-fits to a moving window containing 5-6 ToAs, and the Hidden Markov Model (HMM) glitch detection algorithm developed by \citet{Melatos2020}.
Local $\nu$-$\dot{\nu}$ measurements allowed us to obtain a rough estimate of the glitch size that was needed to obtain both a coherent solution and verify the correct pulse numbering was applied.
The HMM algorithm provided both an independent means of obtaining preliminary measurements of the permanent step changes in spin frequency ($\Delta\nu_{p}$) and spin-down frequency ($\Delta\dot{\nu}_{p}$) associated with a glitch, and an automated means for detecting any additional glitches that were missed by visual inspection of the data. 
The HMM detector did not identify any new glitches in pulsars aside from PSR J1413$-$6141. 
The high rate of glitches in this pulsar meant a phase connected solution was unable to be obtained without the preliminary glitch properties returned by the HMM detector.
Once we had obtained a preliminary solution with confidence in the pulse numbering, we then applied the Bayesian pulsar timing package {\sc TempoNest}~\citep{Lentati2014} to construct posterior probability distributions for both the deterministic and stochastic pulsar properties.
Depending whether the pulsar had glitched or not, this final step followed either the single or multi-stage process that we describe below. 

\subsection{Measuring glitch properties}

\begin{table}
    \centering
    \caption{Prior ranges on intrinsic, stochastic and glitch parameters. $\Delta_{\mathrm{param}}$ is the uncertainty returned by {\sc tempo2}, $T$ is length of each pulsar's data set. The value of $x$ is between $10^{3}$-$10^{5}$ depending on the pulsar.
    }
    \label{tbl:priors}
    \setlength{\tabcolsep}{4pt}
    \resizebox{\linewidth}{!}{
    \begin{tabular}{lc}
        \hline
        Parameter & Prior type (range) \\
        \hline
        RAJ, DecJ, $\nu$, $\dot{\nu}$, $\ddot{\nu}$ ($^{\circ}$, $^{\circ}$, Hz, s$^{-2}$, s$^{-3}$) & Uniform ($\pm x\times\Delta_{\mathrm{param}}$) \\
        Proper motion (mas\,yr$^{-1}$) & Uniform ($-1000$, $1000$) \\
        EFAC & Uniform ($-1$, $2$) \\
        EQUAD (s)  & Log-uniform ($-10$, $1$) \\
        Red noise amplitude (yr$^{3/2}$)  & Log-uniform ($-15$, $-3$) \\
        Red noise spectral index & Uniform ($0$, $20$) \\
        Low frequency cutoff (Hz) & Log-uniform ($-1$, $0$) \\
        Sinusoid amplitude (s) & Log-uniform ($-10$, $0$) \\
        Sinusoid phase (rad) & Uniform ($0$, $2\pi$) \\
        Log-sinusoid frequency (Hz) & Log-uniform ($1/T_{\rm span}$, $100/T_{\rm span}$) \\
        Glitch phase jump (rotations)  & Uniform ($-5$, $5$) \\
        Glitch permanent change in $\nu$ (Hz)  & Log-uniform ($-11$, $-4$) \\
        Glitch change in $\dot{\nu}$ (Hz$^{-2}$) & Uniform ($-10^{-18}$, $-10^{-11}$) \\
        Glitch decaying change in $\nu$ (Hz) & Log-uniform ($-11$, $-4$) \\
        Glitch recovery timescale (days)  & Uniform ($1$, $2000$) \\
        \hline
    \end{tabular}
    }
\end{table}

For pulsars that were found to have glitched, we performed an initial {\sc TempoNest} analysis to measure the properties of the glitch(es).
The simplest glitches can be described by a permanent step-function in the pulsar spin-frequency, $\Delta\nu_{p}$. 
Some glitches required more complex modelling that included step changes in spin-down, $\Delta\dot{\nu}_{p}$, and one or more exponential recoveries of the spin-frequency towards its pre-glitch value ($\Delta\nu_{d}$) over time ($\tau_{d}$) are required to fully describe their phenomenology. 
The combined, initial step-changes in $\nu$ and $\dot{\nu}$ for glitches with recoveries are simply $\Delta\nu_{g} = \Delta\nu_{p} + \Delta\nu_{d}$ and $\Delta\dot{\nu}_{g} = \Delta\dot{\nu}_{p} - \Delta\nu_{g} Q / \tau_{d}$, where $Q = \Delta\nu_{d}/\Delta\nu_{g}$ is the fractional amount by which a glitch has recovered.
Since glitches have spin-frequency rise-times of only a few seconds \citep[e.g.][]{Ashton2019}, few have been detected in the midst of an observation.
As a result, there is often some level of ambiguity in both exact time a glitch occurred ($t_{g}$) and the precise number of times the neutron star has rotated between the last pre-glitch and first post-glitch observations.
This issue can be overcome by adding an unphysical jump in the pulsar rotation phase ($\Delta\phi_{g}$) to the glitch model, thereby ensuring a phase connected solution is maintained across the glitch epoch.
Adding all of these components together, we obtained the standard model for pulsar rotational phase following a glitch
\begin{equation}\label{eqn:glitch}
\begin{aligned}
    \phi_{\mathrm{g}}(t) = \Delta\phi_{g} + \Delta\nu_{p}(t - t_{\mathrm{g}}) + \frac{1}{2}\Delta\dot{\nu}_{p}(t - t_{\mathrm{g}})^{2} \\
     - \Big(\sum_{i=0}^{k} \Delta\nu_{d,i}\tau_{d,i} \Big[ 1 - e^{-(t - t_{g})/\tau_{d,i}} \Big] \Big),
\end{aligned}
\end{equation}
where $k$ is the total number of exponential recoveries. 
As the exponential is a non-linear function, fitting for it with {\sc tempo2} requires some level of a priori knowledge of its value. 
Previous works usually worked around this issue by measuring $\tau_{d}$ from either (or via a combination of) local fits to $\nu$ and $\dot{\nu}$ over time or iterative {\sc tempo2} fits to minimise the $\chi^{2}$ of the post-fit residuals once a good initial estimate is obtained~\citep[e.g.][]{Yu2013}.
This non-linearity is less of an issue as we explicitly restricted the prior range of allowable values for $\tau_{d}$ in our parameter estimation. 
Hence we were able to include the recovery timescale as a free parameter when fitting the ToAs. 

To avoid potential biases in our measurements, we modelled both the pulsar glitch and timing noise parameters simultaneously~\citep{vanHaasteren2013}.
The red noise power spectrum is modelled in {\sc TempoNest} as a simple power law of the form
\begin{equation}\label{eqn:rn_model}
	P(f) = \frac{A_{\rm r}^{2}}{12\pi^{2}} \Big( \frac{f}{\mathrm{yr}^{-1}} \Big)^{-\beta_{\rm r}},
\end{equation}
in which $A_{\rm r}$ and $\beta_{\rm r}$ are the red noise amplitude (in units of yr$^{3/2}$) and spectral index and $f$ is the frequency of the power spectrum.
Excess scatter in the timing residuals relative to their formal uncertainties due to pulse-to-pulse shape variations and radiometer noise was accounted for by modifying the uncertainties of each ToA as
\begin{equation}
	\sigma_{\mathrm{ToA},i}^{2} = \sigma_{Q}^{2} + F\sigma_{i}^{2},
\end{equation}
where, $\sigma_{Q}$ (the parameter EQUAD in {\sc TempoNest}) accounts for any additional time-independent white noise processes, $\sigma_{i}$ is the original uncertainty on the $i$-th ToA and $F$ (EFAC) is a free-parameter that describes unaccounted instrumental distortions. 
To reduce the computation time when fitting the glitch parameters, we treated the astrometric and rotational properties of each pulsar as a set of nuisance parameters that are analytically marginalised over.
We used the prior ranges listed in Table~\ref{tbl:priors} throughout this work.
For pulsars that possessed large inter-glitch slopes in $\dot{\nu}$, we had to include a $\ddot{\nu}$-term when conducting the glitch parameter estimation.
Failing to do so resulted in both the $\dot{\nu}$ gradient as well as any $\Delta\dot{\nu}_{p}$-terms in the glitch model being partially absorbed by the power-law red noise model.
A quick comparison of the Bayesian evidences obtained for pulsars where both timing models were fit revealed an overwhelming preference for the model that included step-changes in the spin-down rate and a $\ddot{\nu}$ term.
As an example, for PSR~J1420$-$6048 we obtained a natural log-Bayes factor of $186$ in favour of the $\ddot{\nu}$-inclusive model.

\begin{figure}
    \centering
    \includegraphics[width=\linewidth]{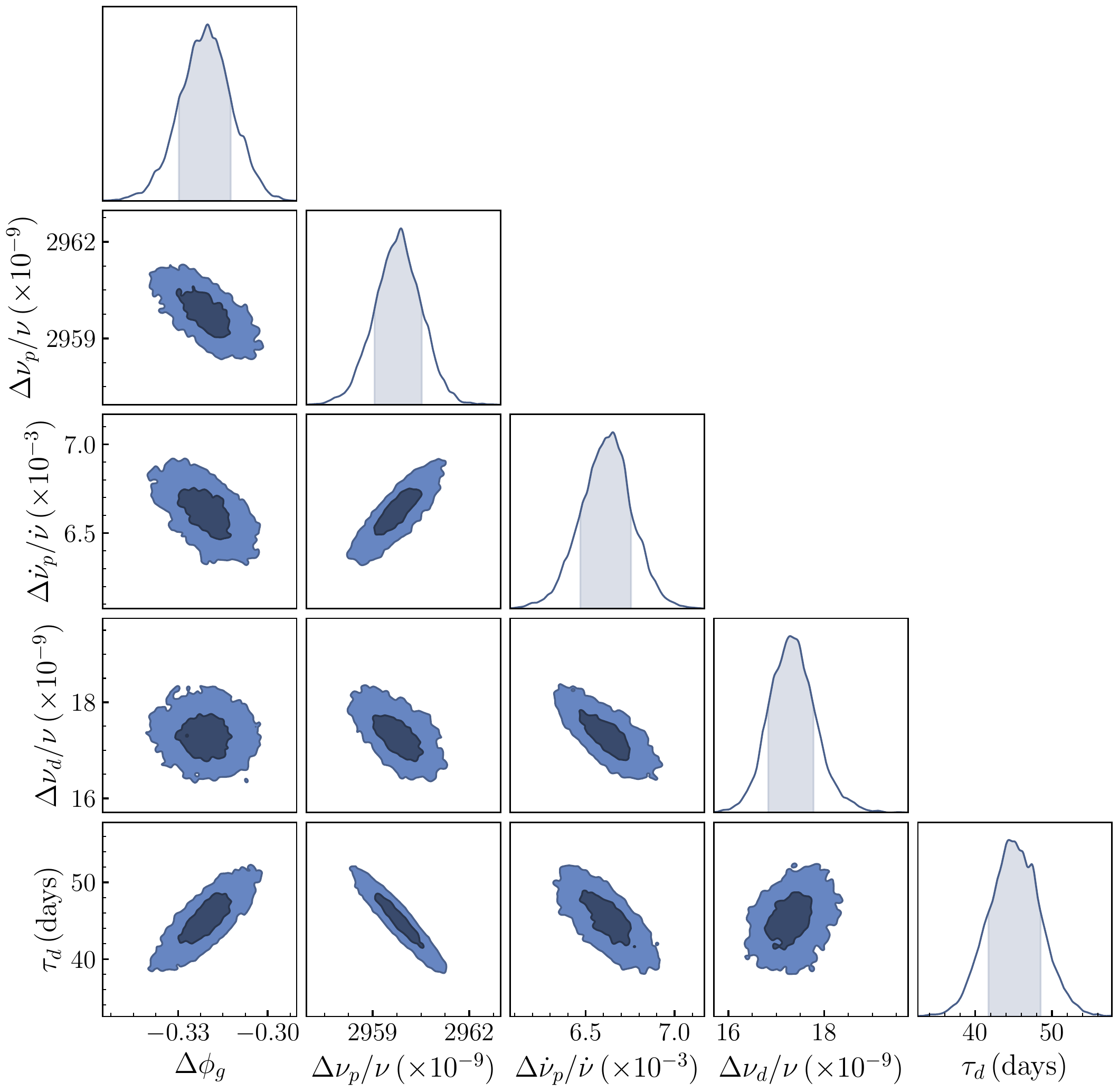}
    \caption{Example corner plot showing the one- and two-dimensional posterior distributions for the MJD 55745 glitch in PSR~J1524$-$5625. Contours in the two-dimensional posteriors indicate the 68\% and 95\% confidence intervals. Shading in the one-dimensional posteriors cover the 68\% confidence region.}
    \label{fig:j1524_gl}
\end{figure}

An example outcome of our methodology is shown in Figure~\ref{fig:j1524_gl}, where the permanent and decaying properties of a glitch in PSR~J1524$-$5625 were well recovered.

\subsection{Pulsar properties and extended noise models}

After obtaining timing models with solutions for the glitches, we then moved on to refining the deterministic properties of each pulsar and searching for additional, previously unmeasured pulsar second spin-period derivatives and proper motion.
We generally did not attempt re-fitting glitches in the archival data sets when searching for these new deterministic pulsar properties. 
Instead, we opted to marginalise over the glitch parameters as many of these glitches had been well characterised in previous works. 
Our standard approach to measuring the intrinsic, non-glitch properties of each pulsar in our sample included fitting for the pulsar position, rotational and stochastic properties.
For pulsars that have glitched, we used the post-fit ephemeris where the glitch properties were set to be analytically marginalised over during the refitting process.
Aside from measuring the standard positional, spin/spin-down and red noise parameters (termed the PL model), we also searched for evidence of additional, unmodelled pulsar properties by iteratively adding them to our timing models.
This included fitting for a braking index via including a $\ddot{\nu}$ term (the PL+F2 model) and searching for pulsar proper-motions (PL+PM model) with component values ranging between $\pm 1000$\,mas\,yr$^{-1}$.
Following the methodology of~\citet{Parthasarathy2019}, we also compared various extensions of the standard timing noise model implemented in {\sc TempoNest}. This included low-frequency components with fluctuation cycles that are longer than our data sets (LFC model), and a variant of the red noise model in Equation~\ref{eqn:rn_model} that incorporates a spectral break in the red power-spectrum (BPL model) as
\begin{equation}
    P_{r}(f) = \frac{A_{r}^{2}}{12\pi^{2}} \frac{(f_{c}/{\mathrm{yr}^{-1}})^{-\beta_{\rm r}}}{[1 + (f/f_{c})^{-\beta_{\rm r}/2}]^{2}},
\end{equation}
where $f_{c}$ is the turnover frequency. 
Lastly, we checked for the presence of periodic signals that could be induced by the presence of an unmodelled binary companion or quasi-periodic spin-down state switching \citep[e.g.][]{Lyne2010} by including a sinusoid to the timing model (SIN model). 
In cases where evidence for more than one process was favoured over the simplest PL model, we conducted more complex joint fits for multiple additional processes.
As an example, we might fit for $\ddot{\nu}$, proper-motion and low-frequency components in addition to the other pulsar properties simultaneously -- i.e. a PL+F2+PM+LFC model. 

We used Bayesian model selection to assess which model best describes the data, specifically through comparing the ratio of Bayesian evidences between competing models, often referred to as the Bayes factor ($\mathcal{B}$).
In this work, we used a relatively conservative Bayes factor threshold of $\ln(\mathcal{B}) = 3$.
Under the~\citet{Kass1995} interpretation of the Bayes factor, $\ln(\mathcal{B}) > 3$ indicates a significant preference for one model over the other.
If $\ln(\mathcal{B}) < 3$, then we consider there to be insufficient evidence to distinguish one model from the other.

\section{Glitches and timing results}\label{sec:glitch}

We identified and characterised 124 glitches in 52 of the 74 pulsars in our sample. Our measurements of their properties are presented in Table~\ref{tbl:glitches}. This includes 37 previously published glitches, references for which are given in the final column of the table and a further 13 glitches that are publicly listed in the Jodrell Bank glitch catalogue\footnote{\href{www.jb.man.ac.uk/pulsar/glitches/gTable.html}{jb.man.ac.uk/pulsar/glitches/gTable.html}}~\citep{Espinoza2011a} but have not been published elsewhere. 
\begin{table*}
    \centering
    \begin{threeparttable}
    \caption{Median recovered glitch parameters and associated $68\%$ credible intervals (indicated by uncertainties with a $+$ or $-$). Glitch epochs correspond either to previous reported values, or were computed via the method detailed in Section 3.3 of \citet{Espinoza2011a}. Values in parentheses represent the 1-$\sigma$ uncertainties on the last digit. Glitches with multiple recoveries have additional recovery timescales and fractions listed in rows underneath the main glitch parameters. Note the results listed for PSRs J1341$-$6220 and J1740$-$3015 were obtained via {\sc Tempo2} least-squares fitting, where the uncertainties represent to 1-$\sigma$ uncertainties returned by {\sc Tempo2}. See Section \ref{subsec:multi_glitch} for details regarding glitches 14 and 15 in PSR~J1341$-$6220. 
    References are given to where a glitch was first reported. Those with a $^{\dagger}$ are listed in the Jodrell Bank glitch catalogue but not published elsewhere.}
    \label{tbl:glitches}
    \renewcommand{\arraystretch}{1.2}
    \begin{tabular}{lccccccccc}
        \hline
        PSRJ & Gl. no. & $t_{\mathrm{g}}$ & $\Delta\nu_{g}/\nu$ & $\Delta\dot{\nu}_{g}/\dot{\nu}$ & $\Delta\dot{\nu}_{p}/\dot{\nu}$ & $\tau_{d}$ & $Q$ & Ref \\
         & & (MJD) & ($\times 10^{-9}$) & ($\times 10^{-3}$) & ($\times 10^{-3}$) & (days) &  & \\
        \hline
        J0631$+$1036 & 1 & $58352.14(4)$ & $120 \pm 2$ & $0.8^{+0.7}_{-0.5}$ & $0.8^{+0.7}_{-0.5}$ & $-$ & $-$ & This work$^{\dagger}$ \\
        J0729$-$1448 & 1 & $54697(3)$ & $6646^{+13}_{-9}$ & $37^{+25}_{-11}$ & $11.8^{+0.3}_{-0.2}$ & $40^{+36}_{-12}$ & $0.006(2)$ & \citet{Weltevrede2010} \\
        J0742$-$2822 & 1 & $55020.469(4)$ & $100.9 \pm 0.3$ & $0.2^{+0.3}_{-0.2}$ & $0.2^{+0.3}_{-0.2}$ & $-$ & $-$ & \citet{Espinoza2011a} \\
        J0742$-$2822 & 2 & $56727.7(0.1)$ & $2.6 \pm 0.2$ & $\lesssim 0.2$ & $\lesssim 0.2$ & $-$ & $-$ & \citet{Basu2020} \\
        J0835$-$4510 & 1 & $55408.802$ & $1902.4 \pm 0.5$ & $7 \pm 1$ & $6.99 \pm 0.09$ & $13 \pm 2$ & $0.00548(8)$ & \citet{Buchner2010ATel} \\
        J0835$-$4510 & 2 & $56555.871$ & $3057 \pm 2$ & $4.6 \pm 0.3$ & $4.69 \pm 0.06$ & $148 \pm 8$ & $0.0270(4)$ & \citet{Buchner2013ATel} \\
        J0835$-$4510 & 3 & $57734.484991(29)$ & $1439.8 \pm 0.5$ & $11.2 \pm 0.3$ & $6.30 \pm 0.03$ & $5.8 \pm 0.1$ & $0.00546(8)$ & \citet{Palfreyman2018} \\
        J0901$-$4624 & 1 & $57179(6)$ & $0.9 \pm 0.1$ & $0.02^{+0.04}_{-0.01}$  & $0.02^{+0.04}_{-0.01}$ & $-$ & $-$ & This work \\
        J0908$-$4913 & 1 & $58765.06(5)$ & $22.2 \pm 0.2$ & $\lesssim 0.5$ & $\lesssim 0.5$ & $-$ & $-$ & \citet{Lower2019} \\
        J0940$-$5428 & 1 & $55346(8)$ & $1573.9^{+1.1}_{-0.8}$ & $11 \pm 2$ & $4.9 \pm 0.2$ & $49^{+16}_{-10}$ & $0.0068(8)$ & This work \\
        J0940$-$5428 & 2 & $58322(16)$ & $1100.5^{+0.6}_{-0.5}$ & $4.0^{+0.2}_{-0.1}$ & $4.0^{+0.2}_{-0.1}$ & $-$ & $-$ & This work \\
        J1015$-$5719 & 1 & $56695(6)$ & $3232.3 \pm 0.6$ & $11 \pm 2$ & $3.7 \pm 0.2$ & $103^{+19}_{-17}$ & $0.0078(7)$ & This work \\
        J1016$-$5857 & 1 & $55030(9)$ & $1919.8^{+1.1}_{-0.9}$ & $6 \pm 1$ & $3.4 \pm 0.2$ & $64^{+37}_{-25}$ & $0.005(1)$ & \citet{Yu2013} \\
        J1016$-$5857 & 2 & $55595(10)$ & $1464.4^{+1.1}_{-0.9}$ & $4^{+5}_{-1}$ & $2.3^{+0.3}_{-0.2}$ & $88^{+337}_{-50}$ & $0.005(4)$ & This work \\
        J1016$-$5857 & 3 & $56975(8)$ & $6.21^{+2.35}_{-1.18}$ & $-$ & $-$ & $-$ & $-$ & This work \\
        J1019$-$5749 & 1 & $55595(10)$ & $1.33 \pm 0.4$ & $0.12^{+0.22}_{-0.09}$ & $0.12^{+0.22}_{-0.09}$ & $-$ & $-$ & This work \\
        J1019$-$5749 & 2 & $55981(10)$ & $377.8 \pm 0.4$ & $0.51 \pm 0.3$ & $0.51 \pm 0.3$ & $-$ & $-$ & This work \\
        J1028$-$5819 & 1 & $57881(14)$ & $2296.5^{+0.5}_{-0.4}$ & $35^{+1}_{-2}$ & $3.5 \pm 0.3$ & $54.4^{+2}_{-3}$ & $0.0114(2)$ & This work \\
        J1048$-$5832 & 1 & $54495(4)$ & $3044.1 \pm 0.9$ & $5.2^{+0.5}_{-0.4}$ & $5.2^{+0.5}_{-0.4}$ & $-$ & $-$ & This work \\
        J1048$-$5832 & 2 & $56756(4)$ & $2963^{+4}_{-2}$ & $9^{+9}_{-4}$ & $4.0 \pm 0.5$ & $33^{+52}_{-17}$ & $0.004(2)$ & This work \\
        J1052$-$5954 & 1 & $54493.695(1)$ & $6778 \pm 1$ & $70^{+7}_{-7}$ & $6.5^{+0.8}_{-0.7}$ & $64 \pm 6$ & $0.0057(3)$ & \citet{Weltevrede2010} \\
        J1055$-$6028 & 1 & $57035(10)$ & $105.7^{+0.6}_{-0.5}$ & $1.3^{+0.3}_{-0.2}$ & $1.3^{+0.3}_{-0.2}$ & $-$ & $-$ & This work \\
        J1105$-$6107 & 1 & $54711(21)$ & $35^{+1}_{-2}$ & $24 \pm 4$ & $24 \pm 4$ & $-$ & $-$ & \citet{Weltevrede2010} \\
        J1105$-$6107 & 2 & $55300(16)$ & $949^{+2}_{-1}$ & $17 \pm 4$ & $17 \pm 4$ & $-$ & $-$ & \citet{Yu2013} \\
        J1112$-$6103 & 1 & $55288(7)$ & $1793 \pm 1$ & $6 \pm 2$ & $3.61^{+1.1}_{-0.8}$ & $313^{+237}_{-175}$ & $0.014(11)$ & This work \\
        J1112$-$6103 & 2 & $57922(6)$ & $1283 \pm 1$ & $4.9^{+0.5}_{-0.4}$ & $4.9^{+0.5}_{-0.4}$ & $-$ & $-$ & This work \\
        J1248$-$6344 & 1 & $56043(5)$ & $1.7 \pm 0.2$ & $\lesssim 0.13$ & $\lesssim 0.13$ & $-$ & $-$ & This work \\
        J1301$-$6305 & 1 & $55124(10)$ & $4169^{+3}_{-2}$ & $5.8^{+0.7}_{-0.5}$ & $5.8^{+0.7}_{-0.5}$ & $-$ & $-$ & This work \\
        J1301$-$6305 & 2 & $57718(6)$ & $658 \pm 3$ & $6.0 \pm 0.5$ & $6.0 \pm 0.5$ & $-$ & $-$ & This work \\
        J1320$-$5359 & 1 & $56534(10)$ & $10.5 \pm 0.1$ & $0.2^{+0.2}_{-0.1}$ & $0.2^{+0.2}_{-0.1}$ & $-$ & $-$ & This work \\
        J1320$-$5359 & 2 & $56737(14)$ & $246.8 \pm 0.1$ & $0.08^{+0.14}_{-0.06}$ & $0.08^{+0.14}_{-0.06}$ & $-$ & $-$ & This work \\
        J1341$-$6220 & 1 & $54468(18)$ & $313 \pm 1$ & $0.65 \pm 0.08$ & $0.65 \pm 0.08$ & $-$ & $-$ & \citet{Weltevrede2010} \\
        J1341$-$6220 & 2 & $54871(11)$ & $307.2 \pm 0.6$ & $1.43 \pm 0.06$ & $1.43 \pm 0.06$ & $-$ & $-$ & \citet{Weltevrede2010} \\
        J1341$-$6220 & 3 & $55042(16)$ & $1528 \pm 4$ & $63 \pm 8$ & $0.36 \pm 0.09$  & $9 \pm 1$ & $0.042(3)$ & \citet{Yu2013} \\
        J1341$-$6220 & 4 & $55484(11)$ & $2.4 \pm 0.3$ & $0.49 \pm 0.01$ & $0.49 \pm 0.01$ & $-$ & $-$ & This work \\
        J1341$-$6220 & 5 & $55835(7)$ & $329.0 \pm 0.4$ & $0.98 \pm 0.02$ & $0.98 \pm 0.02$ & $-$ & $-$ & This work \\
        J1341$-$6220 & 6 & $56098(12)$ & $151.2 \pm 0.4$ & $0.22 \pm 0.04$ & $0.22 \pm 0.04$ & $-$ & $-$ & This work \\
        J1341$-$6220 & 7 & $56386(5)$ & $96 \pm 1$ & $\lesssim 0.6$ & $\lesssim 0.6$ & $-$ & $-$ & This work \\
        J1341$-$6220 & 8 & $56479(9)$ & $37 \pm 2$ & $\lesssim 0.6$ & $\lesssim 0.6$ & $-$ & $-$ & This work \\
        J1341$-$6220 & 9 & $56602(12)$ & $1709.7 \pm 0.3$ & $1.06 \pm 0.1$  & $1.06 \pm 0.1$ & $-$ & $-$ & This work \\
        J1341$-$6220 & 10 & $57357(9)$ & $28.7 \pm 0.7$ & $0.25 \pm 0.09$ & $0.25 \pm 0.09$ & $-$ & $-$ & This work \\
        J1341$-$6220 & 11 & $57492(1)$ & $21.7 \pm 0.1$ & $0.8 \pm 0.1$ & $0.8 \pm 0.1$ & $-$ & $-$ & This work \\
        J1341$-$6220 & 12 & $57647(13)$ & $706 \pm 1$ & $0.77 \pm 0.09$ & $0.77 \pm 0.09$ & $-$ & $-$ & This work \\
        J1341$-$6220 & 13 & $57880(14)$ & $60 \pm 3$ & $\lesssim 0.6$ & $\lesssim 0.2$ & $86 \pm 47$ & $0.102(3)$ & This work \\
        J1341$-$6220 & 14 & $58178(15)$ & $*$ & $*$ & $*$ & $-$ & $-$ & This work \\
        J1341$-$6220 & 15 & $58214(4)$ & $*$ & $*$ & $*$ & $-$ & $-$ & This work \\
        J1357$-$6429 & 1 & $54803(17)$ & $2332^{+4}_{-3}$ & $13 \pm 1$ & $13 \pm 1$ & $-$ & $-$ & \citet{Weltevrede2010} \\
    \end{tabular}
    \end{threeparttable}
    \renewcommand{\arraystretch}{}
\end{table*}
\begin{table*}
    \centering
    \contcaption{}
    \begin{threeparttable}
    \renewcommand{\arraystretch}{1.3}
    \begin{tabular}{lccccccccc}
        \hline
        PSRJ & Gl. no. & $t_{\mathrm{g}}$ & $\Delta\nu_{g}/\nu$ & $\Delta\dot{\nu}_{g}/\dot{\nu}$ & $\Delta\dot{\nu}_{p}/\dot{\nu}$ & $\tau_{d}$ & $Q$ & Ref \\
         &  & (MJD) & ($\times 10^{-9}$) & ($\times 10^{-3}$) & ($\times 10^{-3}$) & (days)s &  & \\
        \hline
        J1357$-$6429 & 2 & $55576(10)$ & $4860^{+3}_{-2}$ & $14.7^{+0.7}_{-0.8}$ & $14.7^{+0.7}_{-0.8}$ & $-$ & $-$ & This work \\
        J1357$-$6429 & 3 & $57795(22)$ & $2250 \pm 11$ & $7 \pm 2$ & $7 \pm 2$ & $-$ & $-$ & This work \\
        J1357$-$6429 & 4 & $58148(15)$ & $1930^{+5}_{-4}$ & $\lesssim 0.9$ & $\lesssim 0.9$ & $-$ & $-$ & This work \\
        J1406$-$6121 & 1 & $56193(10)$ & $143.5^{+1.6}_{-0.8}$ & $2^{+4}_{-2}$ & $0.4^{+0.3}_{-0.2}$ & $107^{+206}_{-62}$ & $0.03(2)$ & This work \\
        J1410$-$6132 & 1 & $54652(19)$ & $263 \pm 2$ & $\lesssim 0.23$ & $\lesssim 0.23$ & $-$ & $-$ & \citet{Weltevrede2010} \\
        J1413$-$6141 & 1 & $54303(1)$ & $2412 \pm 3$ & $\lesssim 0.6$ & $\lesssim 0.6$ & $-$ & $-$ & \citet{Yu2013} \\
        J1413$-$6141 & 2 & $55744(7)$ & $235 \pm 2$ & $\lesssim 0.7$ & $\lesssim 0.7$ & $-$ & $-$ & This work \\
        J1413$-$6141 & 3 & $56147(12)$ & $200^{+3}_{-2}$ & $\lesssim 0.3$ & $\lesssim 0.3$ & $-$ & $-$ & This work \\
        J1413$-$6141 & 4 & $56567(5)$ & $371^{+2}_{-1}$ & $\lesssim 0.8$ & $\lesssim 0.8$ & $-$ & $-$ & This work \\
        J1413$-$6141 & 5 & $56975(8)$ & $30 \pm 2$ & $\lesssim 0.4$ & $\lesssim 0.4$ & $-$ & $-$ & This work \\
        J1413$-$6141 & 6 & $57236(4)$ & $266 \pm 2$ & $\lesssim 0.9$ & $\lesssim 0.9$ & $-$ & $-$ & This work \\
        J1413$-$6141 & 7 & $57509(10)$ & $356 \pm 2$ & $\lesssim 0.4$ & $\lesssim 0.4$ & $-$ & $-$ & This work \\
        J1413$-$6141 & 8 & $57838(6)$ & $2137 \pm 2$ & $\lesssim 0.5$  & $\lesssim 0.5$ & $-$ & $-$ & This work \\
        J1420$-$6048 & 1 & $54652(9)$ & $927.6^{+0.7}_{-0.6}$ & $6 \pm 1$ & $4.19 \pm 0.4$ & $45^{+20}_{-16}$ & $0.007(4)$ & \citet{Weltevrede2010} \\
        J1420$-$6048 & 2 & $55410(9)$ & $1352.8^{+0.5}_{-0.4}$ & $5.4 \pm 0.2$ & $5.4 \pm 0.2$ & $-$ & $-$ & \citet{Yu2013} \\
        J1420$-$6048 & 3 & $56267(6)$ & $1954.2 \pm 0.3$ & $5.7 \pm 0.2$ & $5.7 \pm 0.2$ & $-$ & $-$ & This work \\
        J1420$-$6048 & 4 & $57210(8)$ & $1210^{+2}_{-1}$ & $9^{+7}_{-4}$ & $3.5^{+0.4}_{-0.3}$ & $19^{+25}_{-12}$ & $0.009(2)$ & This work \\
        J1452$-$6036 & 1 & $55055.22(4)$ & $28.95 \pm 0.03$ & $2.5^{+1.0}_{-0.9}$ & $\lesssim 0.7$ & $2340^{+822}_{-672}$ & $0.12(5)$ & \citet{Yu2013} \\
        J1452$-$6036 & 2 & $57115(6)$ & $0.13 \pm 0.03$ & $\lesssim 0.2$ & $\lesssim 0.2$ & $-$ & $-$ & This work \\
        J1452$-$6036 & 3 & $58600.292(3)$ & $270.61 \pm 0.03$ & $1.2 \pm 0.3$ & $1.2 \pm 0.3$ & $-$ & $-$ & \citet{Lower2020} \\
        J1524$-$5625 & 1 & $55745(7)$ & $2977.0^{+0.7}_{-0.5}$ & $15.5^{+0.9}_{-0.7}$ & $6.6^{+0.2}_{-0.1}$ & $45^{+4}_{-3}$ & $0.0058(2)$ & This work \\
        J1614$-$5048 & 1 & $55734(2)$ & $4098 \pm 3$ & $3 \pm 1$ & $3 \pm 1$ & $-$ & $-$ & This work \\
        J1614$-$5048 & 2 & $56443(11)$ & $5949^{+9}_{-7}$ & $\lesssim 1$ & $\lesssim 1$ & $-$ & $-$ & This work \\
        J1617$-$5055 & 1 & $54747(7)$ & $334 \pm 3$ & $9^{+11}_{-2}$ & $0.48^{+5}_{-1}$ & $227^{+262}_{-38}$ & $0.975(6)$ & This work \\
        J1617$-$5055 & 2 & $55164(9)$ & $11 \pm 2$ & $0.8 \pm 0.6$ & $0.8 \pm 0.6$ & $-$ & $-$ & This work \\
        J1617$-$5055 & 3 & $55316(6)$ & $68 \pm 2$ & $2.2^{+0.6}_{-0.5}$ & $2.2^{+0.6}_{-0.5}$ & $-$ & $-$ & This work \\
        J1617$-$5055 & 4 & $55873(11)$ & $55 \pm 2$ & $1.1 \pm 0.6$ & $1.1 \pm 0.6$ & $-$ & $-$ & This work \\
        J1617$-$5055 & 5 & $56267(6)$ & $2068 \pm 2$ & $13.2^{+0.6}_{-0.7}$ & $13.2^{+0.6}_{-0.7}$ & $-$ & $-$ & This work \\
        J1644$-$4559 & 1 & $56600(14)$ & $717.4^{+0.3}_{-0.2}$ & $0.4^{+0.5}_{-0.3}$ & $0.4^{+0.5}_{-0.3}$ & $-$ & $-$ & This work \\
        J1646$-$4346 & 1 & $55288(7)$ & $8591^{+6}_{-5}$ & $16^{+9}_{-5}$ & $8.1^{+1.2}_{-0.9}$ & $126^{+137}_{-62}$ & $0.005(2)$ & This work \\
        J1650$-$4502 & 1 & $57778(8)$ & $12767^{+2}_{-1}$ & $290^{+80}_{-60}$ & $27^{+7}_{-6}$ & $82^{+21}_{-16}$ & $0.0061(8)$ & This work \\
        J1702$-$4128 & 1 & $57719(6)$ & $3090 \pm 1$ & $10^{+2}_{-1}$ & $4.77 \pm 0.09$ & $88^{+32}_{-19}$ & $0.0040(4)$ & This work \\
        J1702$-$4310 & 1 & $57510(10)$ & $3129^{+4}_{-1}$ & $5^{+2}_{-1}$ & $3.4^{+0.1}_{-0.2}$ & $50^{+72}_{-36}$ & $0.002(1)$ & This work \\
        J1705$-$3950 & 1 & $58236(14)$ & $9331^{+6}_{-4}$ & $61^{+11}_{-10}$ & $5.9^{+0.3}_{-0.2}$ & $55^{+10}_{-9}$ & $0.0053(4)$ & This work \\
        J1709$-$4429 & 1 & $54711(22)$ & $2752.5 \pm 0.2$ & $13.8^{+0.9}_{-1.0}$ & $7.4 \pm 0.09$ & $55^{+6}_{-7}$ & $0.010(1)$ & \citet{Weltevrede2010} \\
        J1709$-$4429 & 2 & $56354(13)$ & $2951.9 \pm 0.6$ & $8 \pm 1$ & $4.2^{+0.4}_{-0.3}$ & $54^{+11}_{-9}$ & $0.006(1)$ & This work \\
        J1709$-$4429 & 3 & $58178(6)$ & $2432.8^{+0.7}_{-0.6}$ & $8.5 \pm 0.9$ & $4.6^{+0.4}_{-0.3}$ & $49 \pm 8$ & $0.0061(9)$ & \citet{Lower2018} \\
        J1718$-$3825 & 1 & $54911(2)$ & $2.2 \pm 0.2$ & $\lesssim 0.08$ & $\lesssim 0.08$ & $-$ & $-$ & \citet{Yu2013} \\
        J1718$-$3825 & 2 & $57950(7)$ & $7.1 \pm 0.1$ & $\lesssim 0.07$ & $\lesssim 0.07$ & $-$ & $-$ & This work \\
        J1730$-$3350 & 1 & $55926(6)$ & $2250.7^{+1.0}_{-0.9}$ & $7^{+3}_{-2}$ & $5 \pm 2$ & $151.01^{+199}_{-81}$ & $0.007(1)$ & This work$^{\dagger}$ \\
        J1731$-$4744 & 1 & $55735.18(14)$ & $52.7 \pm 0.4$ & $3 \pm 1$ & $0.51^{+0.5}_{-0.4}$ & $151^{+35}_{-57}$ & $0.10(5)$ & \citet{Shternin2019} \\ 
        J1731$-$4744 & 2 & $56239.86(77)$ & $11.0 \pm 0.3$ & $\lesssim 0.4$ & $\lesssim 0.4$ & $-$ & $-$ & \citet{Shternin2019} \\
        J1731$-$4744 & 3 & $56975(8)$ & $6.4 \pm 0.3$ & $\lesssim 0.2$ & $\lesssim 0.2$ & $-$ & $-$ & This work \\
        J1731$-$4744 & 4 & $57978.17(2)$ & $3145^{+2}_{-4}$ & $1.2^{+0.4}_{-0.5}$ & $1.2^{+0.4}_{-0.5}$ & $-$ & $-$ & \citet{Jankowski2017ATel} \\ 
        J1734$-$3333 & 1 & $56351(12)$ & $86.7^{+7.9}_{-7.1}$ & $0.3^{+0.3}_{-0.2}$ & $0.3^{+0.3}_{-0.2}$ & $-$ & $-$ & This work$^{\dagger}$ \\
        J1737$-$3137 & 1 & $54348(4)$ & $1341.8 \pm 0.6$ & $3.0^{+2.0}_{-0.8}$ & $1.7^{+0.2}_{-0.3}$ & $152^{+224}_{-75}$ & $0.004(1)$ & \citet{Espinoza2011a} \\
        J1737$-$3137 & 2 & $57147(8)$ & $8.1 \pm 0.4$ & $\lesssim 0.1$ & $\lesssim 0.1$ & $-$ & $-$ & This work \\
    \end{tabular}
    \begin{tablenotes}
      \item
    \end{tablenotes}
    \end{threeparttable}
    \renewcommand{\arraystretch}{}
\end{table*}
\begin{table*}
    \centering
    \contcaption{}
    \begin{threeparttable}
    \renewcommand{\arraystretch}{1.3}
    \begin{tabular}{lccccccccc}
        \hline
        PSRJ & Gl. no. & $t_{\mathrm{g}}$ & $\Delta\nu_{g}/\nu$ & $\Delta\dot{\nu}_{g}/\dot{\nu}$ & $\Delta\dot{\nu}_{p}/\dot{\nu}$ & $\tau_{d}$ & $Q$ & Ref \\
         &  & (MJD) & ($\times 10^{-9}$) & ($\times 10^{-3}$) & ($\times 10^{-3}$) & (days) &  & \\
        \hline
        J1737$-$3137 & 3 & $58207(29)$ & $4494.1^{+0.9}_{-1.4}$ & $1.20 \pm 0.08$ & $1.20 \pm 0.08$ & $-$ & $-$ & This work \\
        J1737$-$3137 & 4 & $58838(24)$ & $15 \pm 2$ & $\lesssim 0.1$ & $\lesssim 0.1$ & $-$ & $-$  & This work \\
        J1740$-$3015 & 1 & $54450.19(1)$ & $45 \pm 2$ & $\lesssim 2$ & $\lesssim 0.7$ & $54 \pm 9$ & $0.15(1)$ & \citet{Weltevrede2010} \\
        J1740$-$3015 & 2 & $54695.19(2)$ & $2 \pm 1$ & $\lesssim 0.5$ & $\lesssim 0.5$ & $-$ & $-$ & \citet{Yuan2010} \\
        J1740$-$3015 & 3 & $54810.9(1)$ & $4 \pm 1$ & $\lesssim 0.2$ & $\lesssim 0.2$ & $-$ & $-$ & \citet{Espinoza2011a} \\
        J1740$-$3015 & 4 & $54928.6(1)$ & $3 \pm 0.7$ & $\lesssim 0.1$ & $\lesssim 0.1$ & $-$ & $-$ & \citet{Espinoza2011a} \\
        J1740$-$3015 & 5 & $55220(14)$ & $2659 \pm 4$ & $1.2 \pm 0.2$ & $0.7 \pm 0.1$ & $258 \pm 31$ & $0.008(1)$ & \citet{Yu2013} \\
        J1740$-$3015 & 6 & $55936.2(1)$ & $16 \pm 6$ & $1.1 \pm 0.3$ & $1.1 \pm 0.3$ & $-$ & $-$ & This work$^{\dagger}$ \\
        J1740$-$3015 & 7 & $57499.371(4)$ & $228 \pm 3$ & $2.2 \pm 0.5$ & $0.97 \pm 0.03$ & $89 \pm 13$ & $0.057(9)$ & \citet{Jankowski2016ATel} \\
        J1740$-$3015 & 8 & $58232.4(4)$ & $835.5 \pm 0.4$ & $0.26 \pm 0.04$ & $0.26 \pm 0.04$ & $-$ & $-$ & \citet{Basu2020} \\
        J1757$-$2421 & 1 & $55702(6)$ & $7800.5^{+2.1}_{-0.4}$ & $66^{+31}_{-29}$ & $3.1 \pm 0.2$ & $15^{+10}_{-9}$ & $0.0003(2)$ & \citet{Yuan2017} \\
                     & & & & & & $92^{+23}_{-12}$ & $0.0014(1)$ & \\
                     & & & & & & $618^{+215}_{-108}$ & $0.0021(3)$ & \\
        J1801$-$2304 & 1 & $55371.1(2)$ & $3.0 \pm 0.5$ & $\lesssim 0.18$ & $\lesssim 0.18$ & $-$ & $-$ & \citet{Yu2013} \\
        J1801$-$2304 & 2 & $55851.7(5)$ & $1.6^{+0.6}_{-0.7}$ & $\lesssim 0.16$ & $\lesssim 0.16$ & $-$ & $-$ & This work$^{\dagger}$ \\
        J1801$-$2304 & 3 & $56158.360(2)$ & $513.6 \pm 0.5$ & $\lesssim 0.25$ & $\lesssim 0.25$ & $-$ & $-$ & This work$^{\dagger}$ \\
        J1801$-$2304 & 4 & $57586.4(1)$ & $96.2 \pm 0.5$ & $\lesssim 0.21$ & $\lesssim 0.21$ & $-$ & $-$ & This work$^{\dagger}$ \\
        J1801$-$2451 & 1 & $54661(2)$ & $3083.7 \pm 0.7$ & $6.5 \pm 0.5$ & $6.5 \pm 0.5$ & $-$ & $-$ & \citet{Yu2013} \\
        J1801$-$2451 & 2 & $56943(7)$ & $2423.5 \pm 0.9$ & $5.9^{+0.5}_{-0.4}$ & $5.9^{+0.5}_{-0.4}$ & $-$ & $-$ & This work \\
        J1803$-$2137 & 1 & $55775(2)$ & $4785.9^{+1.2}_{-0.9}$ & $17 \pm 1$ & $7.3 \pm 0.3$ & $40^{+5}_{-4}$ & $0.0071(5)$ & This work$^{\dagger}$ \\
        J1825$-$0935 & 1 & $53734.6(1)$ & $6 \pm 2$ & $1.2^{+1.9}_{-0.9}$ & $1.2^{+1.9}_{-0.9}$ & $-$ & $-$ & \citet{Espinoza2011a} \\
        J1825$-$0935 & 2 & $54115.78(4)$ & $117^{+13}_{-11}$ & $1^{+2}_{-1}$ & $1^{+2}_{-1}$ & $-$ & $-$ & \citet{Yuan2010} \\
        J1826$-$1334 & 1 & $56534(10)$ & $129.6 \pm 0.2$ & $1.27^{+0.11}_{-0.08}$ & $1.27^{+0.11}_{-0.08}$ & $-$ & $-$ & This work \\
        J1826$-$1334 & 2 & $56690(9)$ & $2421.2 \pm 0.3$ & $4.5^{+0.7}_{-0.5}$ & $4.0^{+0.5}_{-0.4}$ & $164^{+162}_{-87}$ & $0.002(1)$ & This work \\
        J1837$-$0604 & 1 & $55873(11)$ & $1376 \pm 1$ & $8 \pm 3$ & $1.5^{+1.2}_{-0.8}$ & $328^{+125}_{-100}$ & $0.06(2)$ & This work$^{\dagger}$ \\
        J1837$-$0604 & 2 & $56503(4)$ & $16.7^{+0.8}_{-0.7}$ & $\lesssim 0.7$ & $\lesssim 0.7$ & $-$ & $-$ & This work$^{\dagger}$ \\
        J1841$-$0345 & 1 & $58455(7)$ & $112.3^{+0.8}_{-0.7}$ & $56.3^{+0.83}_{-6.1}$ & $56.3^{+0.83}_{-6.1}$ & $-$ & $-$ & This work \\
        J1841$-$0524 & 1 & $54503(21)$ & $1032.8 \pm 0.5$ & $\lesssim 2$ & $\lesssim 0.1$ & $488^{+162}_{-150}$ & $0.023(9)$ & \citet{Weltevrede2010} \\
        J1841$-$0524 & 2 & $55524(19)$ & $806.2^{+0.6}_{-0.5}$ & $\lesssim 2$ & $\lesssim 0.1$ & $288^{+112}_{-87}$ & $0.018(5)$ & This work$^{\dagger}$ \\
        J1841$-$0524 & 3 & $56567(2)$ & $23 \pm 1$ & $\lesssim 0.5$ & $\lesssim 0.5$ & $-$ & $-$ & This work$^{\dagger}$ \\
        J1847$-$0402 & 1 & $55509(1)$ & $0.47 \pm 0.03$ & $0.02^{+0.03}_{-0.01}$ & $0.02^{+0.03}_{-0.01}$ & $-$ & $-$ & This work$^{\dagger}$ \\
        J1847$-$0402 & 2 & $58244(21)$ & $0.16 \pm 0.04$ & $0.01^{+0.02}_{-0.01}$ & $0.01^{+0.02}_{-0.01}$ & $-$ & $-$ & This work \\
        \hline
    \end{tabular}
    \begin{tablenotes}
      \item
    \end{tablenotes}
    \end{threeparttable}
    \renewcommand{\arraystretch}{}
\end{table*}
For the remaining 74 glitches we report their properties for the first time.
With the addition of this work, there are now more than 600 glitches known across the pulsar population.

The glitch with the largest amplitude in our sample was that of PSR~J1650$-$4502 on MJD 57780, where $\Delta\nu_{g}/\nu = (12767^{+2}_{-1}) \times 10^{-9}$.
In terms of fractional glitch size, this event is the fourth-largest detected in a rotation-powered pulsar and seventh largest overall when including glitches in magnetars. 
The recovered step change in $\Delta\dot{\nu}_{g}/\dot{\nu} = (290^{+80}_{-60}) \times 10^{-3}$ and small fractional glitch recovery of $Q = 0.0061(8)$, are typical of glitches with similar fractional sizes \citep[e.g.][]{Yu2013}.
PSR~J1650$-$4502 does appear to be somewhat of an outlier among pulsars that have exhibited very large fractional glitch sizes.
It has a $\dot{P}$ that is almost an order of magnitude smaller than the cluster of pulsars that exhibit similarly large amplitude glitches (see panel c of figure 14 in \citealt{Yu2013}), yet possesses a similar rotation period.
A search for flux density and polarization variations following the glitch will be the subject of future work.
In contrast, the smallest fractional glitch we observed was found in PSR~J1452$-$6036 on MJD 57115.
With an amplitude of only $\Delta\nu_{g}/\nu = 0.13 \pm 0.03 \times 10^{-9}$, this glitch is among the 16 smallest glitches ever recorded. 

\subsection{Minimum detectable glitch size and sample completeness}

\begin{figure*}
    \centering
    \includegraphics[width=0.95\linewidth]{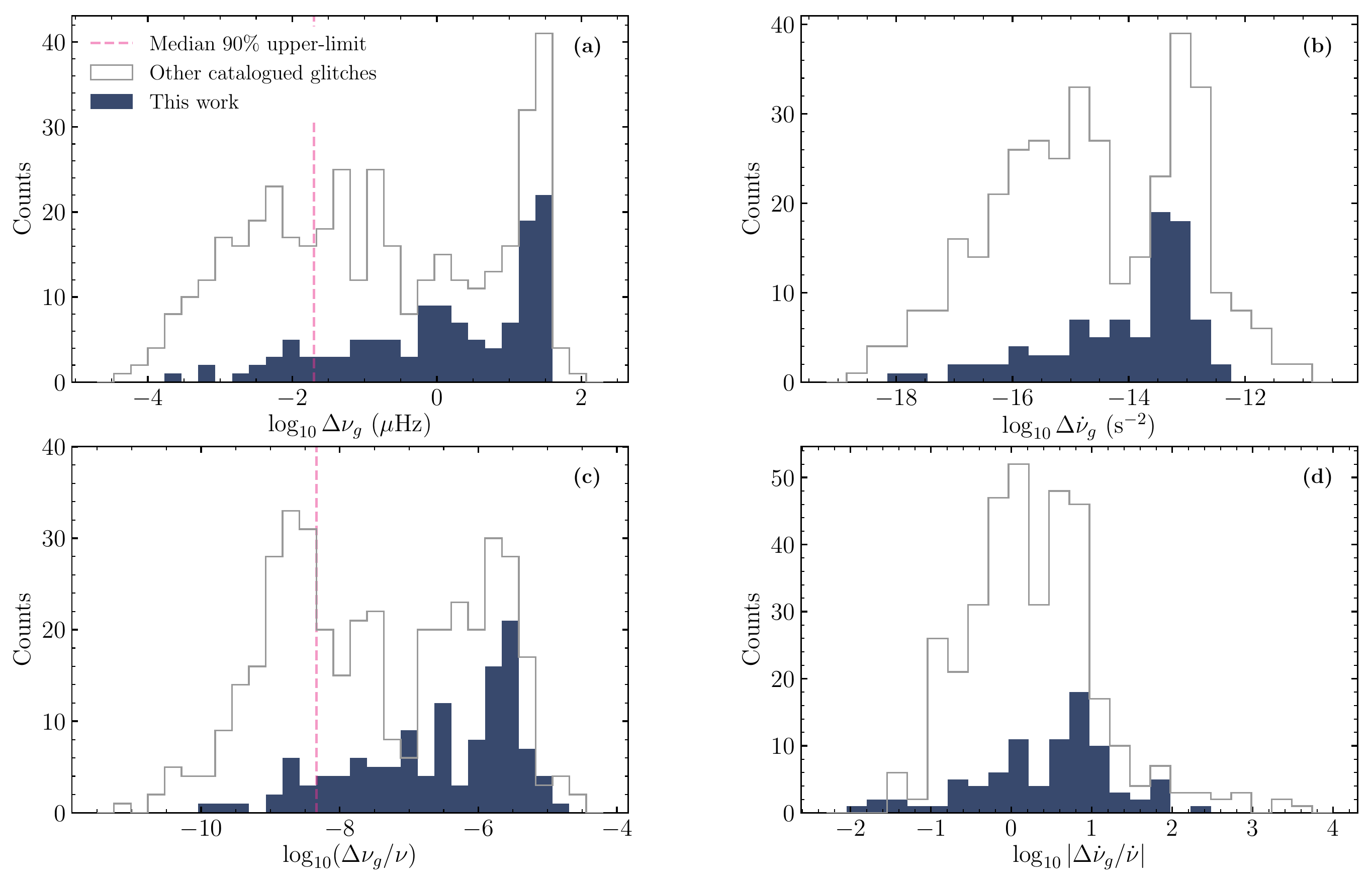}
    \caption{Histograms of $\Delta\nu_{g}$ (a), $\Delta\dot{\nu}_{g}$ (b), $\Delta\nu_{g}/\nu$ (c) and $\Delta\dot{\nu}_{g}/\dot{\nu}$ (d) for the 124 glitches in our sample (`This work'; filled dark blue) and the larger ATNF glitch catalogue after removing those listed in Table \ref{tbl:glitches} (open grey). The dashed magenta lines in panels (a) and (c) indicate the median upper-limit on the glitch size after averaging across our pulsar sample (see text for details).}
    \label{fig:glitch_hist}
\end{figure*}

A comparison of how our sample of 124 glitches fit in with those previously published is shown in Figure~\ref{fig:glitch_hist}, where our measurements of $\Delta\nu_{g}$ and $\Delta\dot{\nu}_{g}$ span nearly the complete spectrum of reported values in v1.64 of the ATNF glitch catalogue \citep{Manchester2005}\footnote{\href{https://www.atnf.csiro.au/research/pulsar/psrcat/glitchTbl.html}{https://www.atnf.csiro.au/research/pulsar/psrcat/glitchTbl.html}}.
The general lack of glitches below $\Delta\nu_{g} \lesssim 10^{-2}$\,$\mu$Hz, or $\Delta\nu_{g}/\nu \lesssim 10^{-9}$ in the bottom panel, is likely the result of a selection effect where smaller glitches become increasingly difficult to detect becaue of limitations in pulsar observing cadence. 
Indeed, the relatively sparse observation cadence of the P574 programme (approximately one observation every month) makes it difficult to differentiate the smallest glitches ($\Delta\nu_{g}/\nu \lesssim 1 \times 10^{-9}$) from stochastic variations in pulse phase due to timing noise. 
As a result, three previously catalogued small glitches were not picked up by the HMM detector in PSR~J1740$-$3015 along with the glitch in PSR~J1825$-$0935. 
Note, the non-detection of the glitch in PSR~J1825$-$0935 may be a result of it occurring within a 651\,d gap in the Parkes timing observations. 
However, targeted parameter estimation with {\sc TempoNest} was able to recover all but one of the small glitches in PSR~J1740$-$3015, and the missed glitch in PSR~J1825$-$0935.

A natural question to ask is: for a given data set, what is the smallest glitch which can be reliably detected?
We tackled this question empirically using synthetic data sets.
For reasons of practicality we could only approach the question this way using the HMM detector -- even for a single data set many hundreds of injections are necessary to give a reasonable estimate of the performance of the detector, and so an automated approach to glitch detection is required.
The essential figure of merit is a 90\% frequentist upper limit on the size of undetected glitches in each data set, denoted $\Delta\nu_g^{90\%}$ (see Appendix \ref{appdx:upper_lims} for details).
We calculated a $\Delta\nu_g^{90\%}/\nu$ for the full timing baseline of nearly every pulsar considered in the sample, as well as a separate $\Delta\nu_g^{90\%}/\nu$ across each stretch of inter-glitch data.
A complete list of the $\Delta\nu_g^{90\%}/\nu$ values can be found in Table~B1.

There is considerable variation in $\Delta\nu_g^{90\%}/\nu$ across the sample with values between $1.4 \times 10^{-10}$ and $7.7 \times 10^{-8}$.
This is attributable to differences in the amount of timing noise in the individual pulsars.
Variations may also arise due to differences in observing cadence, particularly the presence of long gaps in the data.
As a result, we obtain mean and median upper limits of $\Delta\nu_g^{90\%}/\nu = 8.1 \times 10^{-9}$ and $\Delta\nu_g^{90\%}/\nu = 4.6 \times 10^{-9}$ respectively across the entire sample.

\subsection{Vela pulsar timing and pulsars with more than five glitches}\label{subsec:multi_glitch}

\textbf{PSR J0835$-$4510 (B0833$-$45, `Vela')}: The Vela pulsar was the first pulsar observed to glitch \citep{Radhakrishnan1969b} and has been observed for over five decades.
Our attempts to fit the three large glitches experienced by Vela within our timing data via {\sc TempoNest} did not converge, which may be due to the numerical precision issue that also affected the \citet{Shannon2016} analysis of 21\,yrs of Vela pulsar timing.
While \citet{Shannon2016} worked around this issue by implementing long-double precision in a bespoke version of {\sc TempoNest}, implementing a similar correction was not practicable for our analysis. 
Instead, we performed generalised least-square fitting to the Vela pulsar ToAs to measure its glitch properties using {\sc tempo2}.
We added additional glitch parameters to the model in an iterative fashion, where parameters were only kept in the final model if the weighted root-mean-square of the residuals was lower than before they were included.
In Table~\ref{tbl:glitches} we report improved measurements of $\Delta\nu_{g}$ and $\Delta\dot{\nu}_{g}$ for the two glitches on MJDs 55408 and 56555 over the previously reported values in \citet{Buchner2010ATel} and \citet{Buchner2013ATel}.
Only single recovery processes were measured for all three glitches, with decay timescales of $13 \pm 2$\,d, $149 \pm 8$\,d and $5.9 \pm 0.1$\,d respectively.
Our timing cadence was insufficient to resolve the short-term $0.96 \pm 0.17$\,d recovery reported by \citet{Sarkissian2017}.
The additional long-term recovery processes that \citet{Shannon2016} found evidence for could not be constrained, largely due to our inability to model the glitches and timing noise simultaneously via the least-squares fitting of {\sc tempo2}. 
Improved modelling of the Vela glitches is left for future work. 

\textbf{PSR~J1341$-$6220 (B1338$-$62):} This pulsar has a low characteristic age ($\tau_{c} = 12.1$\,kyr) and is potentially associated with the supernova remnant G308.8$-$0.1 \citep{Kaspi1992}.
It is the most prolific glitching pulsar in our sample, with 24 previously published glitches found between MJD 47989 and 55484. In addition to re-analysing three glitches previously found by \citet{Weltevrede2010} and \citet{Yu2013}, we report the discovery of a further 12 new glitches that occurred between MJD 55484 and 58214.
The prevalence of glitches in this pulsar precluded the use of our standard {\sc TempoNest}-based approach to inferring the glitch properties, as the large dimensionality of the timing model resulted in even highly parallelised {\sc TempoNest} runs failing to converge.
Instead, we computed generalised least-squares fits to the ToAs using {\sc tempo2}.
As a result, our recovered glitch parameters will be slightly contaminated by unaccounted red noise.
The 14th and 15th glitches are separated by only a single observing epoch, hence we could only obtain a joint measurement of their properties, where $\Delta\nu_{g}/\nu = (155.4 \pm 0.5) \times 10^{-9}$ and $\Delta\dot{\nu}_{g}/\dot{\nu} = (1.47 \pm 0.03) \times 10^{-3}$
We are confident there are indeed two closely spaced glitches as opposed to a single glitch as there is a clear detection of the pulsar during the intervening epoch, and our attempts to fit for only a single glitch assuming the epoch of glitch 14 failed to whiten the residuals.
Glitch 9 is the largest to be reported in this pulsar to date, and only the fifth found to have a fractional amplitude greater than $\Delta\nu_{g}/\nu = 10^{-6}$, while glitch 4 has the smallest amplitude of any found in this pulsar.

\textbf{PSR~J1413$-$6141:} We report seven new glitches in this pulsar with a variety of amplitudes ranging between $\Delta\nu_{g}/\nu = (30 \pm 2)\times 10^{-9}$ to $(2137 \pm 2) \times 10^{-9}$. 
These glitches are smaller than the largest reported glitch in this pulsar on MJD $54303$ \citep{Yu2013}, where our re-analysis recovered an amplitude of $\Delta\nu_{g}/\nu = (2412 \pm 3) \times 10^{-9}$.
The fifth glitch in our sample is the smallest to have been found in this pulsar to date, smaller than the $\Delta\nu_{g}/\nu = (39 \pm 4) \times 10^{-9}$ glitch on MJD 51290~\citep{Yu2013}.
None of these glitches appear to have induced detectable step-changes in $\dot{\nu}$, nor show evidence for exponential recoveries in $\nu$ or $\dot{\nu}$.

\textbf{PSR~J1740$-$3015 (B1737$-$30):}
As for PSR~J1341$-$6220, the large number of glitches necessitated using the generalised least-squares method implemented in {\sc tempo2} to measure their properties. 
Of the eight previously published glitches, only seven were recovered with values of $\Delta\nu_{g}$ that are inconsistent with zero. 
We only obtained an upper-limit of $\Delta\nu_{g}/\nu \lesssim 0.9 \times 10^{-9}$ for the small glitch reported by \citet{Jankowski2016ATel} as occurring on MJD 57346.
This glitch was not picked up by the HMM glitch detection algorithm.
It was also not recovered in a re-analysis of the UTMOST timing data by \citet{Lower2020}, suggesting this event may have been a misidentified variation in spin-phase caused by timing noise. 
The only unpublished glitch within our timing baseline for PSR~J1740$-$3015 occurred on MJD 55936 with a moderate amplitude of $\Delta\nu_{g}/\nu = (16 \pm 6) \times 10^{-9}$, $\Delta\dot{\nu}_{g}/\dot{\nu} = (1.1 \pm 0.3) \times 10^{-9}$ and no apparent exponential recovery. 

\subsubsection{Statistics of glitches in PSRs J1341$-$6220 and J1413$-$6141}

Recent developments in modelling the statistics of pulsar glitches have focused on microphysics-agnostic meta-models.
Two such meta-models, where stress is accumulated either as a state-dependent Poisson process \citep[SDP;][]{Carlin2019} or as a Brownian process \citep[BSA;][]{Carlin2020}, are predicted to show similar auto- and cross-correlations between their glitch amplitudes and wait-times.
However, under the fast-driven SDP, pulsars are expected to have glitch amplitude and wait-time distributions with the same overall shape \citep{Carlin2019}, whereas the amplitudes and wait-times can be drawn from differing distributions under the BSA. 

Using the sample of 34 well constrained glitches, we find PSR~J1341$-$6220 exhibits Spearman rank correlation coefficients of $\rho_{s+} = 0.67 \pm 0.14$ (p-value, $0.33 \pm 0.01$) and $\rho_{s-} = -0.18 \pm 0.18$ (p-value, $(2^{+1}_{-3}) \times 10^{-5}$) for the cross-correlations between the glitch size and the forward (referring to the previous glitch) and backward (referring to the next glitch) waiting times respectively.
Note, the p-value for the backward cross-correlation is an artefact of the significant scatter and corresponding lack of correlation. 
The moderate forward cross-correlation is consistent with previously reported values \citep{Melatos2018, Fuentes2019, Carlin2020}.
The marginal auto-correlations for the glitch wait time, $\rho_{s,\, \Delta T_{g}} = -0.0048$ (p-value $0.98$), and size, $\rho_{s,\,\Delta\nu} = -0.26$ (p-value $0.15$), are consistent with the BSA process.

The 14 glitches in PSR~J1413$-$6141 show a strong forward cross-correlation between the wait time and sizes of its glitches, with $\rho_{s+} = 0.82 \pm 0.17$ (p-value, $0.29^{+0.02}_{-0.04}$), and a weak backwards anti-correlation of $\rho_{s-} = -0.31 \pm 0.29$ (p-value, $(6^{+2}_{3})\times 10^{-4}$). 
Similar to PSR~J1341$-$6220, the auto-correlations of PSR~J1413$-$6141 are relatively weak, with Spearman coefficients for the wait time and amplitude of $\rho_{s,\, \Delta T_{g}} = -0.0070$ (p-value $0.98$) and $\rho_{s,\,\Delta\nu} = -0.27$ (p-value $0.36$) respectively.

\subsection{Recoveries}

\begin{figure}
    \centering
    \includegraphics[width=\linewidth]{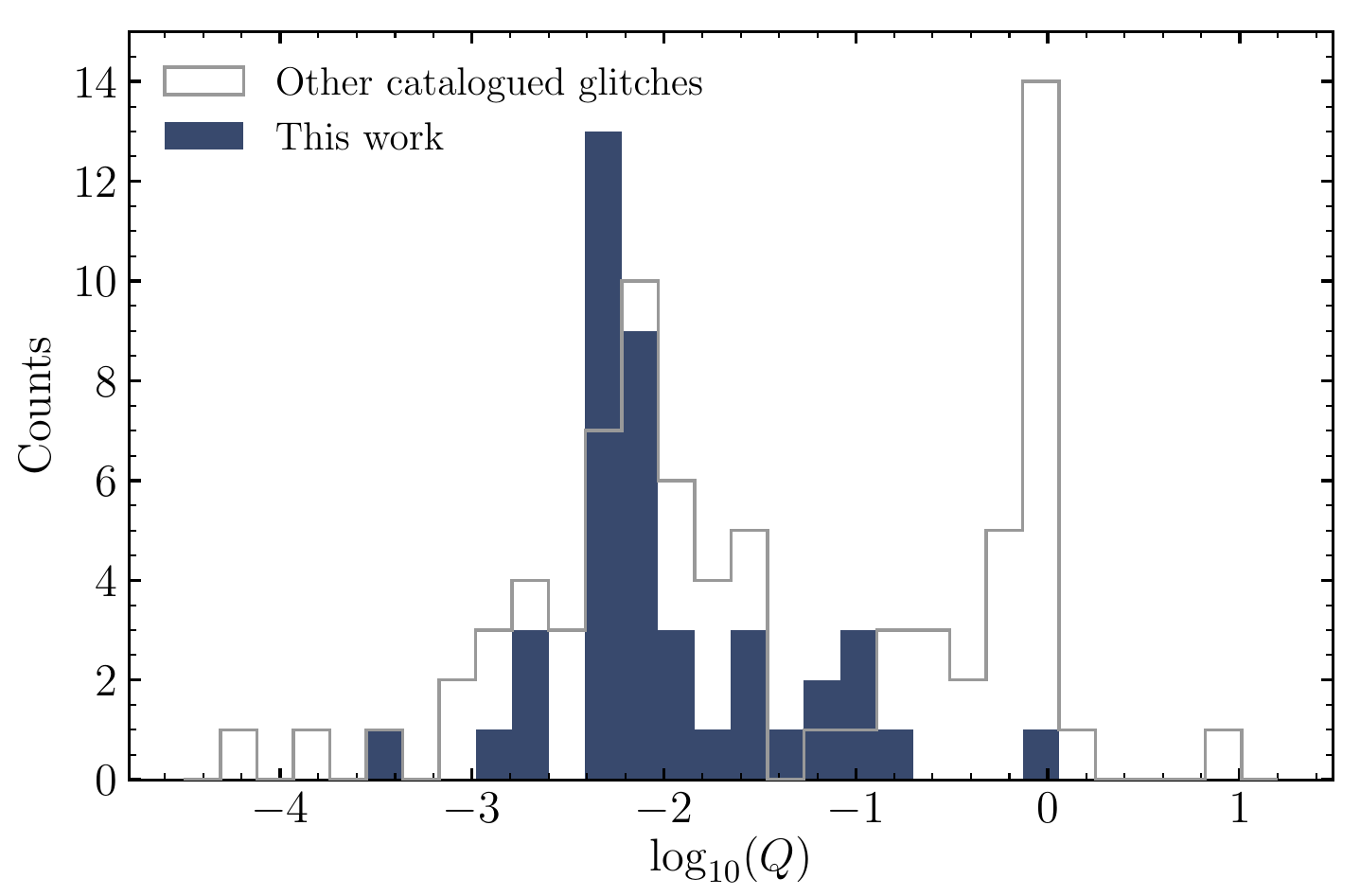}
    \caption{Same as Figure \ref{fig:glitch_hist} but for the distribution of fractional glitch recoveries ($Q$).}
    \label{fig:rcov_hist}
\end{figure}

Of the 124 pulsar glitches listed in Table \ref{tbl:glitches} only 41 were found to have at least one exponential recovery where the recovery parameters are well constrained. 
Figure \ref{fig:rcov_hist} shows the fractional recovery of the spin-frequency following a glitch.
Our sample possesses only a single peak at $Q \sim 0.0069$, unlike the bi-modal distribution seen in the overall glitch sample, but we note that the second peak at high $Q$-values largely comprises of recoveries measured in the Crab pulsar (PSR~J0534$+$2200; \citealt{Lyne2015}), which is not observed by our programme.
The only glitch from our set that has high $Q$ is the MJD 54762 glitch in PSR~J1617$-$5055.

For the large glitch in PSR~J1757$-$2421 on MJD 55702, our parameter estimation recovered the $\sim$15- and $\sim$92-day glitch recovery timescales found by \citet{Yuan2017} in addition to a third, long-term exponential recovery process with $\tau_{d,3} \sim 618$\,days.
This three-component recovery model is strongly preferred over one containing only two-components, with $\ln(\mathcal{B}) = 10.3$.
The presence of this additional recovery process would explain the stronger post-glitch timing noise they measured. 
Each of the three exponential recoveries exhibited sequentially larger recovery fraction, a phenomenon that is seen in the short- and long-term recoveries of the Vela pulsar \citep[e.g.][]{Shannon2016}.
Our total fractional recovery of $Q = 0.0038(2)$ is consistent with the value of $Q = 0.0035(9)$ found by \citet{Yuan2017}, however their measurement only included the first two recoveries. 
This could be due the difference in methods that were applied for measuring the recovery terms, as we fit for all of the glitch parameters simultaneously as opposed to incrementally adding and fitting for additional parameters.

\subsection{Glitch statistics}
An empirical relation for determining the glitch rate for a given pulsar was developed by \citet{Fuentes2017}. They divided glitches into `large' and `small' with a separation at $\Delta\nu_{g} = 10\mu$Hz. As the completeness to small glitches is not well established and the size distribution of large glitches is relatively tight (see Figure~\ref{fig:glitch_hist}), they were able to derive an approximate wait time, $T_g$, between large glitches via
\begin{equation}
    T_g = \frac{1}{420\,{\rm Hz}^{-1} \,\, \dot{\nu}}.
    \label{eqn:rate}
\end{equation}
We can test this on our sample of pulsars. Combining the data sets from \citet{Parthasarathy2019} and this paper, we have observed 159 pulsars over a 10~year span. From Equation~\ref{eqn:rate}, assuming Poisson statistics for the glitch waiting times, our expectation is for 25 pulsars to undergo a total of 45 glitches for an average realisation. This compares well with the actual value of 28 pulsars having experienced 44 large glitches.

When we rank the pulsars by expected glitch activity, PSR J1513-5908 is highest, with the expectation of a glitch every 1.1~yr. In fact, this pulsar has not had a single glitch (large or small) in more than 35 yr of observing \citep{Parthasarathy2020}. Apart from this singular anomaly, 18 of the next 20 pulsars in rank order have all had large glitches. We also note that of the eight pulsars with multiple large glitches all have values of $T_g < 12$\,yr. In our sample of pulsars with a large glitch, PSR~J1757$-$2421 has the longest wait time of 320~yr, or a glitch probability in 10 years of only 0.0028 according to Equation \ref{eqn:rate}.
It is clear that Equation~\ref{eqn:rate} is a simplification of the underlying processes which determine glitch activity (see e.g. \citealt{Melatos2018}), and assumes that large and small glitches arise from different processes.
Nevertheless, it does a reasonable job at predicting wait-times for pulsars with characteristic ages below $10^5$~yr.

\subsection{New and updated pulsar parameters}\label{sec:params}

In addition to fitting for the rotational properties of our pulsars, we also generated posterior distributions with {\sc TempoNest} for their astrometric properties.
Tables containing the recovered astrometric and rotational properties of each pulsar can be found in Appendix \ref{appdx:ephem}.
For completeness, figures showing the whitened timing residuals for each pulsar after removing the maximum a posteriori realisation of their timing models (both deterministic and stochastic components) are available in the supplementary materials.

As a result, we measured the proper-motions of 11 pulsars,
summarized in Table~\ref{tbl:proper_motion}.
\begin{table}
\begin{center}
\caption{Pulsars with significant proper motions. Those with proper motions measured for the first time are highlighted in bold. Unless otherwise specified, the distances reported here are inferred from pulsar dispersion measures. $^{\star}$Distance from parallax measurement. $^{\dagger}$Average distance from electron density models and HI-observations. \label{tbl:proper_motion}}
\renewcommand{\arraystretch}{1.2}
\setlength{\tabcolsep}{4pt}
\resizebox{\linewidth}{!}{
\begin{tabular}{lccccc}
\hline
PSRJ & $\mu_{\alpha}$ & $\mu_{\delta}$ & $\mu_{\rm T}$ & $D$ & $V_{{\rm T}}$ \\
 & (mas yr$^{-1}$) & (mas yr$^{-1}$) & (mas yr$^{-1}$) & (kpc) & (km s$^{-1}$) \\
\hline
J0659$+$1414 & $49 \pm 13$ & $78^{+75}_{-68}$ & $96^{+70}_{-45}$ & $3.47^{\star}$ & $132^{+96}_{-62}$ \\
\textbf{J0908$-$4913} & $-37 \pm 9$ & $31 \pm 10$ & $47 \pm 9$ & $3.0^{\dagger}$ & $674 \pm 127$ \\
\textbf{J1003$-$4747} & $-12 \pm 2$ & $21 \pm 2$ & $24 \pm 3$ & $0.37$ & $42 \pm 3$ \\
J1057$-$5226 & $49 \pm 4$ & $-6 \pm 5$ & $50 \pm 4$ & $0.09$ & $21 \pm 2$ \\
J1320$-$5359 & $13 \pm 2$ & $52 \pm 2$ & $54 \pm 2$ & $2.2$ & $563^{+24}_{-21}$ \\
\textbf{J1359$-$6038} & $-4 \pm 4$ & $10 \pm 5$ & $12^{+5}_{-4}$ & $5.0$ & $278^{+109}_{-100}$ \\
\textbf{J1452$-$6036} & $-5 \pm 3$ & $-5 \pm 3$ & $7 \pm 3$ & $6.1$ & $204^{+96}_{-88}$ \\
J1709$-$4429 & $17 \pm 3$ & $11 \pm 10$ & $17 \pm 4$ & $2.6$ & $210 \pm 45$ \\
J1731$-$4744 & $60^{+11}_{-9}$ & $-178^{+24}_{-22}$ & $183 \pm 23$ & $0.7$ & $607 \pm 75$ \\
J1826$-$1334 & $32^{+8}_{-9}$ & $-$ & $32^{+8}_{-9}$ & $3.6$ & $549^{+144}_{-162}$ \\
\hline
\end{tabular}
}
\renewcommand{\arraystretch}{}
\end{center}
\end{table}
Here the proper-motion in RA and DEC are given by $\mu_{\alpha} \equiv \dot{\alpha} \cos(\delta)$ and $\mu_{\delta} \equiv \dot{\delta}$, which when combined gives the total proper-motion $\mu_{\rm T} = \sqrt{\mu_{\alpha}^{2} + \mu_{\delta}^{2}}$.
The 2-D (or 1-D in the case of a proper-motion in only a single coordinate) transverse velocity is calculated as
\begin{equation}
    V_{\rm T} = 4.74\,{\rm km}\,{\rm s}^{-1} \Big( \frac{\mu_{\rm T}}{{\rm mas}\,{\rm yr}^{-1}} \Big) \Big( \frac{D}{{\rm kpc}} \Big),
\end{equation}
where $D$ is the distance to the pulsar.
For most pulsars we used the median dispersion measure distance returned by the TC93~\citep{Taylor1993}, NE2001~\citep{Cordes2002} and YMW16~\citep{Yao2017} Galactic free-electron density models.
Where available, we used distances inferred from the parallax measurements~\citep[e.g.][]{Deller2019}.

\begin{figure}
    \centering
    \includegraphics[width=\linewidth]{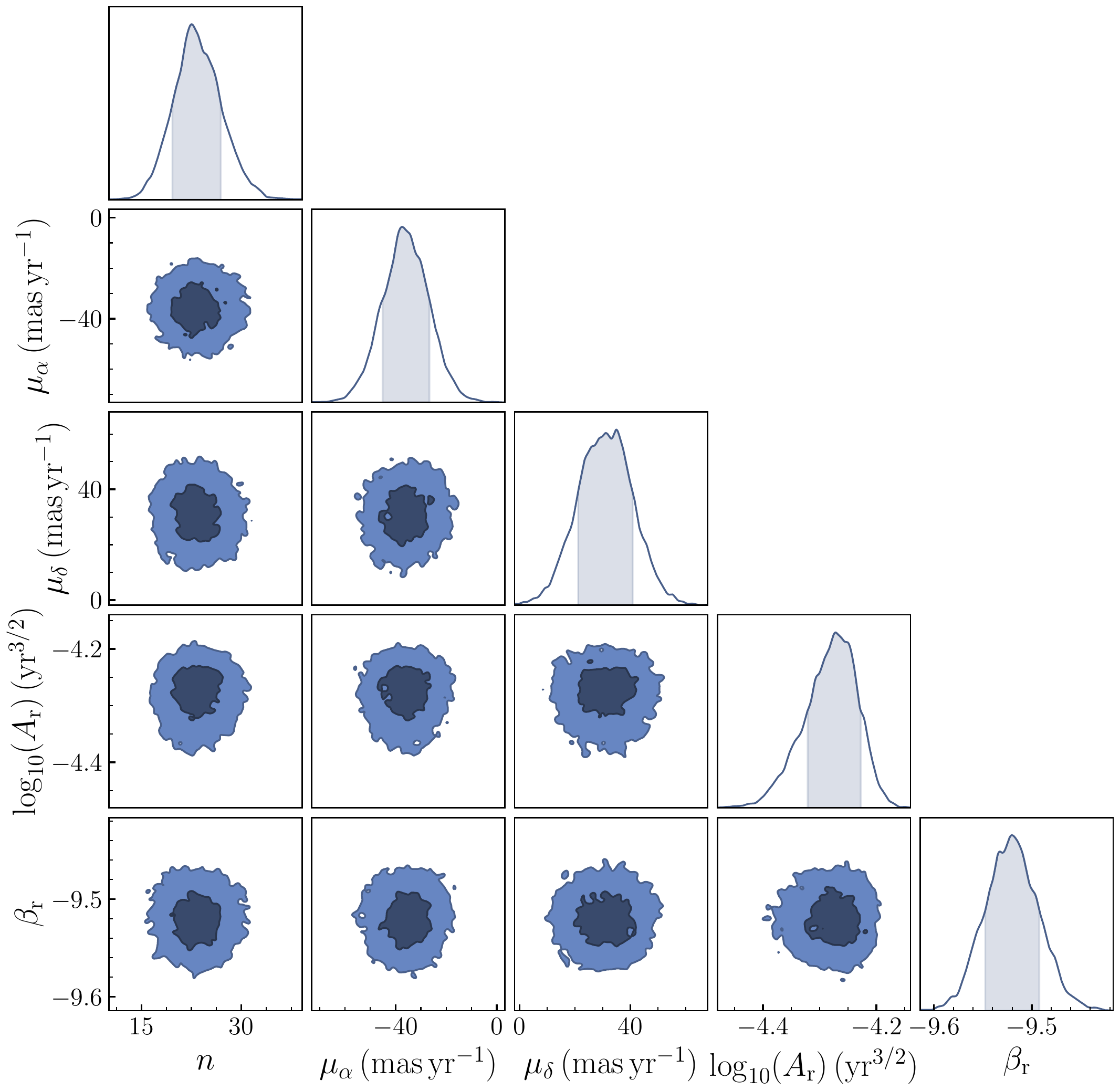}
    \caption{One- and two-dimensional posterior distributions for the braking index, proper motion and red noise parameters of PSR~J0908$-$4913.}
    \label{fig:j0908_post}
\end{figure}

Of the four new measurement of proper motion, those of PSRs J1359$-$6038 and J1452$-$6036 are well constrained in $\mu_\alpha$ and $\mu_\delta$, with both pulsars possessing $V_{\rm T}$ that are consistent with the bulk pulsar population~\citep[see][]{Hobbs2005} at their nominal distance.
For PSR~J0908$-$4913 (B0906$-$49), the posterior distributions for $\mu_\alpha$ and $\mu_\delta$ are shown in Figure~\ref{fig:j0908_post}. The pulsar is moving in a north-westerly direction, which is consistent with the implied direction of motion from radio imaging of its bow-shock nebula~\citep{Gaensler1998}. The implications of this proper motion are discussed further in a companion paper \citep{Johnston2021b}.
Finally for PSR~J1003$-$4747 there is a large disparity in its distance inferred via the NE2001 and YMW16 electron-density models, and hence its implied $V_{\rm T}$. The NE2001 model gives a distance of $2.9$\,kpc with a corresponding $V_{{\rm T}} = 335 \pm 27$\,km\,s$^{-1}$, whereas the YMW16 distance of only $0.37$\,kpc gives $V_{{\rm T}} = 42 \pm 3$\,km\,$^{-1}$.
A velocity measurement via scintillation or a direct distance measurement via parallax would solve this issue.

Our measurements for pulsars with previously reported proper motions are largely consistent with the published values.
This includes PSR~J1709$-$4429, where our radio measurements are consistent with the values of $\mu_{\alpha} = 13 \pm 2$ and $\mu_{\delta} = 1 \pm 2$ obtained from X-ray imaging \citep{deVries2021}.
The proper-motion of PSR~J1731$-$4744 (B1727$-$47) is consistent with the values of $\mu_{\alpha} = 73 \pm 15$\,mas\,yr$^{-1}$ and $\mu_{\delta} = -132 \pm 14$\,mas\,yr$^{-1}$ inferred by \citet{Shternin2019}, adding further credence to the claimed association with the supernova remnant RCW~114.

Finally, PSR~J1617$-$5055 has a small characteristic age ($8$\,kyr) and high rotational kinetic energy loss rate ($\dot{E} = 1.6\times10^{37}$\,ergs\,s$^{-1}$) but no pulsed gamma-ray emission has yet been detected from the pulsar.
One suggestion for this absence was the lack of a coherent timing solution due to the large amount of timing noise~\citep{Abdo2013}.
However, even our phase coherent timing solution was unable to recover pulsed gamma-ray emission in \textit{Fermi} data covering MJD 54220--56708 (D.~A.~Smith 2020, private communication).
As observations of PSR~J1617$-$5055 were discontinued after MJD 56708 our timing model is unable to be used for folding additional \textit{Fermi} photons, as any subsequent glitches are unaccounted for.

\section{Braking indices and long-term evolution}\label{sec:brake}

\begin{figure}
    \centering
    \includegraphics[width=0.95\linewidth]{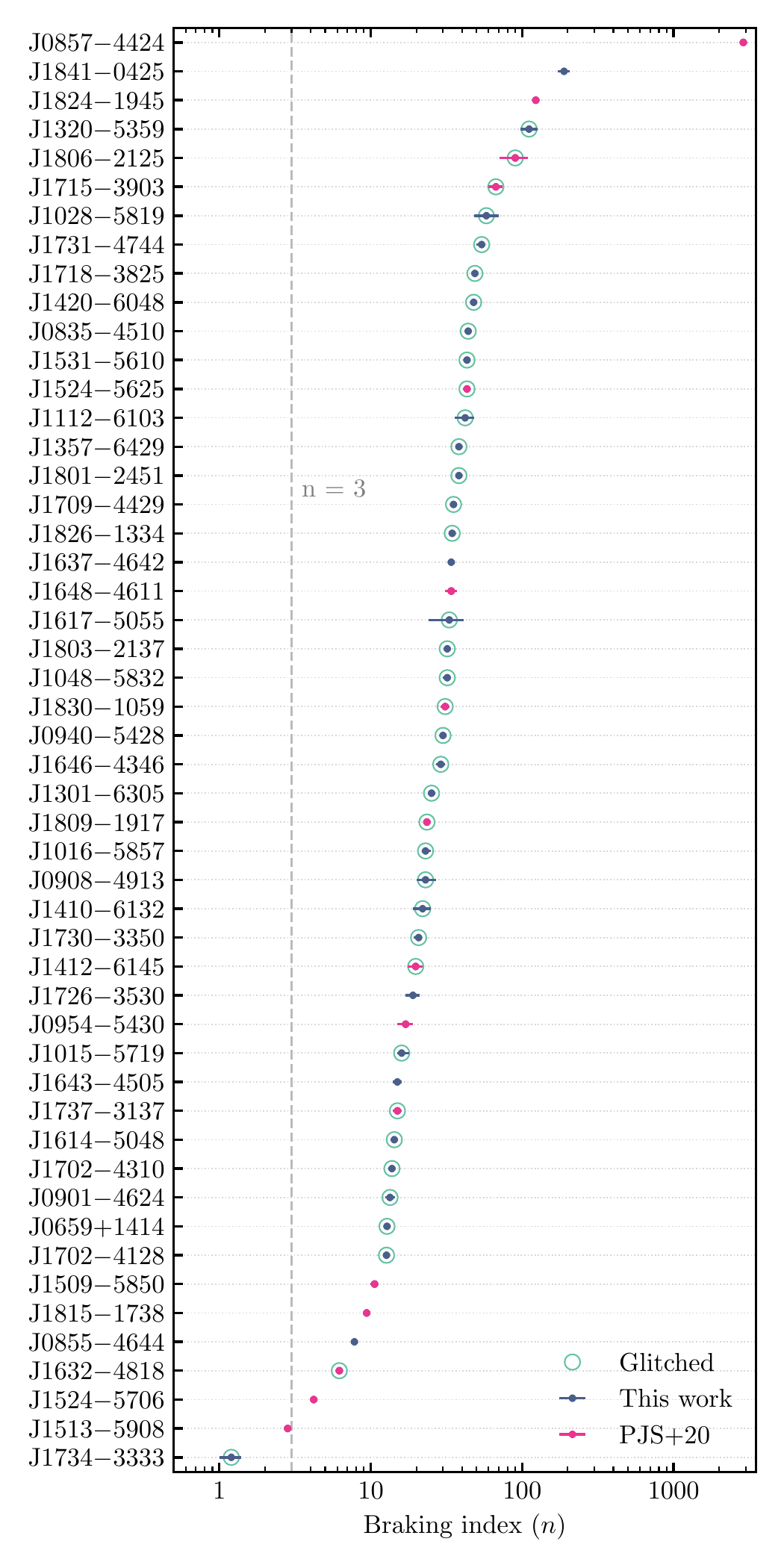}
    \caption{Median recovered braking indices and 68\% confidence intervals for the 33 pulsars in Table~\ref{tbl:brake} (dark blue) and \citet{Parthasarathy2020} (PJS+20; magenta). Glitching pulsars are highlighted by green circles.}
    \label{fig:brake}
\end{figure}

Of the 74 pulsars in our sample, we found 34 were best described by a timing model that included a power-law red noise process and a $\ddot{\nu}$-term.
The resulting Bayes factors in favour of the model containing a $\ddot{\nu}$-term over one where $\ddot{\nu}$ is fixed at zero, along with the inferred $n$ from Equation \ref{eqn:brake} and observing timespan are listed in Table \ref{tbl:brake}.
A graphical comparison of the braking indices and associated 68\,per cent confidence intervals from both this work and \citet{Parthasarathy2020} is depicted in Figure \ref{fig:brake}.

\begin{table}
\begin{center}
\caption{Braking indices for pulsars with $\ln(\mathcal{B}^{\mathrm{F2} \neq 0}_{\mathrm{F2} = 0}) > 3$. Those highlighted with a `${\star}$' have been observed to glitch. Pulsars above the line have either never been seen to glitch, or do not have large values of $\Delta\dot{\nu}_{g}/\dot{\nu}$ ($\lesssim 1 \times 10^{-3}$) associated with their glitches. For pulsars below the line, the listed values of $n$ are representative of their average inter-glitch braking, not their long-term evolution. \label{tbl:brake}}
\renewcommand{\arraystretch}{1.2}
\begin{tabular}{lccc}
\hline
PSRJ & $\ln(\mathcal{B}^{\mathrm{F2} \neq 0}_{\mathrm{F2} = 0})$ & Braking index ($n$) & $T$ (d) \\
\hline
J0659$+$1414$^{\star}$ & $21.0$  & $12.8^{+0.3}_{-0.2}$ & $3964$ \\
J0855$-$4644 & $16.8$  & $7.8^{+0.3}_{-0.2}$ & $4249$ \\
J0901$-$4624$^{\star}$ & $122.2$ & $13.4^{+1.0}_{-0.9}$ & $7620$ \\
J0908$-$4913$^{\star}$ & $11.8$  & $23^{+4}_{-3}$ & $10112$ \\
J1320$-$5359$^{\star}$ & $14.2$  & $111^{+16}_{-14}$ & $7933$ \\
J1410$-$6132$^{\star}$ & $8.2$   & $22 \pm 3$ & $4116$ \\
J1718$-$3825$^{\star}$ & $57.8$  & $48.7^{+0.7}_{-0.8}$ & $7593$ \\
J1726$-$3530 & $30.5$  & $19 \pm 2$ & $6028$ \\
J1734$-$3333$^{\star}$ & $5.4$   & $1.2 \pm 0.2$ & $5986$ \\
J1841$-$0425$^{\star}$ & $14.5$  & $189 \pm 18$ & $4202$ \\
\hline
J0835$-$4510$^{\star}$ & $227.2$ & $44 \pm 2$ & $4209$ \\
J0940$-$5428$^{\star}$ & $49.9$  & $30 \pm 1$ & $8353$ \\
J1015$-$5719$^{\star}$ & $3.6$   & $16^{+2}_{-1}$ & $4249$ \\
J1016$-$5857$^{\star}$ & $49.7$  & $23^{+2}_{-1}$ & $7170$ \\
J1028$-$5819$^{\star}$ & $14.9$  & $58^{+12}_{-10}$ & $3906$ \\
J1048$-$5832$^{\star}$ & $65.0$  & $32 \pm 2$ & $10560$ \\
J1112$-$6103$^{\star}$ & $10.3$  & $42 \pm 6$ & $7620$\\
J1301$-$6305$^{\star}$ & $88.0$  & $25.2 \pm 0.7$ & $7504$ \\
J1357$-$6429$^{\star}$ & $190.5$ & $38.2 \pm 0.5$ & $6978$ \\
J1420$-$6048$^{\star}$ & $255.5$ & $47.8^{+0.9}_{-1.0}$ & $7175$ \\
J1524$-$5625$^{\star}$ & $73.6$  & $43.2 \pm 0.7$ & $4249$ \\
J1614$-$5048$^{\star}$ & $56.1$   & $14.3^{+0.3}_{-0.6}$ & $10140$ \\
J1617$-$5055$^{\star}$ & $19.8$  & $33^{+8}_{-9}$ & $2488$ \\
J1646$-$4346$^{\star}$ & $17.2$  & $29 \pm 2$ & $10557$ \\
J1702$-$4128$^{\star}$ & $8.8$   & $12.7 \pm 0.6$ & $4250$ \\
J1702$-$4310$^{\star}$ & $80.7$  & $13.8^{+0.5}_{-0.4}$ & $7248$ \\ 
J1709$-$4429$^{\star}$ & $187.7$ & $35.2^{+0.7}_{-0.5}$ & $10561$ \\ 
J1730$-$3350$^{\star}$ & $20.7$  & $20.7^{+0.7}_{-1.5}$ & $7783$ \\
J1731$-$4744$^{\star}$ & $29.6$  & $54^{+2}_{-4}$ & $9427$ \\ 
J1737$-$3137$^{\star}$ & $5.5$   & $15 \pm 1$ & $4752$ \\
J1801$-$2451$^{\star}$ & $345.3$ & $38.2^{+1.0}_{-0.9}$ & $9573$ \\
J1803$-$2137$^{\star}$ & $102.6$ & $32.0^{+0.8}_{-0.7}$ & $4250$ \\
J1826$-$1334$^{\star}$ & $33.2$  & $32 \pm 1$ & $4184$ \\
\hline
\end{tabular}
\renewcommand{\arraystretch}{}
\end{center}
\end{table}

\citet{Parthasarathy2020} reported a weak correlation between $n$ and $\tau_{c}$, with a Spearman coefficient of $\rho_{s} = 0.34 \pm 0.01$ amongst their sample of 19 predominately non-glitching pulsars.
We find there is no correlation among our sample of pulsars ($\rho_{s} = 0.09$, p-value $0.61$) nor in a joint analysis of both samples ($\rho_{s} = 0.15$, p-value $0.28$).
Hence the weak correlation of \citet{Parthasarathy2020} may have simply been a result of both a selection effect -- i.e. avoidance of generally younger, glitching pulsars -- and an artefact of their relatively small sample size.

Quantitatively, we assessed whether the two sets of braking index measurements satisfy the null hypothesis, i.e. that they were drawn from the same underlying distribution, by performing a two-sample Kolmogorov-Smirnov (KS) test on the cumulative distributions of $\log_{10}(n)$. 
Omitting PSR~J0857$-$4424, which has a large $n$ perhaps resulting from an unmodelled binary companion in a wide orbit \citep{Parthasarathy2020}, we obtain $\mathcal{D}_{KS} = 0.16$ and a p-value of $0.15$. Here, the KS-statistic that is lower than the critical value of $\mathcal{D}_{0.05} = 0.45$ and p-value that is $> 0.05$ indicates the null hypothesis cannot be rejected, and both distributions are indistinguishable from one another at the $0.05$ level.
If the braking indices of both glitching and non-glitching pulsars are indeed drawn from a common distribution, then the same underlying mechanism may be exerting the torque.

\subsection{Observed versus predicted evolution in $P$-$\dot{P}$ space}\label{subsec:ppdot}

\begin{figure*}
    \centering
    \includegraphics[width=\linewidth]{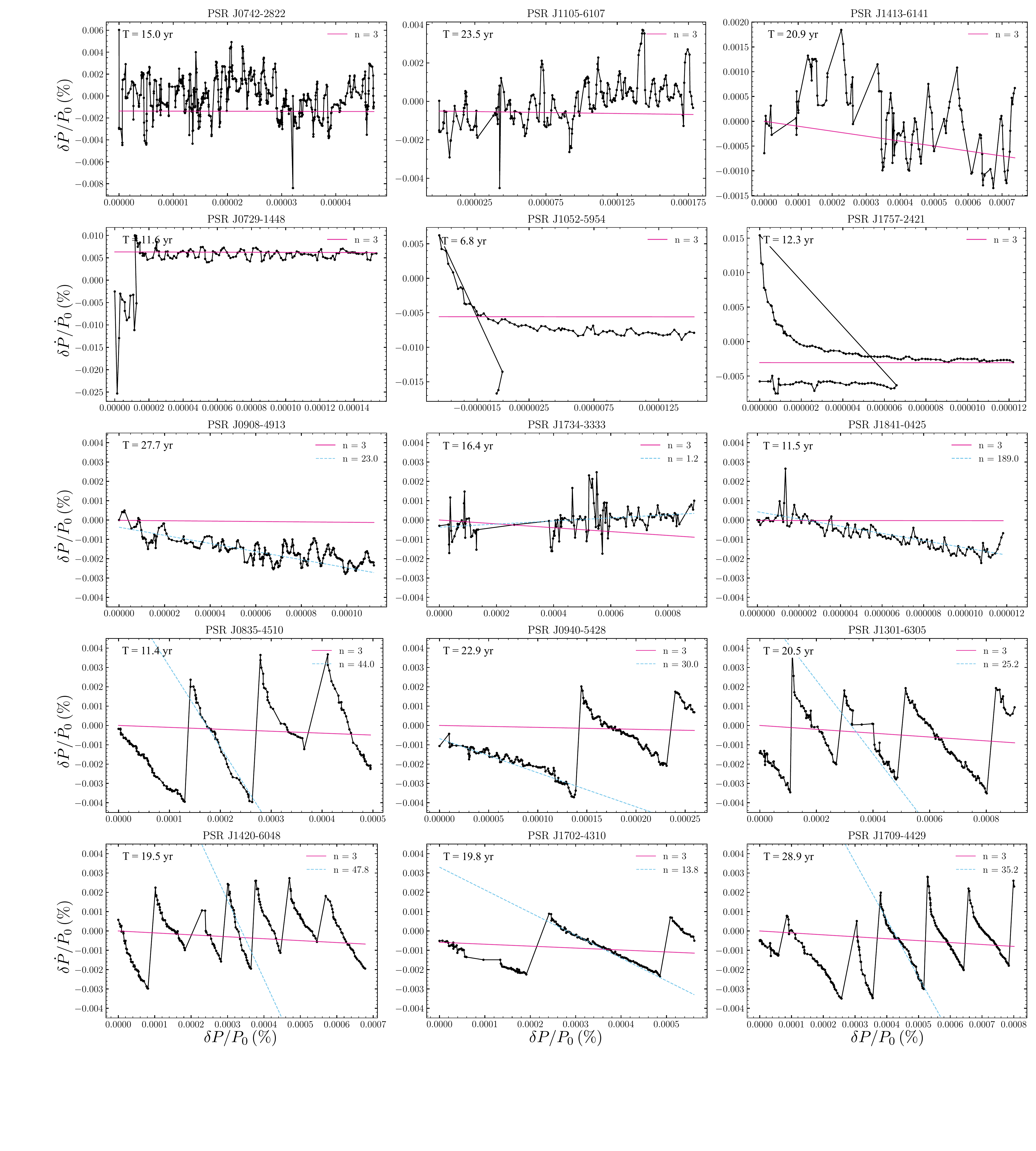}
    \caption{Observed long-term, fractional evolution of $P$ and $\dot{P}$ for a sample of 15 pulsars. The organisation of the panels follows the logic described in Section \ref{subsec:ppdot}. Expected evolutionary paths due to dipole radiation ($n = 3$) and (where possible) the measured braking indices are indicated by the solid magenta and dashed blue lines respectively. Note these lines do not represent fits to the data. See text for further details.}
    \label{fig:ppdot_evol}
\end{figure*}

Measurements of $\nu$, $\dot{\nu}$ and $n$ can be used to make predictions on how the rotation of a pulsar may evolve over long timescales.
Re-writing Equation~\ref{eqn:spin_law} in terms of $P$ and $\dot{P}$, the motion of a pulsar in the $P$-$\dot{P}$ diagram can be described as
\begin{equation}
    \dot{P} = \kappa P^{2 - n},
\end{equation}
where pulsars with $n < 2$ will move towards larger values of both $P$ and $\dot{P}$ over time, while pulsars with increasingly larger values of $n$ evolve more rapidly towards smaller $\dot{P}$ values.
A key assumption here is that $\kappa$ remains constant over time.
If this is not the case, then the measured $n$ from Equation \ref{eqn:brake} would represent the ensemble effect of both the intrinsic braking process and a process that causes $\kappa$ to vary with time.
Additionally, discontinuities and recovery effects induced by glitches also cause pulsars to deviate from expected evolutionary paths.
This kind of behaviour has been clearly seen in young pulsars such as the Crab pulsar, where the jumps in $\dot{\nu}$ associated with its glitches result in a smaller braking index of $n = 2.342 \pm 0.001$ compared to the glitch-corrected value of $n = 2.519 \pm 0.002$~\citep{Lyne2015}.

In Figure \ref{fig:ppdot_evol} we show the observed fractional evolution of $P$ and $\dot{P}$ for 15 pulsars that represent the various evolutionary pathways that are visible throughout our overall sample. 
Each of these fractional $P$-$\dot{P}$ diagrams were produced by performing stride-fits to $\nu$ and $\dot{\nu}$ over discrete windows containing at least five ToAs, and then converting the resulting values to to spin-period ($P = 1/\nu$) and period derivative ($\dot{P} = -\dot{\nu}/\nu^{2}$). 
The timing models used for performing the local fits did not account for the glitches seen in these pulsars.
For pulsars with a measured $\ddot{\nu}$ component, we overlaid both the expected evolutionary lines for $n = 3$ and their observed braking indices. 
Pulsars where $\ddot{\nu}$ could not be distinguished have only the $n = 3$ evolutionary track overlaid.
We separated the pulsars into four broad groups based on the visual appearance of their evolutionary tracks in $P$-$\dot{P}$ space.
\begin{enumerate}
    \item Flat: pulsars with a small, undetected $n$ that appear to evolve with roughly constant $\dot{P}$ over our timing baseline (top row of Figure~\ref{fig:ppdot_evol})
    \item Flat-jumps: pulsars with an unresolved $n$ with jumps in $\dot{P}$ associated with large glitches (second row of Figure~\ref{fig:ppdot_evol}).
    \item Inclined: pulsars that follow an evolutionary track defined by a constant $n$ over the duration of our timing programme (third row of Figure~\ref{fig:ppdot_evol}).
    \item Vela-like: pulsars that exhibit both large, positive inter-glitch braking indices and (quasi-)periodic jumps in $\dot{P}$ due to large glitches (bottom two rows of Figure~\ref{fig:ppdot_evol}).
\end{enumerate}

Pulsars in Table \ref{tbl:brake} that experienced glitches with $\Delta\dot{\nu}_{g}/\dot{\nu} \gtrsim 10^{-3}$ components belong to the Vela-like group.
They evolve with large $n$ between glitches, with the $\Delta\dot{\nu}_{g}$ component of the glitch serving to `reset' much of their $\dot{P}$ evolution towards the value observed immediately after the previous glitch.
Indeed, examination of the panels of Figure \ref{fig:ppdot_evol} shows the expected evolutionary tracks corresponding to values of $n$ from Table \ref{tbl:brake} matches the observed inter-glitch behaviour, while the long-term evolution of these pulsars is consistent with a `small $n$' process, though the precise value of this long-term $n$ is unclear.
This behaviour is in line with previous studies of $\dot{\nu}$ evolution among samples of actively glitching, young pulsars \citep[see figure 1 in][]{Espinoza2017}.
Measurements of the $n$ that dominates their long-term evolution would require simultaneous measurements of both the effects of glitches, pulsar rotation and astrometry, and the addition of new $\Delta\ddot{\nu}_{p}$ parameters to account for their large inter-glitch braking indices. 
Such a high-dimensional problem is incompatible with our current approach to modelling the timing of these pulsars, and is therefore left for future works.

The remaining pulsars from Table \ref{tbl:brake} fall into the inclined group of pulsars. 
Unlike the Vela-like pulsars, these pulsars have yet to undergo a large glitch within our data span, and they therefore follow the evolutionary path given by their measured $n$ over at least the timescales covered by the Parkes timing programme.
The pulsars that have undergone small glitches (such as PSR~J0908$-$4913 and PSR~J1734$-$3333) continue along their pre-glitch paths without interruption over the decades of available timing data.
Note that the y-axis scale of the inclined and Vela-like pulsars in Figure~\ref{fig:ppdot_evol} is identical, and we see that for the glitching pulsars, the maximum deviation of $\dot{P}$ from the $n=3$ line is approximately $0.004\%$. 
The inclined pulsars, however, have not yet had time to reach this critical point of their inter-glitch evolution and so we predict that they will undergo large glitches similar to the Vela-like pulsars at some point in the (near) future.
We surmise that the pulsars with high $n$ in \citet{Parthasarathy2020} fall into the same category.
While there does appear to be some level of variation in $\ddot{\nu}$ between glitches in the Vela-like pulsars shown in Figure \ref{fig:ppdot_evol}, the precise nature of the relationship between the glitches and the measured braking indices is unclear.
This aspect is further explored in Section \ref{subsec:interp_ppdot}. 

Pulsars not listed in Table \ref{tbl:brake} can be categorised as possessing evolutionary paths that fall into either the flat or flat-jump categories. 
In general, these pulsars do not show strong evidence for a $\ddot{\nu}$ term in their timing model over a model with red noise alone, although this can be explained given their relatively small $\dot{\nu}$ values in relation to other pulsars in our sample.
Hence, longer observing time spans may be required to resolve the $\ddot{\nu}$ component of their rotational evolution.
The younger pulsars (where $\tau_{c} \lesssim 10^{4}$\,yr) generally undergo small ($\Delta\nu_{g}/\nu < 10^{-6}$) glitches with negligible $\Delta\dot{\nu}_{g}$ components (with a handful of exceptions), while the older flat-jump pulsars have singular, extremely large amplitude glitches with a significant $\Delta\dot{\nu}_{g}$. 
For instance, PSRs J0729$-$1448 and J1413$-$6141, both of which are presented in Figure \ref{fig:ppdot_evol}, have characteristic ages of $35$ and $14$\,kyr respectively.
Despite experiencing several large glitches (see Table \ref{tbl:glitches} and \citealt{Yu2013}), PSR~J1413$-$6141 did not exhibit any resolved changes in $\dot{\nu}$.
When combined with a lack of a distinguishable $\ddot{\nu}$, it appears to follow a flat evolutionary path like many of the older, non-glitching pulsars.
In contrast, PSR~J0729$-$1448 clearly falls into the flat-jumps category thanks to a large amplitude glitch with a strong $\Delta\dot{\nu}$ component. 
This large glitch is similar to the giant glitches found in PSRs J1052$-$5954, J1650$-$4502 and J1757$-$2421, which have characteristic ages of $143$, $376$ and $285$\,kyr respectively and underwent the three largest glitches listed in Table \ref{tbl:glitches}.
The observed $P$-$\dot{P}$ evolution of both PSRs J1052$-$5954 and J1757$-$2421 are almost entirely dominated by exponential recoveries that occurred following their glitches, whereas the glitch in PSR~J0729$-$1448 did not show any evidence of recovery.

There is also the question of whether the flat and flat-jumps pulsars are related to the inclined and Vela-like pulsars.
Given the comparatively low values of $\dot{\nu}$ associated with most of the flat pulsars, their inferred glitch wait times from Equation \ref{eqn:rate} are significantly longer than their current timing baselines -- much like the inclined pulsars. 
Physically, this can be ascribed to the low spin-down rates of these pulsars requiring longer periods of time to build-up a sufficiently large stress within the neutron star for a glitch to be triggered. 
Under this scenario, our detection of the giant glitches in the flat-jump pulsars can be attributed to both the large size and hundreds of years' worth of pulsar timing accumulated by our sample.
There are of course a number of obvious exceptions, namely PSR~J1413$-$6141 that we discussed earlier in this section. 
The potential implications of these pulsars are further expanded upon in Section \ref{subsec:low_n} below.

\subsubsection{The braking index of PSR J1734--3333}

\begin{figure}
    \centering
    \includegraphics[width=\linewidth]{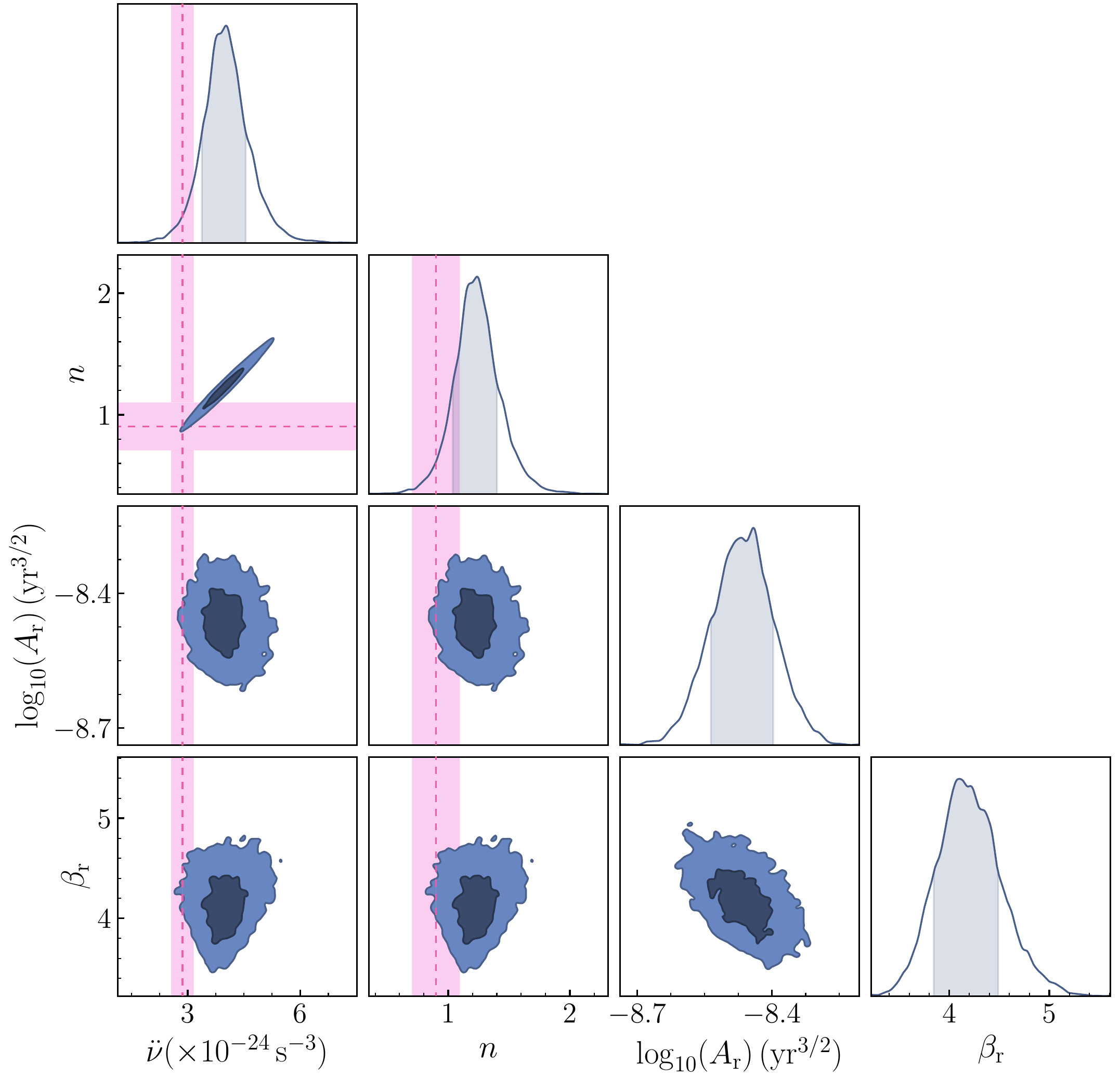}
    \caption{One- and two-dimensional posterior distributions for $\ddot{\nu}$, braking index and red noise parameters of PSR~J1734$-$3333. Dashed magenta lines and shading indicate the $\ddot{\nu}$ and $n$ values and 1-$\sigma$ uncertainties from~\citet{Espinoza2011a}.}
    \label{fig:j1734_post}
\end{figure}

PSR~J1734$-$3333 is a pulsar with a high-magnetic field strength that was found by \citet{Espinoza2011b} to possess an unusually small braking index of $n = 0.9 \pm 0.2$.
They surmised that this small $n$ may be evidence that the magnetic field strength of this pulsar may be growing over time.
Similarly small braking indices can also result from particle outflows \citep[e.g.][]{Michel1969}.
The results of our {\sc TempoNest} analysis of this pulsar are shown in Figure~\ref{fig:j1734_post}, where the recovered $n = 1.2 \pm 0.2$ is consistent with the \citet{Espinoza2011b} value at the 68\,percent confidence interval. 
PSR~J1734$-$3333 did undergo a relatively small glitch on MJD 56350 with no detectable post-glitch recovery.
We tested whether the glitch had any effect on $n$ by conducting a separate {\sc TempoNest} run that excluded the post-glitch ToAs.
The resulting posterior distribution for $n$ was almost identical to what we obtained when including the post-glitch data, with a Jensen-Shannon divergence~\citep{Lin1991}\footnote{Identical probability distributions have a Jensen-Shannon divergence of 0\,bit (i.e. no information gained between distributions) and maximally divergent distributions have 1\,bit.} of $\lesssim 0.002$\,bit.
Hence our measurement is unaffected by any low-level glitch recovery effects that were not modelled. 
As noted in \citet{Espinoza2011b}, and is obvious in our Figure \ref{fig:ppdot_evol}, the pulsar appears to be headed towards the location of the magnetars in $P$-$\dot{P}$ space.
If this braking index were to remain constant over time, then the pulsar would take around $(38 \pm 9)$\,kyr to obtain similar rotational properties to the bulk magnetar population.

\subsection{Connecting large braking indices to glitches in Vela-like pulsars}\label{subsec:interp_ppdot}

In order to understand the relationship between the glitch parameters and our inferred values of $n$ listed in Table \ref{tbl:brake}, we investigated how $n$ varied before, after and between consecutive glitches.
To measure the separate inter-glitch braking indices ($n_{\rm int}$), we followed a variation of the technique employed by \citet{Yu2013}.
The ToAs for the pulsars in Table \ref{tbl:brake} with two or more glitches were separated into multiple sections (pre-first-glitch, inter-glitch and post-final-glitch), while the remaining pulsars that experienced only a single glitch were split into pre- and post-glitch sections. 
Using {\sc TempoNest}, we fitted for $\nu$, $\dot{\nu}$ and $\ddot{\nu}$ along with at least one set of glitch recovery parameters ($\nu_{d}$ and $\tau_{d}$) to each section of ToAs. 
The inter-glitch fits were performed independent of the `global', long-term fits referred to in Section \ref{subsec:interp_ppdot}. Hence, modelling of the white and red noise were also performed during each inter-glitch fit.
Table \ref{tbl:int_glitch} presents the resulting $n_{\rm int}$, wait times between glitches ($T_{g}$) and both the preceding glitch amplitude and fractional change in $\dot{\nu}$.
Lower limits on $T_g$ are listed for entries corresponding to the pre-first-glitch/post-last-glitch and single glitch cases.
As with the inferred braking indices presented in Table \ref{tbl:brake}, the values of $n_{\rm int}$ are all positive because they are robust to the presence of the timing noise, which is fitted simultaneously as a power-law process.
Note the measurements obtained here are representative of how the pulsars are behaving between glitches, and do not necessarily reflect their long-term rotational evolution (cf. Figure \ref{fig:ppdot_evol}).

\begin{table}
\begin{center}
\caption{Non-zero braking indices measured before, after and between subsequent glitches for the 16 pulsars with one or more large glitches, associated waiting times (or lower-limits) until the next glitch, as well as the amplitudes and fractional changes in spin-down of the preceding glitch. \label{tbl:int_glitch}}
\renewcommand{\arraystretch}{1.2}
\setlength{\tabcolsep}{4pt}
\resizebox{\linewidth}{!}{
\begin{tabular}{lcccc}
\hline
PSRJ & $n_{\rm int}$ & $T_{g}$ (days) & $(\Delta\nu_{g}/\nu)_{\rm prev}$ & $(\Delta\dot{\nu}_{g}/\dot{\nu})_{\rm prev}$ \\
\hline
J0835$-$4510 & $33.3^{+0.7}_{-0.6}$ & $>1130$ & N/A & N/A \\
J0835$-$4510 & $51.6 \pm 0.5$ & $1164$ & $1902.4 \pm 0.5$ & $7 \pm 1$ \\
J0835$-$4510 & $52^{+4}_{-3}$ & $1176$ & $3057 \pm 2$ & $4.6 \pm 0.3$ \\
J0835$-$4510 & $62 \pm 2$ & $>734$ & $1908.3 \pm 0.2$ & $11.2 \pm 0.3$ \\
J0940$-$5428 & $21 \pm 1$ & $>4479$ & N/A & N/A \\
J0940$-$5428 & $57^{+5}_{-3}$ & $2925$ & $1573.9^{+1.1}_{-0.8}$ & $11 \pm 2$ \\
J1015$-$5719 & $16 \pm 1$ & $>2462$ & N/A & N/A \\
J1015$-$5719 & $87^{+12}_{-6}$ & $>1761$ & $3232.3 \pm 0.6$ & $11 \pm 2$ \\
J1016$-$5857 & $25 \pm 1$ & $2491$ & $1622.6 \pm 0.3$ & $3.69 \pm 0.05$ \\
J1016$-$5857 & $35 \pm 3$ & $1515$ & $1919.8^{+1.1}_{-0.9}$ & $6 \pm 1$ \\
J1016$-$5857 & $132^{+29}_{-34}$ & $421$ & $1464.4^{+1.1}_{-0.9}$ & $4^{+5}_{-1}$ \\
J1048$-$5832 & $58 \pm 6$ & $1754$ & $2995 \pm 7$ & $3.7 \pm 0.1$ \\
J1048$-$5832 & $47 \pm 5$ & $1945$ & $771 \pm 2$ & $4.62 \pm 0.06$ \\
J1048$-$5832 & $64^{+16}_{-12}$ & $940$ & $1838.4 \pm 0.5$ & $3.7 \pm 0.3$ \\
J1048$-$5832 & $37^{+5}_{-6}$ & $817$ & $28.5 \pm 0.4$ & $0.19 \pm 0.14$ \\
J1048$-$5832 & $32 \pm 3$ & $2260$ & $3044.1 \pm 0.9$ & $5.2^{+0.5}_{-0.4}$ \\
J1320$-$5359 & $104^{+13}_{-14}$ & $>5975$ & N/A & N/A \\
J1320$-$5359 & $1847^{+347}_{-434}$ & $197$ & $10.5 \pm 0.1$ & $0.2^{+0.2}_{-0.1}$ \\
J1320$-$5359 & $151^{+36}_{-26}$ & $>1703$ & $246.8 \pm 0.1$ & $0.08^{+0.14}_{-0.06}$ \\
J1357$-$6429 & $18.1 \pm 0.2$ & $2712$ & $2332^{+4}_{-3}$ & $13 \pm 1$ \\
J1357$-$6429 & $73.4 \pm 0.9$ & $843$ & $4860^{+3}_{-2}$ & $14.7^{+0.7}_{-0.8}$ \\
J1357$-$6429 & $34.8 \pm 0.8$ & $2219$ & $2250 \pm 11$ & $7 \pm 2$ \\
J1420$-$6048 & $49.9^{+0.9}_{-1.0}$ & $1154$ & $1146.2 \pm 0.6$ & $3.83 \pm 0.08$ \\
J1420$-$6048 & $45 \pm 2$ & $971$ & $2019 \pm 10$ & $6.6 \pm 0.8$ \\
J1420$-$6048 & $60^{+7}_{-6}$ & $947$ & $1270 \pm 3$ & $3.9 \pm 0.3$ \\
J1420$-$6048 & $74 \pm 1$ & $757$ & $927.6^{+0.7}_{-0.6}$ & $6 \pm 1$ \\
J1420$-$6048 & $56 \pm 2$ & $838$ & $1352.8^{+0.5}_{-0.4}$ & $5.4 \pm 0.2$ \\
J1420$-$6048 & $48 \pm 3$ & $944$ & $1954.2 \pm 0.3$ & $5.7 \pm 0.2$ \\
J1420$-$6048 & $49 \pm 2$ & $>1241$ & $1210^{+2}_{-1}$ & $9^{+7}_{-4}$ \\
J1524$-$5625 & $43.0 \pm 0.7$ & $>1509$ & N/A & N/A \\
J1524$-$5625 & $43 \pm 1$ & $>2711$ & $2975.9^{+0.7}_{-0.6}$ & $15.5^{+0.9}_{-0.7}$ \\
J1617$-$5055 & $32^{+10}_{-8}$ & $519$ & $68 \pm 2$ & $2.2^{+0.6}_{-0.5}$ \\
J1617$-$5055 & $31^{+24}_{-36}$ & $358$  & $55 \pm 2$ & $1.1 \pm 0.6$ \\
J1646$-$4346 & $75^{+19}_{-22}$ & $1324$ & $885 \pm 3$ & $1.5 \pm 0.3$ \\
J1702$-$4310 & $11^{+2}_{-1}$ & $>2551$ & N/A & N/A \\
J1702$-$4310 & $15.2^{+0.5}_{-0.6}$ & $3562$ & $4810 \pm 27$ & $17 \pm 4$ \\
J1702$-$4310 & $23 \pm 2$ & $>939$ & $3129^{+4}_{-1}$ & $5^{+2}_{-1}$ \\
J1709$-$4429 & $16 \pm 1$ & $>835$ & N/A & N/A \\
J1709$-$4429 & $27.0^{+0.7}_{-0.4}$ & $2713$ & $2057 \pm 2$ & $4.0 \pm 0.1$ \\
J1709$-$4429 & $69 \pm 5$ & $1228$ & $1166.7 \pm 0.2$ & $6.22 \pm 0.03$ \\
J1709$-$4429 & $41.2 \pm 0.6$ & $1977$ & $2872 \pm 7$ & $8.0 \pm 0.7$ \\
J1709$-$4429 & $27.7^{+0.7}_{-0.4}$ & $1661$ & $2755 \pm 1$ & $13.8^{+0.9}_{-1.0}$ \\
J1709$-$4429 & $36.8^{+0.6}_{-0.5}$ & $1824$ & $3027^{+7}_{-4}$ & $8 \pm 1$ \\
J1709$-$4429 & $181 \pm 31$ & $>262$ & $2433.5^{+0.8}_{-0.6}$ & $8.5 \pm 0.9$ \\
J1718$-$3825 & $48.9 \pm 0.4$ & $3004$ & $2.2 \pm 0.2$ & $\lesssim 0.08$ \\
J1718$-$3825 & $26 \pm 2$ & $>505$ & $7.1 \pm 0.1$ & $\lesssim 0.7$ \\
J1730$-$3350 & $22^{+5}_{-6}$ & $>1530$ & N/A & N/A \\
J1730$-$3350 & $16.4^{+0.3}_{-0.6}$ & $3871$ & $3202 \pm 1$ & $5.9 \pm 0.1$ \\
\hline
\end{tabular}
}
\renewcommand{\arraystretch}{}
\end{center}
\end{table}
\begin{table}
\centering
\contcaption{}
\renewcommand{\arraystretch}{1.2}
\setlength{\tabcolsep}{4pt}
\resizebox{\linewidth}{!}{
\begin{tabular}{lcccc}
\hline
PSRJ & $n_{\rm int}$ & $T_{g}$ (days) & $(\Delta\nu_{g}/\nu)_{\rm prev}$ & $(\Delta\dot{\nu}_{g}/\dot{\nu})_{\rm prev}$ \\
\hline
J1730$-$3350 & $47 \pm 4$ & $>2381$ & $2250.7^{+1.0}_{-0.9}$ & $7{+3}_{-2}$ \\
J1801$-$2451 & $48^{+4}_{-3}$ & $1164$ & $1987.9(3)$ & $4.6 \pm 0.1$ \\
J1801$-$2451 & $50^{+7}_{-8}$ & $1384$ & $1247.4(3)$ & $4.7 \pm 0.2$ \\
J1801$-$2451 & $41 \pm 13$ & $921$ & $3755.8(4)$ & $6.8 \pm 0.1$ \\
J1801$-$2451 & $25^{+4}_{-3}$ & $1613$ & $17.4 \pm 0.2$ & $1.4 \pm 0.1$ \\
J1801$-$2451 & $26 \pm 2$ & $2239$ & $3083.7 \pm 0.7$ & $6.5 \pm 0.5$ \\
J1801$-$2451 & $41 \pm 3$ & $>1511$ & $2423.5 \pm 0.9$ & $5.9^{+0.5}_{-0.4}$ \\
\hline
\end{tabular}
}
\renewcommand{\arraystretch}{}
\end{table}

As noted in Section \ref{subsec:ppdot}, the pulsars that display large amplitude glitches with a significant $\Delta\dot{\nu}_{g}$ component all possess large $n$. There are several ways to interpret this behaviour. The glitches could be viewed as a mechanism that serves to reset much of the rapid downward $P$-$\dot{P}$ evolution experienced by these pulsars back toward a longer-term, `low-$n$' evolutionary track.
However, in most microphysical theories, glitches are triggered by stress accumulation in $\nu$ (e.g.\ differential rotation between crust and superfluid) rather than $\dot{\nu}$, so one expects glitches to be triggered by high torque rather than high $n \propto \ddot{\nu}$.
On the other hand, $n_{\rm int}$ could stem from a form of post-glitch recovery. Suggested mechanisms include Ekman circulation in a two-component star \citep{vanEysden2010, vanEysden2012}, changes in the effective moment of inertia from progressive re-coupling of the crust to the superfluid core \citep[e.g.][]{Smith1999, Antonopoulou2018, Pizzochero2020, Montoli2020}, unpinning and re-pinning of vortices between crustal pinning sites due to thermal fluctuations \citep[the `vortex creep' model;][]{Alpar1984a, Alpar1984b, Alpar1993, Alpar2006, Akbal2017} and turbulence within an array of vortices pinned in the superfluid \citep{Melatos2007, vanEysden2012, Melatos2014, Haskell2020}.
Although these theoretical models are difficult to falsify \citep{Haskell2015}, and it is unclear whether the type of assumed vortex creep can be supported within physical neutron stars~\citep{Link2014}, they do provide a set of phenomenological behaviour that we can test.
In the particular instance of the vortex creep model, $\ddot{\nu}_{\rm int}$ reflects the gradient of $\dot{\nu}(t)$ as it undergoes a linear recovery. in the lead-up to the next glitch.
This can be seen in the phenomenological model of \citet{Alpar2006}, as equating their equations 11 and 12 returns
\begin{equation}
    n_{\rm int} = 2 \times 10^{-3} \Big( \frac{\tau_{c}}{T_{g}} \Big) \Big(\frac{\Delta\dot{\nu}_{g} / \dot{\nu}}{10^{-3}} \Big).
\end{equation}
Substituting both our Equation~\ref{eqn:brake} and $\tau_{c} = 0.5\,\nu\,|\dot{\nu}|^{-1}$ into this relation and re-arranging for $\ddot{\nu}_{\rm int}$ gives
\begin{equation}\label{eqn:F2_relation}
    \ddot{\nu}_{\rm int} = \frac{\Delta\dot{\nu}_{g}}{T_{g}}.
\end{equation}
\citet{Akbal2017} assumes a variant of this relation in their modelling of the Vela pulsar, though modified to allow for small permanent shifts in $\dot{\nu}$.
The turbulent vortex array model of \citet{Haskell2020} also does not attempt to directly relate the internal physics to the observed properties of the neutron star. 
However, they do predict that $\ddot{\nu}_{\rm int}$ (and by extension $n_{\rm int}$) should follow a quadratic dependence on $T_{g}$ if the vortices are pinned within a turbulent region of the star, or a linear dependence for a straight vortex array.
\citet{vanEysden2010} showed that Ekman pumping in a cylindrical vessel containing a two-component superfluid can result in a non-linear spin-down after applying an impulsive acceleration, i.e. an effective $n >> 3$ process (see their figure 2). 
An extension of this model was successful in replicating much of the observed post-glitch behaviour of the Crab and Vela pulsars \citep{vanEysden2012}, though no explicit formula relating the model to $n_{\rm int}$ was derived.

\begin{figure}
    \centering
    \includegraphics[width=\linewidth]{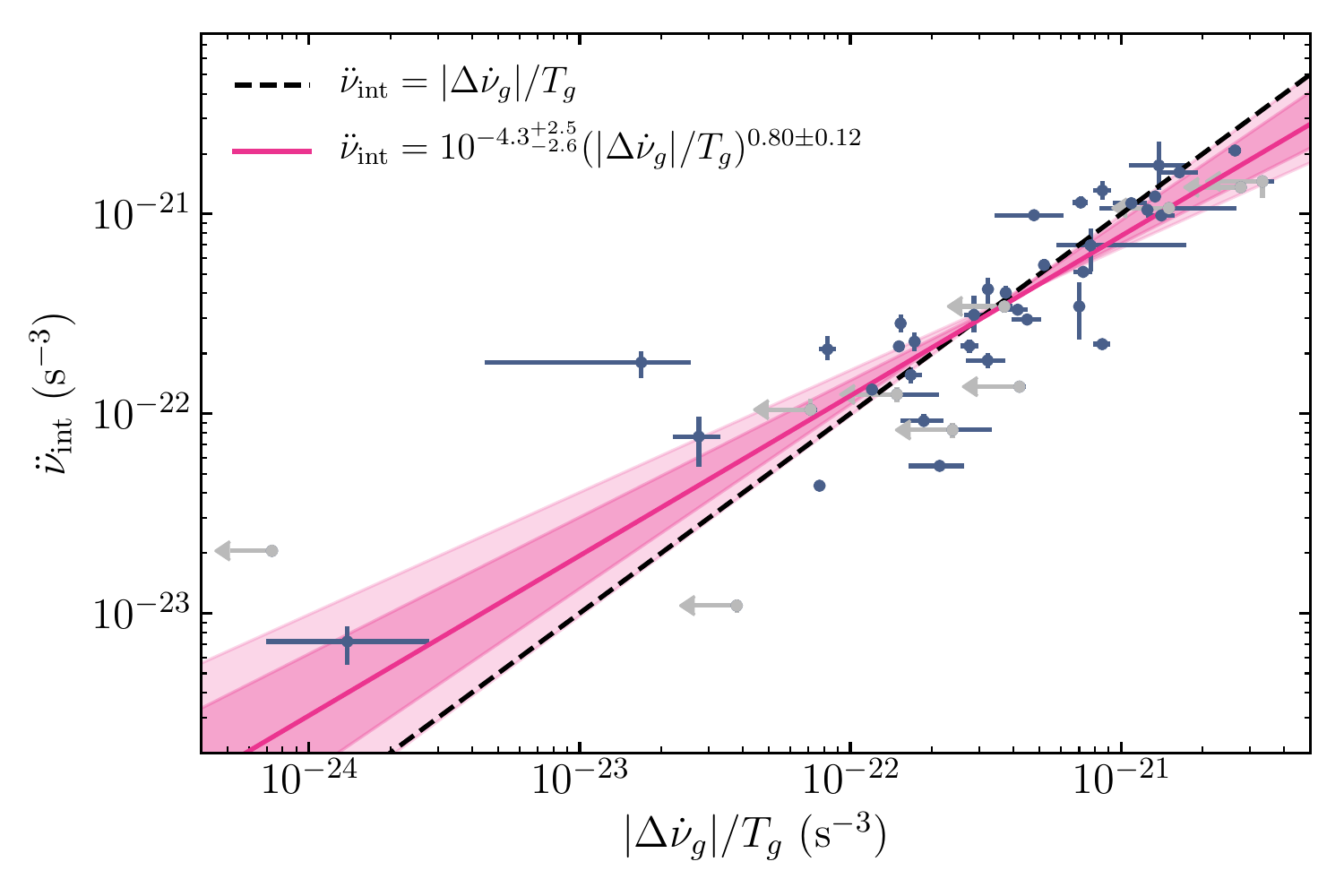}
    \caption{Comparison between $\ddot{\nu}_{\rm int}$ and the $\Delta\dot{\nu}_{g}$ of the previous glitch divided by the inter-glitch wait time for 53 glitches in 16 pulsars. Black-dashed line indicates an exact one-to-one relationship, while the magenta line and shading represents our median power-law fit and the 68 and 95\,percent confidence regions respectively.}
    \label{fig:F2_grad}
\end{figure}

On an individual basis, there was no clear correlation between $n_{\rm int}$ and the glitch properties for the pulsars with more than one inter-glitch measurement.
However, examining at the sample as a whole revealed a weak anti-correlation between $n_{\rm int}$ (and by extension, $\ddot{\nu}_{\rm int}$) and $T_{g}$, with $\rho_{s} = -0.65$ (p-value, $2.8 \times 10^{-5}$). 
This correlation weakens with the inclusion of lower-limits on $T_{g}$, as well as long-term braking indices from the inclined pulsars, where the total observing span for them is taken as a lower-limit on $T_{g}$.
Including both the extra terms from Table \ref{tbl:int_glitch} for the inclined pulsars, and employing a bootstrap approach to sample the lower-limits, returned a Spearman coefficient of $\rho_{s} = -0.37 \pm 0.08$ (p-value, $0.003^{+0.008}_{-0.002}$).
Despite the weakness of these anti-correlations, owing to the significant scatter relative to the sample size, the implied inverse dependence of $n_{\rm int}$ on $T_{g}$ does not match the predictions of \citet{Haskell2020} for straight or turbulent vortices.
The most striking evidence for a potential connection between $n_{\rm int}$ and glitches arises from our comparison of the values $\ddot{\nu}_{\rm int}$ from Table \ref{tbl:int_glitch} with $\Delta\dot{\nu}_{g}$ divided by $T_{g}$ in Figure \ref{fig:F2_grad}. 
It is clear the expected one-to-one relation from the linear recovery process (Equation \ref{eqn:F2_relation}) indicated by the dashed line is largely adhered to. 
Fitting a power-law to only the points with confident measurements of $|\Delta\dot{\nu}_{g}|/T_{g}$ using a projected, bivariate Gaussian likelihood (see equations 26 through 32 of \citealt{Hogg2010}), we obtained the relation
\begin{equation}\label{eqn:actual_F2_relation}
    \ddot{\nu}_{\rm int} = 10^{-4.3^{+2.5}_{-2.6}}(|\Delta\dot{\nu}_{g}|/T_{g})^{0.80 \pm 0.12},
\end{equation}
which when plotted in Figure \ref{fig:F2_grad} is consistent with exact one-to-one relation at the 95\,percent confidence interval.
We also checked whether there was a stronger forwards or backwards correlation between $\ddot{\nu}_{\rm int}$ and $\Delta\dot{\nu}_{g}/\dot{\nu}$ (i.e, comparing to $\Delta\dot{\nu}_{g}$ from the preceding or following glitch).
Using the same bootstrapping technique from earlier, we obtained overlapping Spearman coefficients of $\rho_{s-} = 0.74 \pm 0.04$ for the $\Delta\dot{\nu}_{g}$ of previous glitch and $\rho_{s+} = 0.82^{+0.03}_{-0.06}$ for the next glitch.
The consistency between the two is not surprising given the fractional step-change in spin-down and inter-glitch wait-times are similar among the Vela-like pulsars (see Figure \ref{fig:ppdot_evol} and Table \ref{tbl:glitches}).

\subsection{Pulsars with seemingly small braking indices}\label{subsec:low_n}

While there is a clear link between large glitches and large values of $n_{\rm int}$ in the Vela-like pulsars, the same cannot be said for the pulsars that experienced predominately small glitches ($\Delta\nu_{g}/\nu < 10^{-6}$) and those that have undergone single gigantic glitches ($\Delta\nu_{g}/\nu \sim 10^{-5}$).
The majority of these pulsars did not favour timing models that included a $\ddot{\nu}$ component, and also show little evidence of a significant fractional evolution in $\dot{P}$ as a function of $P$ over our observing span.

\begin{figure}
    \centering
    \includegraphics[width=\linewidth]{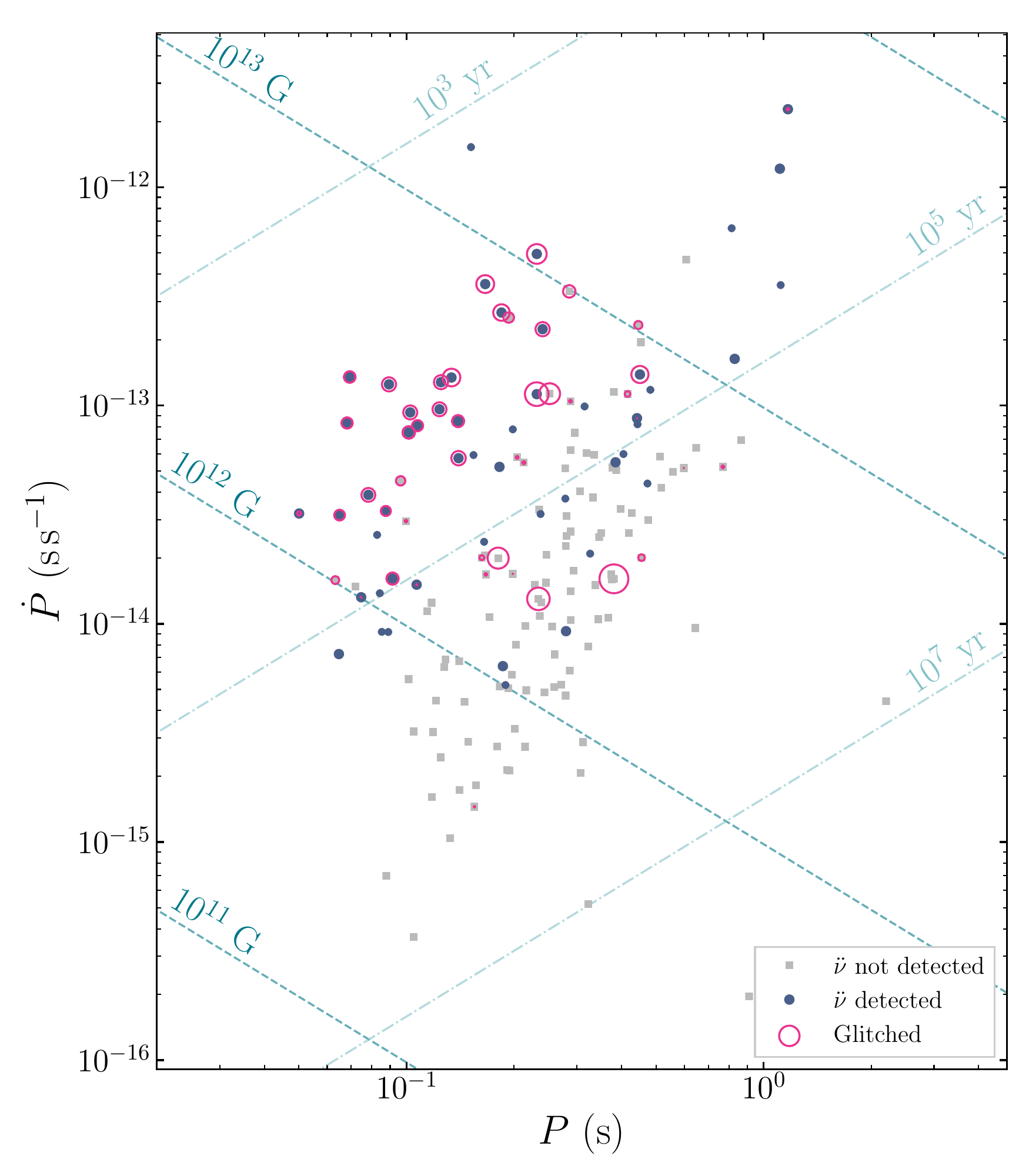}
    \caption{$P$-$\dot{P}$ diagram showing pulsars in both our sample and that of \citet{Parthasarathy2019, Parthasarathy2020} that favour (blue circles) or disfavour (grey squares) a timing model with a $\ddot{\nu}$ term. Pulsars that have glitched are highlighted with open circles, where the size is proportional to the amplitude of their largest glitch.}
    \label{fig:ppdot}
\end{figure}

Figure \ref{fig:ppdot} shows the values of $P$ and $\dot{P}$ for both the pulsars in our sample and those in \citet{Parthasarathy2019, Parthasarathy2020}, where there is a rapid drop-off in the number of pulsars with significant $\ddot{\nu}$ measurements below a characteristic age of $\sim 10^{5}$\,yr. 
Much of this can be put down to $\ddot{\nu}$ being harder to detect in pulsars with smaller values of $\nu$ and $\dot{\nu}$ (see Equation \ref{eqn:brake}), where long timing baselines are needed to distinguish the resulting small fractional change in $\dot{\nu}$ from timing noise \citep{Parthasarathy2020}.
However, there are some young, actively glitching pulsars among our sample that represent obvious outliers. 

According to Equation \ref{eqn:rate}, all three of the most actively glitching pulsars in Table \ref{tbl:glitches} (PSRs J1341$-$6220, J1413$-$6141 and J1740$-$3015) should undergo a large glitch every 18, 11 and 60\,yr respectively.
These pulsars have not only undergone several large glitches in a much shorter span of time than their predicted glitch wait time, they have also exhibited a large number of small glitches (i.e. $\Delta\nu_{g}/\nu \lesssim 10^{-6}$) interspersed between them. 
Glitches of similar amplitudes are seldom seen among the Vela-like pulsars, and indeed Vela itself appears to have a real under-abundance of small amplitude glitches \citep{Howitt2018, Espinoza2021}.
A potential clue as to why these pulsars do not possess detectable large $n$, despite being situated near the Vela-like population in $P$-$\dot{P}$ space, comes from the lack of consistently large $\Delta\dot{\nu}_{g}$ components associated with our glitch measurements in Table \ref{tbl:glitches}.
Our strong observational relationship between $n_{\rm int}$ and $\Delta\dot{\nu}_{g}/T_{g}$ in Equation \ref{eqn:actual_F2_relation} suggests $\ddot{\nu}_{\rm int} \rightarrow 0$ for $\Delta\dot{\nu}_{g} \rightarrow 0$ and $T_{g} \rightarrow \infty$. 
Hence, if there is genuinely no change in spin-down associated with these glitches, then there is no glitch recovery induced large-$n$ process to detect in these pulsars. 
It is therefore possible that the underlying mechanism that drives the linear inter-glitch recoveries of the Vela-like pulsars is not active in these pulsars.

As noted in Section \ref{subsec:interp_ppdot}, the $P$-$\dot{P}$ post-glitch behaviour of the three pulsars with gigantic glitches (PSRs J1052$-$5954, J1650$-$4502 and J1757$-$2421) and PSR~J0729$-$1448 is similar in that they all retained significant changes in their spin-down rates.
A key difference however, is that the pulsars with gigantic glitches all show evidence of at least one exponential recovery that causes some of their $\Delta\dot{\nu}_{g}$ to decay, whereas PSR~J0729$-$1448 showed no evidence of a recovery.
Such behaviour can arise if the core does not decouple from the crust during a glitch, but does so over long timescales after a glitch, leading to the appearance of a permanent increase in the spin-down rate \citep{Haskell2014, Akbal2017}.
Our constraints on the braking index of PSR~J0729$-$1448 returned $n = 2.3 \pm 0.9$ where the $\ddot{\nu}$ model is marginally disfavoured with $\ln(\mathcal{B}^{\rm F2 \neq 0}_{\rm F2 = 0}) = -1.3$, meaning its long-term braking index must be $n < 3.2$ (at 68\% confidence) with no significant glitch-induced, large $\ddot{\nu}$ affecting its timing.
As for why the pulsars with gigantic glitches retain a significant amount of their change in spin-down, it is possible they exhibit the same long-term linear recovery seen in the Vela-like pulsars, just on a much longer timescale than our current observations cover. 
Indeed the posterior distributions for $n$ in these pulsars all peak away from zero, and for PSR~J1052$-$5954 is well constrained to $n = 233 \pm 41$ albeit with a Bayes factor that marginally prefers a model containing a $\ddot{\nu}$-term in addition to power-law red noise ($\ln(\mathcal{B}^{\rm F2 \neq 0}_{\rm F2 = 0}) = 2.7$), but does not exceed our $\ln(\mathcal{B}) > 3$ threshold.
Hence longer-term observations of these pulsars may be able to distinguish their inter-glitch braking indices.

\section{Conclusions}\label{sec:conc}

In this work we presented the results of a search for glitches in the timing of 74 young pulsars, along with inferences on the effects these glitches have on their rotational evolution. 
A total of 124 glitches were identified in these pulsars. 
The overall distributions of glitch properties from our sample largely reflects that of the broader population, and an analysis with the HMM-based glitch detection algorithm of \citet{Melatos2020} suggests our sample is complete to a pulsar averaged 90\,percent upper-limit of $\Delta\nu^{90\%}_{g}/\nu \lesssim 8.1 \times 10^{-9}$.
Having accounted for the glitches, we then conducted a Bayesian model selection study akin to that of \citet{Parthasarathy2019} for each of our pulsars.
This led to the detection of a significant $\ddot{\nu}$ component, and subsequent measurement of the braking index, for 32 pulsars.
Ten of these pulsars have never been seen to undergo a large amplitude glitch that contained a significant $\Delta\dot{\nu}_{g}$-component, whereas the other pulsars exhibited glitches similar to those seen in the Vela pulsar.
The measured values of $n$ are uncorrelated with characteristic age, and the braking index distributions of non-glitching and glitching pulsars are indistinguishable from one another. 

The observed $P$-$\dot{P}$ tracks of these Vela-like pulsars show that they evolve with a high inter-glitch braking index, $n_{\rm int}$, and undergo a large change in $\dot{P}$ at the time of the glitch. 
The glitches seem to occur no later than a change in $\delta\dot{P}/\dot{P}$ of 0.004\%. After accounting for the glitches, the decades-long evolution in $P$-$\dot{P}$ is consistent with a small $n$, similar to what was presented in \citet{Espinoza2017}, in spite of the large $n_{\rm int}$. 
We surmise that the sample of pulsars with large values of $n$ in \citet{Parthasarathy2020} will undergo a large glitch in the near future and that they share similar characteristics to the pulsars in the present sample. 
We show there is a near one-to-one relationship between $\ddot{\nu}_{\rm int}$ and $\Delta\dot{\nu}_{g}/T_{g}$ (Figure \ref{fig:F2_grad}). 

If the physical justification for this relationship is the vortex creep model proposed by \citet{Alpar1984a}, then measurements of $\ddot{\nu}_{\rm int}$ and $\Delta\dot{\nu}_{\rm g}$ following a large glitch could be used to predict when the next large glitch may occur in these pulsars, as was done by \citet{Akbal2017} for the Vela pulsar.
The use of an expanded version of this model that accounts for both over- and under-corrections in $\dot{\nu}$ between glitches (i.e. stochasticity in the fractional $\dot{P}$-jumps of PSRs~J1301$-$6305, J1420$-$6048 and J1709$-$4429 in Figure \ref{fig:ppdot_evol}) could provide useful forecasts for ongoing efforts to observe glitch events in real time with either dedicated search instruments near the glitch epoch (i.e. similar to that performed by \citealt{Dodson2002} and \citealt{Palfreyman2018} with the Mount Pleasant Observatory) or large scale pulsar monitoring programmes at UTMOST \citep{Jankowski2019, Lower2020}, CHIME/Pulsar \citep{CHIME2020} and the future SKA \citep{Watts2015, Stappers2018}.

\section*{Data availability}
The data underlying this study is available from the CSIRO data access portal under the P574 project (\href{https://data.csiro.au/}{data.csiro.au}).
Other data products are available upon reasonable request to the corresponding authors.

\section*{Acknowledgements}
The Parkes radio telescope (\textit{Murriyang}) is part of the Australia Telescope National Facility which is funded by the Australian Government for operation as a National Facility managed by CSIRO. We acknowledge the Wiradjuri people as the traditional owners of the Observatory site.
This work made use of the OzSTAR national HPC facility, which is funded by Swinburne University of Technology and the National Collaborative Research Infrastructure Strategy (NCRIS).
This work was supported by the Australian Research Council (ARC) Laureate Fellowship FL150100148, ARC Discovery Project DP170103625 and the ARC Centre of Excellence CE170100004 (OzGrav).
M.E.L. receives support from the Australian Government Research Training Program and CSIRO.
R.M.S. is supported through ARC Future Fellowship FT190100155.
S.D. is the recipient of an ARC Discovery Early Career Award (DE210101738) funded by the Australian Government.
L.O. acknowledges funding from the UK Science and Technology Facilities Council (STFC) Grant Code ST/R505006/1.
Work at NRL is supported by NASA. 
We thank J.~B.~Carlin for insightful comments and the anonymous referee for their thorough review.
We acknowledge use of the Astronomer's Telegram and the NASA Astrophysics Data Service.
We made use of the following software packages: {\sc astropy}~\citep{astropy2}, {\sc bilby}~\citep{Ashton2019}, {\sc chainconsumer} \citep{chainconsumer}, {\sc libstempo} \citep{libstempo}, {\sc matplotlib} \citep{matplotlib}, {\sc numpy} \citep{numpy}, {\sc pandas} \citep{pandas}, {\sc psrchive} \citep{Hotan2004, vanStraten2011}, {\sc psrqpy} \citep{psrqpy}, {\sc pymultinest} \citep{Feroz2009, Buchner2016}, {\sc scipy} \citep{scipy}, {\sc tempo2} \citep{Hobbs2006, Edwards2006} and {\sc TempoNest} \citep{Lentati2014}.



\bibliographystyle{mnras}
\bibliography{main}



\appendix

\section{HMM recipe and parameters}\label{appdx:hmm_params}
A complete description of the HMM framework and its application to glitch detection is given in \citet{Melatos2020}. Applying the HMM glitch detector involves choosing a number of parameters, many of which vary between data sets. Here we lay out the HMM parameter choices which have been made in the present work.

Three essential groups of parameters are needed: those which specify the allowed $(\nu, \dot{\nu})$ states, those which relate the observations (ToAs) back to the $(\nu, \dot{\nu})$ states, and those for the probabilities of transitions between $(\nu, \dot{\nu})$ states.
The choices of parameters which specify how the ToAs can be related to the $(\nu, \dot{\nu})$ states follow \citet{Melatos2020}; we refer the reader to Section 3.3 and Appendix C of that paper for further discussion.

Specifying the allowed $(\nu, \dot{\nu})$ parameters amounts to specifying a discretisation of a region of $(\nu, \dot{\nu})$ space.
We first note that the hidden $(\nu, \dot{\nu})$ states should be thought of as a deviation away from a secular phase model, where this secular phase model is derived from an initial \textsc{tempo2} fit.
As such, the region of $(\nu, \dot{\nu})$ space to be discretised is typically taken to be a region containing $(0\,\mathrm{Hz}, 0\,\mathrm{Hz}\,\mathrm{s}^{-1})$.
As a default, we take the boundaries in the $\nu$ direction to be $[\nu_-, \nu_+] = [-3 \times 10^{-7}, 3 \times 10^{-7}]\,\mathrm{Hz}$, and the spacing between discrete states to be $\eta_\nu = 10^{-9}\,\mathrm{Hz}$.
The default boundaries and spacing in the $\dot{\nu}$ direction are $[\dot{\nu}_-, \dot{\nu}_+] = [-10^{-15}, 10^{-15}]\,\mathrm{Hz}\,\mathrm{s}^{-1}$ and $\eta_{\dot{\nu}} = 2 \times 10^{-16}\,\mathrm{Hz}\,\mathrm{s}^{-1}$ respectively.
In the course of the analyses presented here, we found it necessary in some cases and advantageous in others to modify the boundaries of the $\dot{\nu}$ region to better account for the timing noise present in each pulsar.
When the timing noise is strong enough that $\dot{\nu}$ wanders outside the default region, we enlarge the $\dot{\nu}$ region to allow the HMM to track the evolution of $\dot{\nu}$.
In contrast, for some pulsars the timing noise is small enough that the scale of wandering of $\dot{\nu}$ is only a small fraction of the default $\dot{\nu}$ region.
In this case it is advantageous (but not necessary) to reduce the size of the $\dot{\nu}$ region, which allows for faster computation and more stringent upper limits on the size of undetected glitches.

Finally, we consider the probabilities of transitions between $(\nu, \dot{\nu})$ states. 
There is a trivial element: over a gap between consecutive ToAs of length $z$, in the absence of noise the state $(\nu, \dot{\nu})$ should transition to $(\nu + \dot{\nu}z, \dot{\nu})$.
However, timing noise is not negligible in these pulsars.
In general timing noise is incorporated into the HMM through a model which assumes the presence of some kind of stochastic term in the phase model, which causes wandering in $(\nu, \dot{\nu})$.
For most of the pulsars in this study we follow the prescription of \citet{Melatos2020}, which assumes that there is a white noise term in the second time derivative of $\nu$: 
\begin{equation} 
    \frac{\mathrm{d}^2\nu}{\mathrm{d}t^2} = \xi(t),\label{eqn:hmm_melatos_langevin} 
\end{equation} 
where $\xi(t)$ satisfies 
\begin{align}
    &\langle \xi(t) \rangle = 0,\label{eqn:hmm_tn_mean}\\ &\langle \xi(t)\xi(t') \rangle = \sigma^2\delta(t-t').
\label{eqn:hmm_tn_autocorr} 
\end{align}
From these equations we may calculate the covariance matrix $\Sigma$ of $\nu$ and $\dot{\nu}$ over a ToA gap of length $z$: 
\begin{equation} 
    \Sigma = \sigma^2
    \begin{pmatrix} 
        z^3/3 & z^2/2 \\ z^2/2 & z 
    \end{pmatrix} .
\end{equation}
Once $\Sigma$ is given, the probabilities of transitions between $(\nu, \dot{\nu})$ states are given by equations (10)--(12) of \citet{Melatos2020}.
It is important to recognise that (\ref{eqn:hmm_melatos_langevin}) is not proposed as a physical model specific to an individual pulsar.
Rather, equation (\ref{eqn:hmm_melatos_langevin}) is simply a generic, mathematically precise means to introduce stochastic fluctuations into the phase model, that are qualitatively consistent with random-walk character of the phase residuals observed in pulsars as a class.
It has been shown to perform reliably on synthetic and real data in previous studies \citep{Melatos2020}.

The $\sigma$ parameter controls the strength of the timing noise which is included in the HMM, and currently we do not have a reliable method of making a measurement of the timing noise in each pulsar and converting that to a suitable value of $\sigma$.
As such, for this work we adopt a rule of thumb in \citet{Melatos2020} which sets a minimum value of $\sigma$ based on the spacing in the $\dot{\nu}$ grid $\eta_{\dot{\nu}}$: \begin{equation} 
    \sigma = \eta_{\dot{\nu}} \langle z\rangle^{-1/2}. 
    \label{eqn:hmm_tn_sigma_rot} 
\end{equation}

In a few cases we find that the above model is not satisfactory: the analysis produces an implausibly large number of glitch candidates, none of which show any signature in the timing residuals.
In these cases, we adopt an alternative form for $\Sigma$, which is predicated on a white noise term in the \emph{first} time derivative of $\nu$: 
\begin{equation} 
    \frac{\mathrm{d}\nu}{\mathrm{d}t} = \zeta(t),
\end{equation} 
with $\zeta(t)$ satisfying equations analogous to (\ref{eqn:hmm_tn_mean}) and (\ref{eqn:hmm_tn_autocorr}).
In this case $\Sigma$ has the simpler form 
\begin{equation}
    \Sigma = \sigma^2 
    \begin{pmatrix} 
        z & 0 \\ 0 & 1
    \end{pmatrix} 
\end{equation}
The choice of $\sigma$ is no longer given by (\ref{eqn:hmm_tn_sigma_rot}), but instead is set by hand to match the scale of variation in $\dot{\nu}$ which is observed in the $\dot{\nu}$ paths recovered by the HMM, which is typically on the order of $10^{-15}\,\mathrm{Hz}\,\mathrm{s}^{-1}$.

\section{Computing glitch upper-limits}\label{appdx:upper_lims}

We define $\Delta\nu_g^{90\%}$ in the following way: if a large number of synthetic data sets are produced, each with a glitch of size $\Delta\nu_g^{90\%}$ injected at a randomly chosen epoch, we expect that the HMM will detect a glitch in 90\% of those cases at an epoch which is not more than one ToA away from the injected epoch.

This definition suggests an empirical method of calculating $\Delta\nu_g^{90\%}$:
\begin{enumerate}
    \item Make an estimate for $\Delta\nu_g^{90\%}$, denoted $\Delta\nu_g^{x\%}$.
    \item Generate a set of 100 synthetic data sets each with a glitch of size $\Delta\nu_g^{x\%}$.
    \item Perform HMM analyses on each of the 100 synthetic dataets, and record the number $n$ of data sets for which the HMM detects a glitch within one ToA of the injected epoch.
    \item If $n = 90$, terminate and take $\Delta\nu_g^{90\%} = \Delta\nu_g^{x\%}$. Otherwise, choose an updated $\Delta\nu_g^{x\%}$ and return to step (ii).
\end{enumerate}

The refinement of $\Delta\nu_g^{x\%}$ proceeds essentially as a binary search: we choose an initial possible range for $\Delta\nu_g^{90\%}$, typically $[\Delta\nu_g^-, \Delta\nu_g^+] = [10^{-9}, 10^{-7}]\,\mathrm{Hz}$ and take $\Delta\nu_g^{x\%}$ to bisect this range logarithmically (i.e. at the first iteration, $\Delta\nu_g^{x\%} = 10^{-8}\,\mathrm{Hz}$).
Then, if the number of detected glitches exceeds 90, the range is refined to be $[\Delta\nu_g^-, \Delta\nu_g^+] = [\Delta\nu_g^-, \Delta\nu_g^{x\%}]$.
Similarly if the number of detected glitches is less than 90, the range is refined to $[\Delta\nu_g^-, \Delta\nu_g^+] = [\Delta\nu_g^{x\%}, \Delta\nu_g^+]$.
A new choice of $\Delta\nu_g^{x\%}$ is then made to bisect the new range logarithmically, and the procedure repeats.

The generation of synthetic data sets in step (ii) is done using \textsc{libstempo} \footnote{\href{https://github.com/vallis/libstempo}{https://github.com/vallis/libstempo}}.
We inject additive Gaussian error at a level commensurate with the reported ToA error, but we do \emph{not} inject red timing noise --- injecting the latter tends to create difficulties in automating step (iii).
While we do expect that timing noise in the data impacts the ability of the HMM to detect small glitches near the threshold of detectability, we do not expect that it makes a significant difference to the 90\% frequentist upper limit.
Results are given in Table \ref{tbl:upplims}.

\begin{table*}
    \centering
    \caption{Pulsar averaged 90\% upper-limits on $\Delta\nu_{g}/\nu$.}
    \label{tbl:upplims}
    \begin{tabular}{lclclclclc}
    \hline
    PSR & $\Delta \nu_g^{90\%}/\nu$ &
    PSR & $\Delta \nu_g^{90\%}/\nu$ &
    PSR & $\Delta \nu_g^{90\%}/\nu$ &
    PSR & $\Delta \nu_g^{90\%}/\nu$ &
    PSR & $\Delta \nu_g^{90\%}/\nu$ \\
& ($\times10^{-9}$) & & ($\times10^{-9}$) & & ($\times10^{-9}$) & & ($\times10^{-9}$) & & ($\times10^{-9}$)\\
    \hline
J0614$+$2229 & $19$ &
J0627$+$0706 & $2.7$ &
J0631$+$1036 & $5.7$ &
J0659$+$1414 & $1.1$ &
J0729$-$1448 & $10$ \\
J0742$-$2822 & $4.6$ &
J0835$-$4510 & $4.0$ &
J0842$-$4851 & $5.1$ &
J0855$-$4644 & $6.0$ &
J0901$-$4624 & $4.3$ \\
J0908$-$4913 & $5.9$ &
J0940$-$5428 & $2.4$ &
J1003$-$4747 & $0.17$ &
J1015$-$5719 & $4.1$ &
J1016$-$5857 & $3.4$ \\
J1019$-$5749 & $1.6$ &
J1028$-$5819 & $0.88$ &
J1048$-$5832 & $2.0$ &
J1052$-$5954 & $5.7$ &
J1055$-$6028 & $3.5$ \\
J1057$-$5226 & $3.4$ &
J1105$-$6107 & $6.3$ &
J1112$-$6103 & $1.2$ &
J1138$-$6207 & $5.6$ &
J1248$-$6344 & $1.4$ \\
J1301$-$6305 & $13$ &
J1320$-$5359 & $2.2$ &
J1327$-$6400 & $8.2$ &
J1341$-$6220 & $15$ &
J1357$-$6429 & $3.8$ \\
J1359$-$6038 & $5.3$ &
J1406$-$6121 & $3.1$ &
J1410$-$6132 & $43$ &
J1413$-$6141 & $6.8$ &
J1420$-$6048 & $3.4$ \\
J1452$-$6036 & $0.34$ &
J1524$-$5625 & $4.1$ &
J1541$-$5535 & $14$ &
J1600$-$5044 & $19$ &
J1602$-$5100 & $30$ \\
J1614$-$5048 & $47$ &
J1617$-$5055 & $4.0$ &
J1626$-$4807 & $5.6$ &
J1627$-$4706 & $13$ &
J1644$-$4559 & $15$ \\
J1646$-$4346 & $12$ &
J1650$-$4502 & $14$ &
J1701$-$4533 & $0.57$ &
J1702$-$4128 & $5.8$ &
J1702$-$4310 & $5.3$ \\
J1705$-$3950 & $3.6$ &
J1709$-$4429 & $1.3$ &
J1716$-$4005 & $3.9$ &
J1718$-$3825 & $0.55$ &
J1721$-$3532 & $2.6$ \\
J1726$-$3530 & $77$ &
J1730$-$3350 & $6.9$ &
J1731$-$4744 & $5.1$ &
J1734$-$3333 & $30$ &
J1737$-$3137 & $4.5$ \\
J1740$-$3015 & $7.3$ &
J1750$-$3157 & $2.2$ &
J1757$-$2421 & $6.0$ &
J1801$-$2304 & $3.1$ &
J1803$-$2137 & $4.2$ \\
J1825$-$0935 & $9.9$ &
J1826$-$1334 & $2.8$ &
J1835$-$0643 & $0.23$ &
J1837$-$0604 & $4.7$ &
J1841$-$0345 & $2.0$ \\
J1841$-$0425 & $1.8$ &
J1841$-$0524 & $4.8$ &
J1844$-$0256 & $6.1$ &
J1847$-$0402 & $1.9$ &\\
\hline
    \end{tabular}
\end{table*}


\section{Pulsar timing solutions}\label{appdx:ephem}
Preferred timing model and associated Bayes factors are listed in Table \ref{tbl:bayes}. 
Timing solutions for pulsars with glitches are given in Table \ref{tbl:params_glitch} and for pulsars without glitches in Table \ref{tbl:params_noglitch}.

\begin{table*}
    \centering
    \caption{Preferred timing models and associated Bayes factors compared to the standard PL model. For pulsars where the PL model is preferred, we list $\ln(\mathcal{B}_{\rm PL+F2}^{\rm PL})$}.
    \label{tbl:bayes}
    \begin{tabular}{lcclcclcc}
    \hline
    PSR & Preferred model & $\ln(\mathcal{B})$ &
    PSR & Preferred model & $\ln(\mathcal{B})$ &
    PSR & Preferred model & $\ln(\mathcal{B})$ \\
    \hline
J0614$+$2229 & PL & 2.9 &
J0627$+$0706 & PL & 5.8 &
J0631$+$1036 & PL & $-$1.6 \\
J0659$+$1414 & PL+F2+PM & 21.0 &
J0729$-$1448 & PL & 1.3 &
J0742$-$2822 & PL & 3.7 \\
J0835$-$4510 & PL+F2 & 227.2 &
J0842$-$4851 & PL & 6.0 &
J0855$-$4644 & PL+F2 & 16.8 \\
J0901$-$4624 & PL+F2 & 122.2 &
J0908$-$4913 & PL+F2+PM & 11.8 &
J0940$-$5428 & PL+F2 & 49.9 \\
J1003$-$4747 & PL+PM & 15.3 &
J1015$-$5719 & PL+F2 & 3.6 &
J1016$-$5857 & PL+F2 & 49.7 \\
J1019$-$5749 & PL & 8.2 &
J1028$-$5819 & PL+F2 & 14.9 &
J1048$-$5832 & PL+F2 & 65.0 \\
J1052$-$5954 & PL & $-$2.7 &
J1055$-$6028 & PL & 1.7 &
J1057$-$5226 & PL+PM & 3.6 \\
J1105$-$6107 & PL & 1.3 &
J1112$-$6103 & PL+F2 & 10.3 &
J1138$-$6207 & PL & 5.0 \\
J1248$-$6344 & PL & 11.4 &
J1301$-$6305 & PL+F2 & 88.0 &
J1320$-$5359 & PL+F2+PM & 14.2 \\
J1327$-$6400 & PL & 4.4 &
J1341$-$6220 & PL & 5.3 &
J1357$-$6429 & PL+F2 & 190.5 \\
J1359$-$6038 & PL+PM & 0.8 &
J1406$-$6121 & PL & 5.8 &
J1410$-$6132 & PL+F2 & 8.2 \\
J1413$-$6141 & PL & $-$0.5 &
J1420$-$6048 & PL+F2+LFC & 255.5 &
J1452$-$6036 & PL & 9.0 \\
J1524$-$5625 & PL+F2 & 73.6 &
J1541$-$5535 & PL & $-$0.9 &
J1600$-$5044 & PL & 0.8 \\
J1614$-$5048 & PL+F2 & 56.1 &
J1617$-$5055 & PL+F2 & 19.8 &
J1626$-$4807 & PL & 40.7 \\
J1627$-$4706 & PL & 10.7 &
J1644$-$4559 & PL & $-$1.2 &
J1646$-$4346 & PL+F2 & 17.2 \\
J1650$-$4502 & PL & 2.8 &
J1701$-$4533 & PL & 89.1 &
J1702$-$4128 & PL+F2 & 8.8 \\
J1702$-$4310 & PL+F2 & 80.7 &
J1705$-$3950 & PL & 8.9 &
J1709$-$4429 & PL+F2+PM & 187.7 \\
J1716$-$4005 & PL & 9.7 &
J1718$-$3825 & PL+F2+PM & 57.8 &
J1721$-$3532 & PL & 0.4 \\
J1726$-$3530 & PL+F2 & 30.5 &
J1730$-$3350 & PL+F2 & 20.7 &
J1731$-$4744 & PL+F2 & 5.4 \\
J1734$-$3333 & PL+F2 & 29.6 &
J1737$-$3137 & PL+F2 & 5.5 &
J1740$-$3015 & PL & 3.0 \\
J1750$-$3157 & PL & 65.2 &
J1757$-$2421 & PL & 5.5 &
J1801$-$2304 & PL & 2.9 \\
J1801$-$2451 & PL+F2 & 345.5 &
J1803$-$2137 & PL+F2 & 102.6 &
J1825$-$0935 & PL & 7.9 \\
J1826$-$1334 & PL+F2+PM & 33.2 &
J1835$-$0643 & PL & 13.1 &
J1837$-$0604 & PL & $-$1.4 \\
J1841$-$0345 & PL & 55.0 &
J1841$-$0425 & PL+F2 & 14.5 &
J1841$-$0524 & PL & 7.4 \\
J1844$-$0256 & PL & 9.2 &
J1847$-$0402 & PL & 5.4 &\\
\hline
    \end{tabular}
\end{table*}

\begin{landscape}
\begin{table}
    \centering
    \caption{Inferred astrometric (RAJ, DECJ, $\mu_{\alpha}$, $\mu_{\delta}$) and rotational ($\nu$, $\dot{\nu}$, $\ddot{\nu}$) parameters for all 51 glitching pulsars in our sample. 
    All values are in reference to the MJD listed under PEPOCH. 
    Uncertainties in parentheses indicate the $68$\,per cent confidence intervals scaled to the last significant figure. Asymmetric confidence intervals are individually listed. Lower- and upper-limits on $\ddot{\nu}$, $\mu_{\alpha}$ or $\mu_{\delta}$ are given by two comma-separated values in parentheses. $^{\star}$Proper-motion fixed to value from \citet{Dodson2003}.}
    \label{tbl:params_glitch}
    \renewcommand{\arraystretch}{1.1}
    \resizebox{\linewidth}{!}{
    \begin{tabular}{lccccccccccc}
        \hline
        PSRJ & RAJ & DECJ & PEPOCH & $\nu$ & $\dot{\nu}$ & $\ddot{\nu}$ & $\mu_{\alpha}$ & $\mu_{\delta}$ & $N_{\mathrm{ToA}}$ & $T$ & MJD range \\
        
         & (hh:mm:ss) & ($\degr$:$\arcmin$:$\arcsec$) & (MJD) & (Hz) & ($10^{-14}$ s$^{-2}$) & ($10^{-24}$ s$^{-3}$) & (mas\,yr$^{-1}$) & (mas\,yr$^{-1}$) & & (yr) & (MJD) \\
        \hline
        J0631$+$1036 & $06$:$31$:$28(1)$ & $10$:$37$:$11(4)$ & $54750$ & $3.47463459(3)$ & $-126.249(9)$ & $(-15, 21)$ & $(-300, 700)$ & $(-600, 600)$ & $41$ & $3.7$ & $57165$--$58531$ \\
        J0729$-$1448 & $07$:$29$:$16(1)$ & $-14$:$48$:$40(2)$ & $55297$ & $3.97302972(3)$ & $-177.76(4)$ & $(1.1, 1.5)$ & $(-320, 220)$ & $(0, 700)$ & $172$ & $11.6$ & $54220$--$58469$ \\
        J0742$-$2822 & $07$:$42$:$49.0(3)$ & $-28$:$22$:$43.8(4)$ & $55352$ & $5.99624469(2)$ & $-60.44(2)$ & $(-2.3, 8.8)$ & $(-40, 110)$ & $(-290, -130)$ & $774$ & $15.0$ & $52988$--$58469$ \\
        J0835$-$4510 & $08$:$35$:$20.6^{+0.8}_{-0.7}$ & $-45$:$10$:$33.5^{+0.5}_{-0.6}$ & $56364$ & $11.1882333(4)$ & $-1545.8(6)$ & $950(5)$ & $-49.68(6)^{\star}$ & $29.9(1)^{\star}$ & $414$ & $11.5$ & $54260$--$58469$ \\
        J0901$-$4624 & $09$:$01$:$40.11(4)$ & $-46$:$24$:$48.45(3)$ & $55268$ & $2.262304515(1)$ & $-44.737(1)$ & $1.19(9)$ & $(-3, 7)$ & $(-3, 9)$ & $234$ & $20.9$ & $50849$--$58469$ \\
        J0908$-$4913 & $09$:$08$:$35.5(3)$ & $-49$:$13$:$06.4(2)$ & $55332$ & $9.36627103(2)$ & $-132.56(1)$ & $4.3(7)$ & $-37(9)$ & $31(10)$ & $375$ & $27.7$ & $48860$--$58972$ \\
        J0940$-$5428 & $09$:$40$:$58.3(2)$ & $-54:28:40.2(1)$ & $55335$ & $11.42109977(7)$ & $-427.41(7)$ & $47.7 \pm 2.8$ & $(-20, 12)$ & $(-13, 19)$ & $303$ & $22.9$ & $50849$--$59202$ \\
        J1015$-$5719 & $10$:$15$:$38.0(3)$ & $-57$:$19$:$12.1(2)$ & $55332$ & $7.147923371(5)$ & $-292.619(7)$ & $20(2)$ & $(-80, 0)$ & $(30, 130)$ & $154$ & $11.6$ & $54220$--$58469$ \\
        J1016$-$5857 & $10$:$16$:$21.3(5)$ & $-58$:$57$:$11.3(2)$ & $55369$ & $9.3105496(3)$ & $-695.5(2)$ & $121(5)$ & $(40, 120)$ & $(0, 80)$ & $352$ & $19.6$ & $51299$--$58469$ \\
        J1019$-$5749 & $10$:$19$:$52.1^{+0.2}_{-0.1}$ & $-57$:$49$:$06.22(7)$ & $55434$ & $6.153628421(6)$ & $-75.97(2)$ & $(3.1, 4.5)$ & $(-35, 17)$ & $(-51, 5)$ & $153$ & $11.4$ & $54302$--$58469$ \\
        J1028$-$5819 & $10$:$28$:$27.9(1)$ & $-58$:$19$:$06.21(6)$ & $55459$ & $10.94038305(1)$ & $-192.82(2)$ & $20(4)$ & $(-47, -9)$ & $(-36, 4)$ & $137$ & $10.7$ & $54563$--$58469$ \\
        J1048$-$5832 & $10$:$48$:$12.5(9)$ & $-58$:$32$:$04.2(5)$ & $55453$ & $8.083519(2)$ & $-617.9(5)$ & $152(9)$ & $(40, 140)$ & $(70, 190)$ & $523$ & $28.9$ & $47909$--$58469$ \\
        J1052$-$5954 & $10$:$52$:$38.1(1)$ & $-59$:$54$:$44.25^{+0.06}_{-0.07}$ & $55292$ & $5.53716823(3)$ & $-61.07(4)$ & $(13, 19)$ & $(-10, 70)$ & $(-110, 30)$ & $101$ & $6.8$ & $54220$--$56708$ \\
        J1055$-$6028 & $10$:$55$:$39.3(6)$ & $-60$:$28$:$35.5(3)$ & $55397$ & $10.033766397(7)$ & $-297.08(3)$ & $(6, 20)$ & $(-60, 140)$ & $(-30, 150)$ & $171$ & $10.9$ & $54505$--$58469$ \\
        J1105$-$6107 & $11$:$05$:$26.1(7)$ & $-61$:$07$:$49.7(3)$ & $55303$ & $15.8230056(9)$ & $-395.9(2)$ & $(-9, 9)$ & $(-100, 0)$ & $(50, 150)$ & $393$ & $23.5$ & $49868$--$58469$ \\
        J1112$-$6103 & $11$:$12$:$14.8^{+0.7}_{-0.6}$ & $-61$:$03$:$30.9(3)$ & $55456$ & $15.390826(2)$ & $-739.8(7)$ & $149(21)$ & $(-50, 70)$ & $(-80, 40)$ & $312$ & $20.9$ & $50849$--$58469$ \\
        J1248$-$6344 & $12$:$48$:$46.4(2)$ & $-63$:$44$:$09.37(7)$ & $55392$ & $5.0418245518(3)$ & $-42.999(2)$ & $(-2.1, 0.1)$ & $(-120, -40)$ & $(-110, -10)$ & $90$ & $6.8$ & $54219$--$56709$ \\
        J1301$-$6305 & $13$:$01$:$45.7(7)$ & $-63$:$05$:$34.5(3)$ & $55370$ & $5.4164292(8)$ & $-773.2(3)$ & $278(7)$ & $(0, 120)$ & $(-50, 90)$ & $274$ & $20.5$ & $50940$--$58444$ \\
        J1320$-$5359 & $13$:$20$:$53.92(2)$ & $-53$:$59$:$05.39(1)$ & $55408$ & $3.5747990403(6)$ & $-11.812(1)$ & $0.43(6)$ & $13(2)$ & $52(2)$ & $282$ & $21.7$ & $50536$--$58469$ \\
        J1341$-$6220 & $13$:$41$:$42^{+4}_{-5}$ & $-62$:$20$:$17^{+3}_{-2}$ & $56345$ & $5.1690544(9)$ & $-676.7(5)$ & $(1.3, 3.1)$ & $(-500, 500)$ & $(-400, 600)$ & $195$ & $11.6$ & $54220$--$58469$ \\
        J1357$-$6429 & $13$:$57$:$02.5^{+0.9}_{-1.0}$ & $-64$:$29$:$30.2(5)$ & $55000$ & $6.0178443(6)$ & $-1279.32(3)$ & $1039(14)$ & $(-100, 60)$ & $(-210, -10)$ & $293$ & $19.1$ & $51491$--$58469$ \\
        J1406$-$6121 & $14$:$06$:$49.9^{+0.3}_{-0.2}$ & $-61$:$21$:$27.8^{+0.1}_{-0.2}$ & $55390$ & $4.692757029(1)$ & $-120.535(6)$ & $(3.1, 7.5)$ & $(-50, 110)$ & $(-180, 20)$ & $97$ & $6.8$ & $54220$--$56708$ \\
        J1410$-$6132 & $14$:$10$:$22(1)$ & $-61$:$32$:$00.5^{+0.9}_{-0.8}$ & $55433$ & $19.9780635(5)$ & $-1266.0(6)$ & $180(26)$ & $(-300, 140)$ & $(-430, 130)$ & $155$ & $11.3$ & $54353$--$58469$ \\
        J1413$-$6141 & $14$:$13$:$10(2)$ & $-61$:$41$:$15(1)$ & $56011$ & $3.4995041(6)$ & $-408.2(2)$ & $(-30, 56)$ & $(-30, 390)$ & $(-230, 250)$ & $282$ & $20.9$ & $50849$--$58469$ \\
        J1420$-$6048 & $14$:$20$:$08.2(5)$ & $-60$:$48$:$17.5(3)$ & $55404$ & $14.661265(2)$ & $-1748.7(9)$ & $997(18)$ & $(-20, 60)$ & $(-80, 20)$ & $344$ & $19.5$ & $51333$--$58469$ \\
        J1452$-$6036 & $14$:$52$:$51.89(2)$ & $-60$:$36$:$31.37(2)$ & $55370$ & $6.4519531394(5)$ & $-6.0339(8)$ & $(-0.1, 0.5)$ & $-5(3)$ & $-5(3)$ & $150$ & $11.6$ & $54220$--$58469$ \\
        J1524$-$5625 & $15$:$24$:$49.82^{+0.03}_{-0.04}$ & $-56$:$25$:$24.07(3)$ & $55441$ & $12.78268099(1)$ & $-637.03(3)$ & $137(2)$ & $(-6, 12)$ & $(-10, 18)$ & $161$ & $11.6$ & $54220$--$58469$ \\
        J1614$-$5048 & $16$:$14$:$11^{+6}_{-7}$ & $-50:48:02(7)$ & $54359$ & $4.313262(5)$ & $-905(1)$ & $272(5)$ & $(-600, 200)$ & $(-600, 400)$ & $513$ & $27.8$ & $48329$--$58469$ \\
        J1617$-$5055 & $16$:$17$:$29(5)$ & $-50$:$55$:$11(5)$ & $54450$ & $14.4093408(3)$ & $-2837.3(9)$ & $2010(300)$ & $(-800, 400)$ & $(-600, 600)$ & $137$ & $6.8$ & $54220$--$56708$ \\
        J1644$-$4559 & $16$:$44$:$49.3(1)$ & $-45$:$59$:$09.8(2)$ & $57600$ & $2.197422934(5)$ & $-9.701(1)$ & $(0.27, 0.43)$ & $(-14, 6)$ & $(-38, 14)$ & $366$ & $29.1$ & $47913$--$58534$ \\
        J1646$-$4346 & $16$:$46$:$50(1)$ & $-43$:$45$:$53(2)$ & $55388$ & $4.3165103(1)$ & $-208.43(8)$ & $29(2)$ & $(-150, 30)$ & $(-230, 270)$ & $380$ & $28.9$ & $47912$--$58469$ \\
        J1650$-$4502 & $16$:$50$:$32.5^{+0.8}_{-0.9}$ & $-45$:$02$:$31(1)$ & $55389$ & $2.625528094(5)$ & $-11.124(9)$ & $(-1, 9)$ & $(-390, -30)$ & $(-800, 0)$ & $138$ & $11.6$ & $54220$--$58469$ \\
        J1702$-$4128 & $17$:$02$:$52.5(2)$ & $-41$:$28$:$48.2(5)$ & $55366$ & $5.489885905(1)$ & $-157.685(2)$ & $5.8(3)$ & $(-70, 30)$ & $(-250, 50)$ & $141$ & $11.6$ & $54220$--$58470$ \\
        J1702$-$4310 & $17$:$02$:$26.94(6)$ & $-43$:$10$:$41.5(1)7$ & $55375$ & $4.1563312(1)$ & $-385.36(6)$ & $49(2)$ & $(-26, -10)$ & $(-44, -2)$ & $216$ & $19.8$ & $51222$--$58470$ \\
        J1705$-$3950 & $17$:$05$:$29.8(4)$ & $-39$:$50$:$58(1)$ & $56462$ & $3.1351208446(5)$ & $-59.445(1)$ & $(-0.45, -0.11)$ & $(-300, -160)$ & $(-250, 250)$ & $158$ & $13.7$ & $53951$--$58972$\\
        J1709$-$4429 & $17$:$09$:$42.75(6)$ & $-44$:$29$:$08.3(1)$ & $55315$ & $9.755973(1)$ & $-869.3(3)$ & $273(4)$ & $17(4)$ & $11(10)$ & $441$ & $28.9$ & $47909$--$58470$ \\
        J1718$-$3825 & $17$:$18$:$13.558(7)$ & $-38$:$25$:$17.83(2)$ & $55374$ & $13.391416606(7)$ & $-236.311(5)$ & $21.3(2)$ & $-11(1)$ & $(-2, 6)$ & $250$ & $20.8$ & $50877$--$58470$ \\
    \end{tabular}
    }
    \renewcommand{\arraystretch}{}
\end{table}
\end{landscape}
\begin{landscape}
\begin{table}\
    \centering
    \contcaption{$^{\star}$Position fixed to values from \citet{Dexter2017} where POSEPOCH is MJD 57259. $^{\dagger}$Position and proper-motion fixed to values from \citet{Zeiger2008} where POSEPOCH is MJD 53348.}
    \renewcommand{\arraystretch}{1.1}
    \resizebox{\linewidth}{!}{
    \begin{tabular}{lccccccccccc}
        \hline
        PSRJ & RAJ & DECJ & PEPOCH & $\nu$ & $\dot{\nu}$ & $\ddot{\nu}$ & $\mu_{\alpha}$ & $\mu_{\delta}$ & $N_{\mathrm{ToA}}$ & $T$ & MJD range \\
        
         & (hh:mm:ss) & ($\degr$:$\arcmin$:$\arcsec$) & (MJD) & (Hz) & ($10^{-15}$ s$^{-2}$) & ($10^{-24}$ s$^{-3}$) & (mas\,yr$^{-1}$) & (mas\,yr$^{-1}$) & & (yr) & (MJD) \\
        \hline
        J1730$-$3350 & $17$:$30$:$32.4(5)$ & $-33$:$50$:$34(2)$ & $55400$ & $7.1685489(5)$ & $-433.8(1)$ & $54(2)$ & $(-200, -80)$ & $(-820, -180)$ & $274$ & $21.3$ & $50538$--$58321$ \\
        J1731$-$4744 & $17$:$31$:$42.2(1)$ & $-47$:$44$:$37.1(2)$ & $55402$ & $1.20497647(3)$ & $-23.622(8)$ & $2.5(1)$ & $60(11)$ & $-178(24)$ & $317$ & $25.8$ & $49043$--$58470$ \\
        J1734$-$3333 & $17$:$34$:$27(2)$ & $-33$:$33$:$31^{+8}_{-7}$ & $55341$ & $0.854866535(3)$ & $-166.63(1)$ & $3.9(6)$ & $(-400, 400)$ & $(-500, 700)$ & $173$ & $16.4$ & $50686$--$56672$ \\
        J1737$-$3137 & $17$:$37$:$04.3(2)$ & $-31$:$37$:$26.4^{+1.0}_{-0.9}$ & $55292$ & $2.21984993(1)$ & $-68.30(2)$ & $3.1(2)$ & $(20, 100)$ & $(190, 710)$ & $166$ & $13.0$ & $54220$--$58972$ \\
        J1740$-$3015 & $17$:$40$:$33.8^{+0.5}_{-0.4}$ & $-30$:$15$:$45(3)$ & $55472$ & $1.64767845(5)$ & $-126.57(5)$ & $(8, 20)$ & $(-140, 120)$ & $(180, 820)$ & $195$ & $11.6$ & $54220$--$58470$ \\
        J1757$-$2421 & $17$:$57$:$29.33(3)$ & $-24$:$22$:$04(1)$ & $55433$ & $4.2715558534(8)$ & $-23.570(1)$ & $(0.24, 0.76)$ & $-$ & $-$ & $155$ & $12.3$ & $53974$--$58470$ \\
        J1801$-$2304 & $18$:$01$:$19.8149(6)^{\star}$ & $-23$:$04$:$44.63(1)^{\star}$ & $57259$ & $2.40466187(2)$ & $-652.95(9)$ & $(2, 4)$ & $(-33, 21)$ & $(-825, 625)$ & $157$ & $12.3$ & $53974$--$58470$ \\
        J1801$-$2451 & $18$:$01$:$00.016(8)^{\dagger}$ & $-24$:$51$:$27.5(2)^{\dagger}$ & $55297$ & $8.002151(1)$ & $-7979(4)$ & $304(7)$ & $-11(9)^{\star}$ & $-1(15)^{\star}$ & $156$ & $26.2$ & $48896$--$58470$ \\
        J1803$-$2137 & $18$:$03$:$51.4(2)$ & $-21$:$37$:$03(5)$ & $55430$ & $7.48029082(3)$ & $-748.63(9)$ & $239(6)$ & $(20, 100)$ & $(-900, 500)$ & $151$ & $11.6$ & $54220$--$58470$ \\
        J1825$-$0935 & $18$:$25$:$30.6(4)$ & $-09$:$35$:$21(2)$ & $56550$ & $1.30036246(9)$ & $-8.88(3)$ & $(-0.8, 2.2)$ & $(-180, -40)$ & $(-990, -490)$ & $179$ & $18.3$ & $51844$--$58533$ \\
        J1826$-$1334 & $18$:$26$:$13.19(6)$ & $-13$:$34$:$46.5^{+0.4}_{-0.3}$ & $54286$ & $9.85345601(3)$ & $-730.50(3)$ & $172(8)$ & $32(9)$ & $(-80, 40)$ & $141$ & $11.5$ & $54220$--$58404$ \\
        J1837$-$0604 & $18$:$37$:$43.4(6)$ & $-06$:$04$:$49(2)$ & $55558$ & $10.38323993(8)$ & $-485.9(1)$ & $(-40, 240)$ & $(-310, 230)$ & $(-1000, 0)$ & $106$ & $7.5$ & $53968$--$56708$ \\
        J1841$-$0345 & $18$:$41$:$38.7^{+0.2}_{-0.1}$ & $-03$:$48$:$43.8^{+0.5}_{-0.6}$ & $54867$ & $4.899957507(7)$ & $-139.000(3)$ & $(5, 15)$ & $(-520, 20)$ & $(-500, 300)$ & $45$ & $3.7$ & $57165$--$58531$ \\
        J1841$-$0524 & $18$:$41$:$49.3(1)$ & $-05$:$24$:$30.4(3)$ & $55362$ & $2.24311056(1)$ & $-117.55(1)$ & $(-1, 9)$ & $(-60, 80)$ & $(-340, 200)$ & $91$ & $6.7$ & $54268$--$56708$ \\
        J1847$-$0402 & $18$:$47$:$22.84(5)$ & $-04$:$02$:$14.6^{+0.1}_{-0.2}$ & $55448$ & $1.6728026921(1)$ & $-14.4652(2)$ & $(0.08, 1.2)$ & $(-6, 0)$ & $(2, 18)$ & $135$ & $11.5$ & $54268$--$58470$ \\
        \hline
    \end{tabular}
    }
    \renewcommand{\arraystretch}{}
\end{table}
\end{landscape}
\begin{landscape}
\begin{table}
    \centering
    \caption{Astrometric and rotational parameters for the 23 non-glitching pulsars in our sample. Asymmetric confidence intervals are individually listed. Lower- and upper-limits on $\ddot{\nu}$, $\mu_{\alpha}$ or $\mu_{\delta}$ are given by two comma-separated values in parentheses. $^{\star}$Position and proper-motion fixed to values from very-long baseline interferometry~\citep{Deller2019}.}
    \label{tbl:params_noglitch}
    \renewcommand{\arraystretch}{1.1}
    \resizebox{\linewidth}{!}{
    \begin{tabular}{lccccccccccc}
        \hline
        PSRJ & RAJ & DECJ & PEPOCH & $\nu$ & $\dot{\nu}$ & $\ddot{\nu}$ & $\mu_{\alpha}$ & $\mu_{\delta}$ & $N_{\mathrm{ToA}}$ & $T$ & MJD range \\
        
         & (hh:mm:ss) & ($\degr$:$\arcmin$:$\arcsec$) & (MJD) & (Hz) & ($10^{-15}$ s$^{-2}$) & ($10^{-24}$ s$^{-3}$) & (mas\,yr$^{-1}$) & (mas\,yr$^{-1}$) & & (yr) & (MJD) \\
        \hline
        J0614$+$2229 & $06$:$14$:$17.0058(1)^{\star}$ & $22$:$29$:$56.848(1)^{\star}$ & $56000$ & $2.9851428242(6)$ & $-526.51(1)$ & $(0, 5)$ & $-0.23(5)^{\star}$ & $-1.22(7)^{\star}$ & $125$ & $10.9$ & $54505$--$58469$ \\
        J0627$+$0706 & $06$:$27$:$44.17(4)$ & $07$:$06$:$33.4(2)$ & $55382$ & $2.1013735312(3)$ & $-13.155(3)$ & $(-0.5, 0.1)$ & $(-21, 8)$ & $(14, 115)$ & $126$ & $10.7$ & $54548$--$58470$ \\
        J0659$+$1414 & $06$:$59$:$48.19(5)$ & $14$:$14$:$21.3(3)$ & $55586$ & $2.59794880620(3)$ & $-37.09365(8)$ & $0.68(2)$ & $49(13)$ & $78^{+75}_{-68}$ & $129$ & $10.9$ & $54505$--$58469$ \\
        J0842$-$4851 & $08$:$42$:$05.33(5)$ & $-48$:$51$:$20.65(3)$ & $56022$ & $1.55191806425(5)$ & $-2.30338(6)$ & $(-0.46, -0.14)$ & $(-13, 15)$ & $(-21, 9)$ & $104$ & $8.7$ & $55363$--$58531$ \\
        J0855$-$4644 & $08$:$55$:$36.16(2)$ & $-46$:$44$:$13.46(1)$ & $55288$ & $15.4587515567(1)$ & $-173.5507(4)$ & $1.53(7)$ & $(-7, 1)$ & $(-3, 5)$ & $118$ & $11.6$ & $54220$--$58469$ \\
        J1003$-$4747 & $10$:$03$:$21.54(2)$ & $-47$:$47$:$01.40(1)$ & $55344$ & $3.25654603313(2)$ & $-2.19707(2)$ & $0.04(2)$ & $-12(2)$ & $21(2)$ & $199$ & $12.3$ & $53973$--$58469$ \\
        J1057$-$5226 & $10$:$57$:$59.012^{+0.10}_{-0.09}$ & $-52$:$26$:$56.49(7)$ & $55311$ & $5.07321840599(8)$ & $-15.01619(6)$ & $(-0.28,-0.02)$ & $49(4)$ & $-6(5)$ & $397$ & $26.4$ & $48814$--$58469$ \\
        J1138$-$6207 & $11$:$38$:$21.7(3)$ & $-62$:$07$:$59.0(1)$ & $55355$ & $8.5056937000(1)$ & $-90.303(1)$ & $(0.1, 0.4)$ & $(-160, -40)$ & $(20, 140)$ & $89$ & $6.8$ & $54220$--$56708$ \\
        J1327$-$6400 & $13$:$27$:$10(2)$ & $-64$:$00$:$13.2(9)$ & $55411$ & $3.56265606571(9)$ & $-40.327(2)$ & $(-5, 11)$ & $(-17,442)$ & $(61,550)$ & $86$ & $6.8$ & $54220$--$56708$ \\
        J1359$-$6038 & $13$:$59$:$59.33(1)$ & $-60$:$38$:$17.998^{+0.002}_{-0.001}$ & $55426$ & $7.84268967507968(1)$ & $-38.97175768(1)$ & $(0.14,0.46)$ & $-4(4)$ & $10(5)$ & $808$ & $28.9$ & $47913$--$58469$ \\
        J1541$-$5535 & $15$:$41$:$47(1)$ & $-55$:$34$:$07.5^{+0.9}_{-1.0}$ & $55374$ & $3.3799534494(6)$ & $-84.0607(8)$ & $(6, 13)$ & $(-370, 10)$ & $(-510, 70)$ & $146$ & $11.6$ & $54220$--$58469$ \\
        J1600$-$5044 & $16$:$00$:$53.03(3)$ & $-50$:$44$:$21.0153^{+0.03}_{-0.04}$ & $57600$ & $5.19197591041(8)$ & $-13.64417(5)$ & $(0.9, 0.39)$ & $(-5, 3)$ & $(-5, 11)$ & $185$ & $14.9$ & $53040$--$58469$ \\
        J1626$-$4807 & $16$:$26$:$42.5(7)$ & $-48$:$07$:$56(1)$ & $55292$ & $3.4021201284(3)$ & $-20.2284(9)$ & $(-9, 11)$ & $-$ & $-$ & $94$ & $8.0$ & $54220$--$57129$ \\
        J1627$-$4706 & $16$:$27$:$28.79^{+0.08}_{-0.07}$ & $-47$:$06$:$49.3(1)$ & $53165$ & $7.1050074462(2)$ & $-8.7355(2)$ & $(-0.14, -0.02)$ & $(-7, 33)$ & $(-40, 60)$ & $134$ & $10.7$ & $52807$--$56708$ \\
        J1701$-$4533 & $17$:$01$:$29.13(2)$ & $-45$:$33$:$49.18(4)$ & $48360$ & $3.0968518767(3)$ & $-0.49801(5)$ & $(-5.5, -0.7)$ & $-$ & $-$ & $56$ & $5.1$ & $56682$--$58531$ \\
        J1716$-$4005 & $17$:$16$:$42.0(5)$ & $-40$:$05$:$27(1)$ & $54942$ & $3.207052805(4)$ & $-2.921(1)$ & $(-13, -5)$ & $-$ & $-$ & $38$ & $3.4$ & $57276$--$58531$ \\
        J1721$-$3532 & $17$:$21$:$32.76(7)$ & $-35$:$32$:$48.3(3)$ & $55456$ & $3.56591744853(3)$ & $-32.0185(1)$ & $(0.22, 0.34)$ & $(-14, 10)$ & $(-60, 60)$ & $252$ & $18.0$ & $51879$--$58470$ \\
        J1726$-$3530 & $17$:$26$:$07.6(1)$ & $-35$:$29$:$58(4)$ & $55432$ & $0.900460470(1)$ & $-98.25(3)$ & $20(2)$ & $(-300, 60)$ & $-$ & $201$ & $16.5$ & $50681$--$56709$ \\
        J1750$-$3157 & $17$:$50$:$47.31(2)$ & $-31$:$57$:$44.3(1)$ & $50271$ & $1.09846293993(2)$ & $-0.023700(4)$ & $(-1,-0.8)$ & $-$ & $-$ & $42$ & $12.4$ & $53974$--$58500$ \\
        J1835$-$0643 & $18$:$35$:$05.56(8)$ & $-06$:$43$:$06.9(3)$ & $55365$ & $3.2695714904(2)$ & $-43.2078(2)$ & $(1.2, 2.0)$ & $(-22, 24)$ & $(-120, 40)$ & $113$ & $11.3$ & $54268$--$58404$ \\
        J1841$-$0425 & $18$:$41$:$05.7(1)$ & $-04$:$25$:$20.3(5)$ & $55402$ & $5.3720208212(3)$ & $-18.459(1)$ & $1.2(1)$ & $(-21, 6)$ & $(0, 50)$ & $139$ & $11.5$ & $54268$--$58470$ \\
        J1844$-$0256 & $18$:$44$:$30.1(3)$ & $-02$:$54$:$09.7^{+0.8}_{-0.7}$ & $54919$ & $3.66337798586(4)$ & $-38.3135(2)$ & $(-0.7,0.1)$ & $-$ & $-$ & $61$ & $4.3$ & $54268$--$55822$ \\
        \hline
    \end{tabular}
    }
    \renewcommand{\arraystretch}{}
\end{table}


\bsp	
\label{lastpage}
\end{landscape}
\end{document}